\documentclass[useAMS,usenatbib]{mn2e}
\usepackage{graphicx,url,multirow,mathrsfs,amssymb,amsmath,color,textcomp}
\def\fa{Fe K$\alpha$~}
\def\rg{$r_{\rm g}$}
\def\tg{$t_{\rm g}$}
\def\ms{M$_{\rm \sun}$}
\def\apj{ApJ}%
\def\apjl{ApJ}%
\def\apjs{ApJS}%
\def\aap{A\&A}%
\def\mnras{MNRAS}%
\def\pasj{PASJ}%
\def\ssr{Space~Sci.~Rev.}%
\def\nat{Nature}%
\usepackage[normalem]{ulem}
\newcommand{\eqb}{\begin{eqnarray}}
\newcommand{\eqe}{\end{eqnarray}}
\makeatletter
\newtoks\FTN@ftn
\def\pushftn{%
\let\@footnotetext\FTN@ftntext\let\@xfootnotenext\FTN@xftntext
\let\@xfootnote\FTN@xfootnote}
\def\popftn{%
\global\FTN@ftn\expandafter{\expandafter}\the\FTN@ftn}
\long\def\FTN@ftntext#1{%
\edef\@tempa{\the\FTN@ftn\noexpand\footnotetext[\the\csname c@\@mpfn\endcsname]}%
\global\FTN@ftn\expandafter{\@tempa{#1}}}%
\long\def\FTN@xftntext[#1]#2{%
\global\FTN@ftn\expandafter{\the\FTN@ftn\footnotetext[#1]{#2}}}
\def\FTN@xfootnote[#1]{%
\begingroup
\csname c@\@mpfn\endcsname #1\relax
\unrestored@protected@xdef\@thefnmark{\thempfn}%
\endgroup
\@footnotemark\FTN@xftntext[#1]}
\makeatother
\title[GR modelling of NXTL]{General relativistic modelling of the negative reverberation X-ray time delays in AGN\thanks{Based on observations obtained with \textit{XMM}-\textit{Newton}, an ESA science mission with instruments and contributions directly funded by ESA Member States and NASA.}}
\author[D.~Emmanoulopoulos]{D.~Emmanoulopoulos,$^{1}$\thanks{E-mail: D.Emmanoulopoulos@soton.ac.uk} I.~E.~Papadakis,$^{2,3}$ M.~Dov\v{c}iak$^{4}$ and I.~M.~M\textsuperscript{c}Hardy$^{1}$\\
$^{1}$Physics and Astronomy, University of Southampton, Southampton, SO17 1BJ, UK\\
$^{2}$Department of Physics and Institute of Theoretical and Computational Physics, University of Crete, 71003 Heraklion, Greece\\
$^{3}$IESL, Foundation for Research and Technology, 71110 Heraklion, Greece\\
$^{4}$Astronomical Institute of the Academy of Sciences, Bo\v{c}n\'{\i} II 1401, CZ-14100 Praha 4, Czech Republic}
\begin{document}

\date{Accepted 2014 February 4. Received 2014 January 28; in original form 2013 December 17}
\pagerange{\pageref{firstpage}--\pageref{lastpage}} \pubyear{2002}
\maketitle

\label{firstpage}
\begin{abstract}
We present the first systematic physical modelling of the time-lag spectra between the soft (0.3--1 keV) and the hard (1.5--4 keV) X-ray energy bands, as a function of Fourier frequency, in a sample of 12 active galactic nuclei which have been observed by \textit{XMM}-\textit{Newton}. We concentrate particularly on the negative X-ray time-lags (typically seen above $10^{-4}$ Hz) i.e.\ soft band variations lag the hard band variations, and we assume that they are produced by reprocessing and reflection by the accretion disc within a lamp-post X-ray source geometry. We also assume that the response of the accretion disc, in the soft X-ray bands, is adequately described by the response in the neutral \fa line at 6.4 keV for which we use fully general relativistic ray-tracing simulations to determine its time evolution. These response functions, and thus the corresponding time-lag spectra, yield much more realistic results than the commonly-used, but erroneous, top-hat models. Additionally we parametrize the 
positive part of the time-lag spectra (typically seen below $10^{-4}$ Hz) by a power-law. We find that the best-fitting BH masses, $M$, agree quite well with those derived by other methods, thus providing us with a new tool for BH mass determination. We find no evidence for any correlation between $M$ and the BH spin parameter, $\alpha$, the viewing angle, $\theta$, or the height of the X-ray source above the disc, $h$. Also on average, the X-ray source lies only around 3.7 gravitational radii above the accretion disc and $\theta$ is distributed uniformly between 20 and 60\degr. Finally, there is a tentative indication that the distribution of $\alpha$ may be bimodal above and below 0.62. 

\end{abstract}


\section{INTRODUCTION}
\label{sect:intro}
In the current paradigm, active galactic nuclei (AGN) contain a central black hole (BH) which is fed, in most cases, by an optically thick and geometrically thin accretion disc as a result of matter transportation inwards and angular momentum outwards. This disc radiates as a series of black-body components \citep{shakura73} with peak emission at optical-ultraviolet wavelengths \citep{malkan83}. Part of this radiation is assumed to be Compton up-scattered within a mildly relativistic hot electron cloud, often called the \textit{X-ray corona}. This medium is often approximated as a point source (representing the centroid of the X-ray emitting source), lying above the central BH, on the axis of symmetry of the system (i.e.\ BH spin axis). This arrangement is known as the `\textit{lamp-post geometry}'.\par
The photons, from the X-ray source, form a power-law spectrum, and depending on the location of the X-ray source, a substantial part of them may illuminate the accretion disc. In this case, they are either Compton scatted by free or bound electrons, or photoelectrically absorbed followed by fluorescent line emission, or by Auger de-excitation. This yields the so-called `reflection spectrum' consisting of a number of emission and absorption lines (mainly below 1 keV), together with the 6.4 keV \fa emission line from neutral material, which is the strongest X-ray spectral feature \citep{george91}.\par
Depending on the proximity of the reflection process to the BH, the various lines can undergo relativistic broadening \citep{fabian89,laor91}. These relativistically broadened lines may account for the observed \textit{soft X-ray excess} \citep{crummy06}, where the X-ray spectrum below 1 keV (soft band) lie above the power-law extrapolation of the continuum, usually measured in the 1.5--4 keV (hard band). However, alternative interpretations for the soft excess are also possible e.g.\ Comptonisation of disc photons from the X-ray source \citep{page04}, Comptonisation of disc photons within a hot layer on the accretion disc \citep{wang03,done12}.\par
In addition to the primary and the reflected components, several AGN exhibit absorption features in their X-ray spectra as well. These feature can arise by: i) outflows which are either thermally or radiatively or magnetically driven winds \citep{cappi06} and can reach mildly relativistic speeds \citep{tombesi10}, ii) inflows \citep{krug10}, and iii) X-ray absorption clouds \citep{risaliti02}. The position of these structures varies from tens up to thousands of gravitational radii\footnote{For a BH mass of $2\times10^{6}$ \ms\ photons need $t_{\rm g}=r_{\rm g}/c=9.8$ s to travel a distance of 1 \rg.}, $r_{\rm g}=G M_{\rm BH}/c^2$.\par    
In this framework, and taking advantage of the highly variable behaviour of AGN, detection of time delays, as a function of the Fourier frequency, between the soft band and the hard band photons, can shed light on the X-ray emission mechanism and the geometry of these systems. Currently the observed negative X-ray time-lags (i.e.\ soft-band variations lagging the hard-band variations; NXTL hereafter), has triggered a great deal of scientific interest on interpreting their nature. After their first tentative detection in the AGN Ark\;564 \citep{mchardy07}, where an origin in reflection from the accretion disc was first proposed, the first statistical significant detection came from \citet{fabian09} for the AGN 1H\;0707-495. Then, \citet{emmanoulopoulos11b} found that this time delayed X-ray behaviour is much more common than was initially thought by analysing the data sets from two widely studied bright AGN, MCG--6-30-15 and Mrk\;766. Subsequently, \citet{demarco13} in a systematic analysis of 32 AGN, found a 
total of 15 AGN  exhibiting NXTLs with high statistical confidence.\par
The opposite time delayed behaviour i.e.\ positive X-ray time delays (hard-band variations lag the soft-band variations), has been known for quite some time in both AGN \citep[e.g.][]{papadakis01,mchardy04,arevalo06,sriram09} and X-ray binaries \citep[XRBs; e.g.][]{miyamoto89,nowak96,nowak99}. Although positive time-lags are expected in the standard Comptonisation process within the X-ray source \citep{nowak99}, they can also be produced by diffusive propagation of perturbations in the accretion flow \citep{kotov01}.  \par
The origin of NXTL still remains unclear. Assuming a lamp-post geometry, in which the reflection of the hard X-rays occurs very close to the BH (few \rg), the NXTL arise from the difference in the path-lengths between the soft (reflected photons) and the hard X-ray photons (coming directly from the X-ray source) to the observer \citep{fabian09,zoghbi10,zoghbi11,cackett13,fabian13}. On the other hand  based on the large-scale distant scattering scenario \citep{miller10_1h0707,legg12}, NXTLs might be caused by scattering of X-rays as they pass through, or are scattered from, a distant (tens up to hundred \rg) absorbing clumpy medium (e.g.\ wind, outflow, cloud), that partially covers the X-ray source, and whose opacity decreases with increasing energy.\par
Currently modelling of NXTLs is usually done in terms of simple top-hat impulse response functions \citep[THIRF, hereafter;][]{miller10_1h0707,zoghbi11,emmanoulopoulos11b}. This approach is, however, just a parametrisation of the NXTL spectra and does not carry any physical information about the geometry and the physical properties of the BH (i.e.\ mass, spin). In order to properly model the impulse response function, one has to take into account in detail the various general relativistic (GR) effects which affect the geometric path of both the hard and reflected X-rays. A first attempt at such a modelling of NXTLs was performed by \citet{chainakun12} for the AGN 1H\;0707-495 using a GR light-bending model and a moving X-ray source in order to take account of the observed X-ray variability. Their conclusion was that a more complex physical model is required in order to explain both the source's geometry and intrinsic variability. More recently, \citet{wilkins13} using the full GR treatment with a variety 
of different geometries of corona, for the same AGN 1H\;0707-495, found the the X-ray source  extends radially outwards to around 35 \rg\ and at a height of around 2 \rg\ above the plane of the accretion disc and that propagating fluctuations might account for the positive low frequency time delays.\par
In this paper we perform the first systematic analysis of the time-lag spectra of a sample of 12 AGN using fully general relativistic impulse response functions (GRIRFs, hereafter). These functions are generated using a lamp-post model with variable BH mass, BH spin parameter, viewing angle and height of the X-ray source above the disc. The paper is organised as follows: Initially, in Section~\ref{sect:obs} we present the observations and data reduction procedures. Then, in Section~\ref{sect:lampPost_irf} we describe the lamp-post model and the method that we use to derive the GRIRFs. In the next section, we outline the procedure for the estimation of the time-lag spectra models from the corresponding GRIRFs, and in Section~\ref{sect:fits} we describe the fitting methodology. Section~\ref{sect:res} contains the results of the best-fitting time-lag spectral models and finally in the last section we give a summary of our results and we discuss our conclusions. Throughout the paper the error estimates for the 
various 
physical 
parameters correspond to the 68.3 per cent confidence intervals unless otherwise stated. Similarly, the error bars of the plot points in all the figures indicate the 68.3 per cent confidence intervals.

\section{OBSERVATIONS AND DATA REDUCTION}
\label{sect:obs}
\begin{table*}
\begin{minipage}{180mm}
\caption{\textit{XMM}-\textit{Newton} observations. The first column, (1), gives the names of the AGN and their BH masses. Each BH mass estimate comes with a footnote (at the end of the table) indicating the corresponding literature reference. Those BH masses that have been estimated using the reverberation mapping technique are followed by the indication (r). The second column, (2), gives the \textit{XMM}-\textit{Newton} observation IDs together with the corresponding observing mode of the EPIC-pn camera: small window (sw), large window (lw) and full window (fw) mode. The third column, (3), gives the net exposure time of the observations i.e\ duration after background subtraction and data screening.} \vspace{0em}
\label{tab:obs}
\begin{tabular}{@{}cccc}
\hline
(1) & (2) & (3) \\
\underline{AGN name} & Obs ID  (PN mode)    & Net exp. \\
 BH Mass  ($\times 10^6$ \ms)    &    &       (ks)    \\
\hline
\underline{NGC\;4395}   \\
$0.36\pm0.11$\footnote{\citet{peterson05}} (r)  &  0142830101 (fw) & 108.7& \\
\underline{NGC\;4051}  \\
$1.73^{+0.55}_{-0.52}$\footnote{\label{tblfn:denney10}{\citet{denney10}}} (r)  &  0109141401 (sw)  &  117.0 &  \\
                       &  0157560101 (lw)  &  50.0  &	  \\
			&  0606320101 (sw)  &  45.3  &	  \\
			&  0606320201 (sw)  &  44.4  &	  \\
			&  0606320301 (sw)  &  31.3  &   \\
			&  0606320401 (sw)  &  28.8  &   \\
			&  0606321301 (sw)  &  30.1  &   \\
			&  0606321401 (sw)  &  39.2  &   \\
			&  0606321501 (sw)  &  38.8  &   \\
			&  0606321601 (sw)  &  41.5  &   \\
			&  0606321701 (sw)  &  38.3  &   \\
			&  0606321801 (sw)  &  39.9  &   \\
			&  0606321901 (sw)  &  6.2   &   \\
			&  0606322001 (sw)  &  36.9  &   \\
			&  0606322101 (sw)  &  37.7  &   \\
			&  0606322201 (sw)  &  41.1  &   \\
			&  0606322301 (sw)  &  42.3  &   \\
\underline{Mrk\;766}	         \\
$1.76^{+1.56}_{-1.40}$\footnote{\citet{bentz09}} (r) &  0304030101 (sw) &   95.1  &   \\
                       &  0109141301 (sw)  &   128.6   &   \\
                       &  0304030301 (sw)  &   98.5    &   \\
                       &  0304030401 (sw)  &   98.5    &   \\
                       &  0304030501 (sw)  &   95.1    &   \\
                       &  0304030601 (sw)  &   98.5    &   \\
                       &  0304030701 (sw)  &   34.6    &   \\
\underline{MCG--6-30-15}   \\
$2.14\pm0.36$\footnote{Estimated from equation 3 in \citet{gultekin09} using the stellar velocity dispersion value of \citet{mchardy05}.}	   & 0111570101 (sw)  &  43.2   &  \\
                       & 0111570201 (sw)   &   55.0    &   \\
                       & 0029740101 (sw)   &   83.5    &   \\
                       & 0029740701 (sw)   &   127.4   &   \\
                       & 0029740801 (sw)   &   125.0   &   \\
\underline{Ark\;564}      \\
$2.32\pm0.41$\footnote{Estimated from equation 5 in \citet{vestergaard06} using the mean values of ${\rm FWHM(H_\beta)}$ and $\lambda{\rm L}_\lambda(5100$ \AA$)$ of \citet{romano04}.}     & 0006810101 (sw)   &   10.6   &    \\ 
                       & 0206400101 (sw)   &   98.9   &    \\
                       & 0670130201 (sw)   &   59.1   &    \\
                       & 0670130301 (sw)   &   55.5   &    \\
                       & 0670130401 (sw)   &   62.5   &    \\
                       & 0670130501 (sw)   &   66.9   &    \\
                       & 0670130601 (sw)   &   60.5   &    \\
                       & 0670130701 (sw)   &   55.3   &    \\
                       & 0670130801 (sw)   &   57.8   &    \\
                       & 0670130901 (sw)   &   55.5   &    \\
\hline
\end{tabular}
\hspace*{0em}\parbox{1\linewidth}{\vspace*{-4.91em}\par
\begin{tabular}{@{}cccc}
\hline
(1) & (2) & (3) \\
Name & Obs ID  (PN mode)    & Net exp. \\
BH Mass  ($\times 10^6$ \ms) &                      &  (ks)    \\
\hline
\underline{1H\;0707-495}    \\
$2.34\pm0.71$\footnote{\label{tblfn:zhou05}\citet{zhou05}}	   &  0110890201 (fw) & 40.7  &   \\       
                       &  0148010301 (fw)   &   78.0   &   \\
                       &  0506200201 (lw)   &   38.7   &   \\     
                       &  0506200301 (lw)   &   38.7   &   \\
                       &  0506200401 (lw)   &   40.6   &   \\
                       &  0506200501 (lw)   &   40.9   &   \\
                       &  0511580101 (lw)   &   121.6  &   \\   
                       &  0511580201 (lw)   &   102.1  &   \\   
                       &  0511580301 (lw)   &   104.2  &   \\    
                       &  0511580401 (lw)   &   101.8  &   \\
                       &  0554710801 (lw)   &   96.1   &   \\
                       &  0653510301 (lw)   &   113.9  &   \\   
                       &  0653510401 (lw)   &   125.8  &   \\   
                       &  0653510501 (lw)   &   117.0  &   \\
\underline{IRAS\;13224-3809}  \\
$5.75\pm0.82$\textsuperscript{\small\ref{tblfn:zhou05}} &  0110890101 (fw) & 60.9    &     \\
                       &  0673580101 (lw)   &  126.1   &   \\
                       &  0673580201 (lw)   &  125.1   &   \\
                       &  0673580301 (lw)   &  125.0   &   \\
                       &  0673580401 (lw)   &  127.5   &   \\
\underline{ESO\;113-G010} \\               
$6.96\pm0.24$\footnote{Estimated from equation 5 in \citet{vestergaard06} using the mean values of 
${\rm FWHM(H_\beta)}$ and the $\lambda{\rm L}_\lambda(5100$ \AA$)$ estimated by \citet{cackett13} using data of \citet{pietsch98}.}&  0301890101 (fw)  &  92.6 &  \\		          
\underline{NGC\;7469}          \\
$12.2\pm1.4$\footnote{\citet{peterson04}} (r)	& 0112170101 (sw)    & 17.6   & \\
                       &  0112170301 (sw)   &   23.1   &   \\
                       &  0207090101 (sw)   &   84.6   &   \\
                       &  0207090201 (sw)   &   78.7   &   \\
\underline{Mrk\;335}     \\   
$26\pm8$\footnote{\citet{grier12}}	 &  0306870101 (sw)  &  132.8 &  \\
                       &  0600540501 (fw)   &   80.7   &   \\
                       &  0600540601 (fw)   &   130.3  &   \\
\underline{NGC\;3516} \\
$31.7^{+2.8}_{-4.2}$\textsuperscript{\small\ref{tblfn:denney10}} (r) & 0107460601 (sw) & 79.3 & \\
                       &  0107460701 (sw)   &   120.5  & \\
                       &  0401210401 (sw)   &   51.7   & \\
                       &  0401210501 (sw)   &   62.6   & \\
                       &  0401210601 (sw)   &   61.6   & \\
                       &  0401211001 (sw)   &   42.2   & \\
\underline{NGC\;5548} & & &\\
$44.2^{+9.9}_{-13.8}$\textsuperscript{\small\ref{tblfn:denney10}} (r) & 0089960301 (sw)&  85.2 & \\
\end{tabular}}\vspace{-1.5em}
\end{minipage}
\end{table*}

\begin{figure*}
\hspace*{10em}\parbox{1\linewidth}{\vspace*{4em}\includegraphics[trim = 20mm 100mm 20mm 100mm, clip, width=10cm]{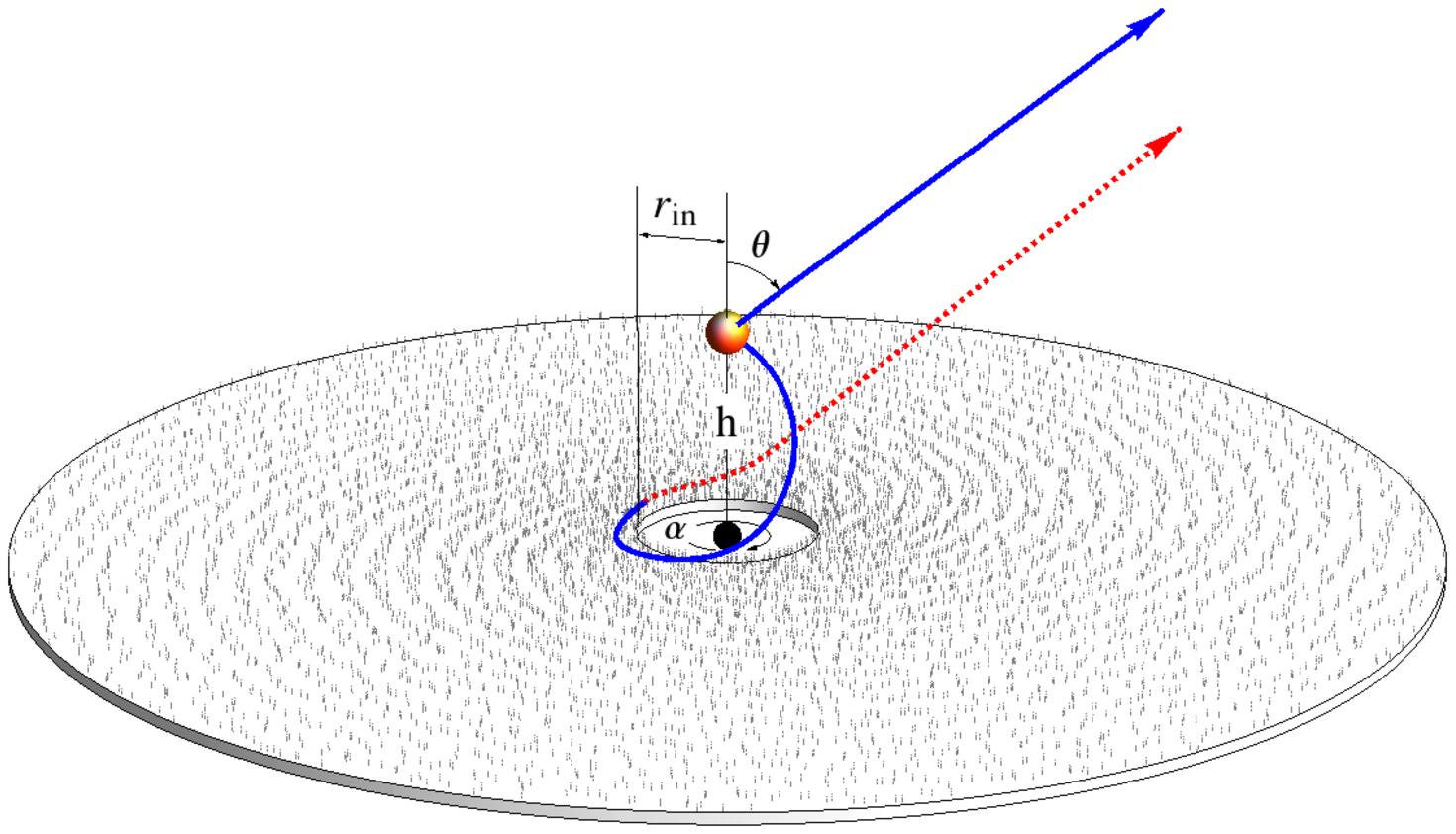}}\\
\hspace*{35em}\parbox{1\linewidth}{\vspace*{-37em}\includegraphics[trim = -50mm -50mm -10mm -10mm, width=1.4in]{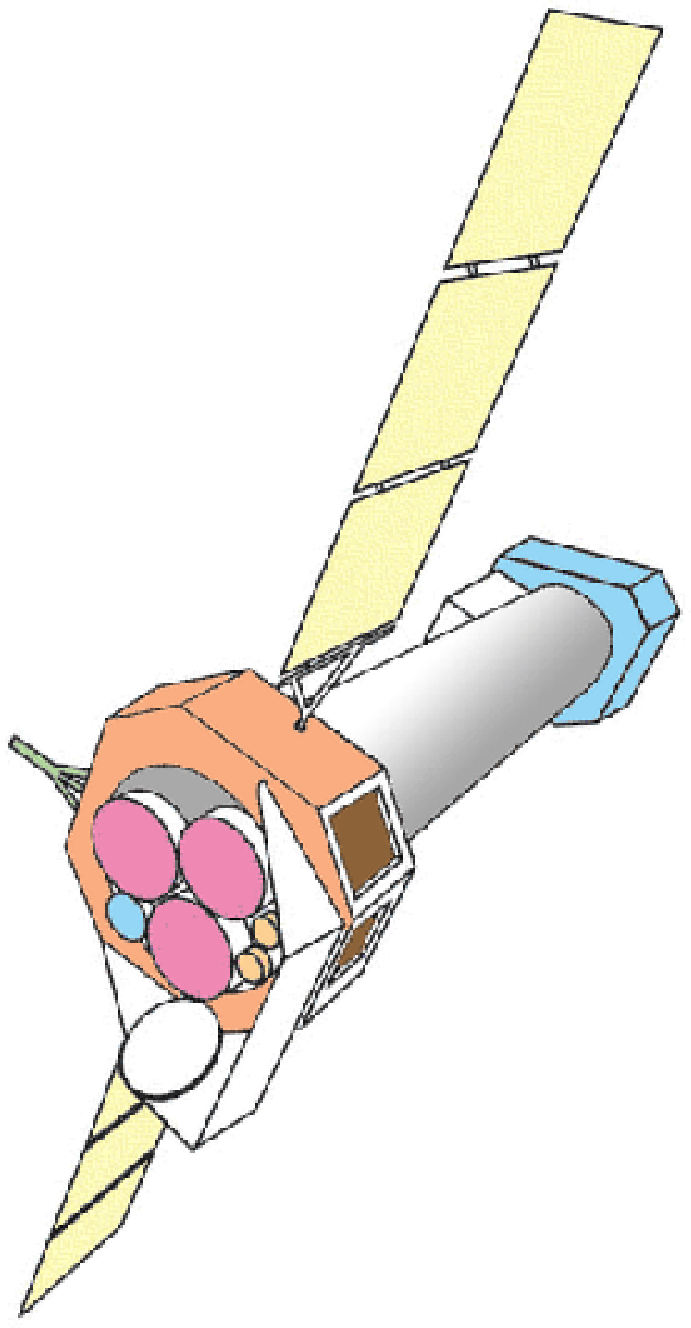}}
\caption{Geometrical layout of the lamp-post model. The accretion disc is an optically thick, geometrically thin, Keplerian cold (i.e.\ neutral) disc that extends from the ISCO, $r_{\rm in}$, which is defined by the BH spin parameter, $\alpha$, to 1000 \rg. The X-ray source is depicted by the orange sphere and it is situated at a height $h$ above the BH. The blue solid lines correspond to the trajectories of two example hard X-ray photons leaving the X-ray source, one heading towards the observer and another one towards the accretion disc. The latter follows a curved trajectory, due to the GR effects, and the trajectory of the corresponding soft X-ray photon from the accretion disc (reflected product) is shown with the red dotted line. The whole system is viewed from an angle $\theta$. --A colour version of this figure is available in the online version of the journal--}
\label{fig:accrDisc}
\end{figure*}

In Table~\ref{tab:obs} we list the details of the \textit{XMM}-\textit{Newton} observations which we used in this work. In the second column we list the Observation ID for each object. The letters in parenthesis refer to the pn observation mode; fw, lw, and sw refer to the {\tt PrimeFullWindow}, {\tt PrimeLargeWindow} and {\tt PrimeSmallWindow} modes, respectively.\par
The \textit{XMM}-\textit{Newton} data were processed using {\sc scientific analysis system} ({\sc sas}) \citep{gabriel04} version 12.0.1. We consider only the EPIC pn data \citep{struder01} as they have a higher count rate and lower pile-up distortion than the MOS data. The source counts for each AGN were accumulated from a circular aperture of radius 40\arcsec\ centered on the source. The background counts were accumulated from a source-free circular region on the same CCD chip as the source. For the type of events, we selected only single and double pixel events, i.e.\ {\tt PATTERN==0-4}, and we rejected bad pixels and events too close to the edges of the CCD by using the standard quality criterion {\tt FLAG==0}.\par 
Source and background light curves were extracted using {\tt evselect} in 10 s time bins.  We checked all light curves for pile-up using the task {\tt epatplot}. We found that only three observations of Ark\;564 were affected by moderate pile-up with Obs. IDs: 0670130201, 0670130501 and 0670130901. In these cases, we used an annular region to exclude the innermost source emission having a radius, in pixels, of: 280, 200 and 250, respectively. The source light curves were screened for high background and flaring activity. The light curve parts with high background activity, usually at the beginning and/or end of an observation, were removed from the final data products. The total net {\tt pn} exposure time, i.e.\ after the exclusion of the high background activity periods, is listed in the third column of Table~\ref{tab:obs}.\par
Since we are interested in studying the time-lags between the soft excess and the X-ray continuum, the source and background light curves were extracted in those energy bands where the soft excess contribution is maximized and the continuum emission is dominating, respectively. Thus, for all the sources, we use for the soft excess and the continuum the energy bands of 0.3--1 and 1.5--4 keV, respectively. The background subtracted light curves were produced using the ({\sc sas}) task {\tt epiclccorr}. Note that in Section~\ref{sssect:sele_energyBands} we discuss why for the purposes of our paper detailed fine-tuning of the selected energy bands is not necessary.\par
The resulting data sets are continuously sampled suffering only from very few data gaps and count rate drops (as a consequence of telemetry drop-outs). With respect to the data gaps, on average there are at most 3--6 missing observations which we filled up by linear interpolation. For the count rate drops, that are also very few (on average between 2--5 points), we rescale the count rate, within each time bin, according to the corresponding fractional exposure time.

\section{THE LAMP-POST MODEL}
\label{sect:lampPost_irf}
In this section we describe the basic physical and geometrical properties of the lamp-post model. Then, we outline the process that we use to estimate the GRIRFs. All the time scales are estimated in normalised time units, of \tg, and thus they scale linearly with the BH mass, $M$, via the relation 
\eqb
t_{{\rm g},M}=4.9255 M\;\rm{s}
\label{eq:mass_scaling}
\eqe
where $M$ is given in units of $10^6$ \ms.\par

\subsection{The geometrical layout and parameter space}
\label{ssect:model}
The lamp-post model that we consider in this paper consists of the following three physical components: a central supermassive BH with an accretion disc illuminated by a point-like X-ray source located on the axis of the system (Fig.~\ref{fig:accrDisc}). This system is characterised by the following parameters: spin parameter and mass of the central BH, height of the X-ray source, and viewing angle.\par
The accretion disc extends from the inner most stable circular orbit (ISCO), $r_{\rm in}$, to 1000 \rg\ and it is an optically thick, geometrically thin, Keplerian cold (i.e.\ neutral) disc. The X-ray source lies above the BH at height $h$ and it comprises the primary source of X-rays, which we assume to be static and to be emitting isotropically with a power-law spectrum of photon index $\Gamma=2$.\par 
The central BH is characterised by its spin parameter, $\alpha$, and its mass, $M$. For the former, which defines the ISCO, we consider three values: 0 (Schwarzschild BH, $r_{\rm in}=6$ \rg), 0.676 (intermediate spin BH, $r_{\rm in}=3.5$ \rg) and 1 (maximally rotating Kerr BH, $r_{\rm in}=1$ \rg).\par
We have chosen a variety of heights for the X-ray source, depending on the spin of the BH. For the case of a Schwarzschild BH, we select an ensemble of 18 heights: $\{2.3,2.9,3.6,4.5,5.7,7,8.8,11,13.7,17.1,21.3,26.5,33.1,41.3,$ $51.5,64.3,80.2,100\}$ \rg. For the intermediate case we add to the ensemble a lower height of 1.9 \rg, and for the Kerr BH we add yet another height of 1.5 \rg, respectively.\par
Finally, the system is observed by a distant observer at a viewing angle of $\theta$ i.e.\ $\theta=0$ or $90\degr$ if the disc is face on or edge on, respectively. For each one of the three BH spin parameters and each one height we consider three angles: 20 40 and $60\degr$.\par
This wide parameter space, consisting of the variables $\alpha$, $h$ and $\theta$, yields a total of $(20+19+18)\times 3=171$ different geometrical layouts of the lamp-post model. The BH mass is not an additional variable in our estimation of GRIRFs as all time-scales and frequencies scale linearly with it (equation~\ref{eq:mass_scaling}). In each geometry the photons will follow different trajectories from the X-ray source to the disc and from the disc to the observer, and thus the response of the system will be different.

\subsection{Estimation of the GRIRFs}
\label{ssect:relativ_respFunct}
In order to compute the response of the accretion disc to the primary illumination from the X-ray source (described by the power-law) we use a flare with a step function profile that has very short duration of 1 \tg. The primary intrinsic spectrum has a normalisation of unity. Then we estimate the response of the disc by measuring the flux \textit{only} of the neutral fluorescent \fa line, at 6.4 keV in the rest frame of the accretion disc.\par

\begin{figure*}
\parbox{0.5\linewidth}{\hspace{-15em}\vspace*{0.9em}\scalebox{0.481}{\includegraphics[trim = 2.3mm 229mm 200.4mm 190mm clip width=2cm]{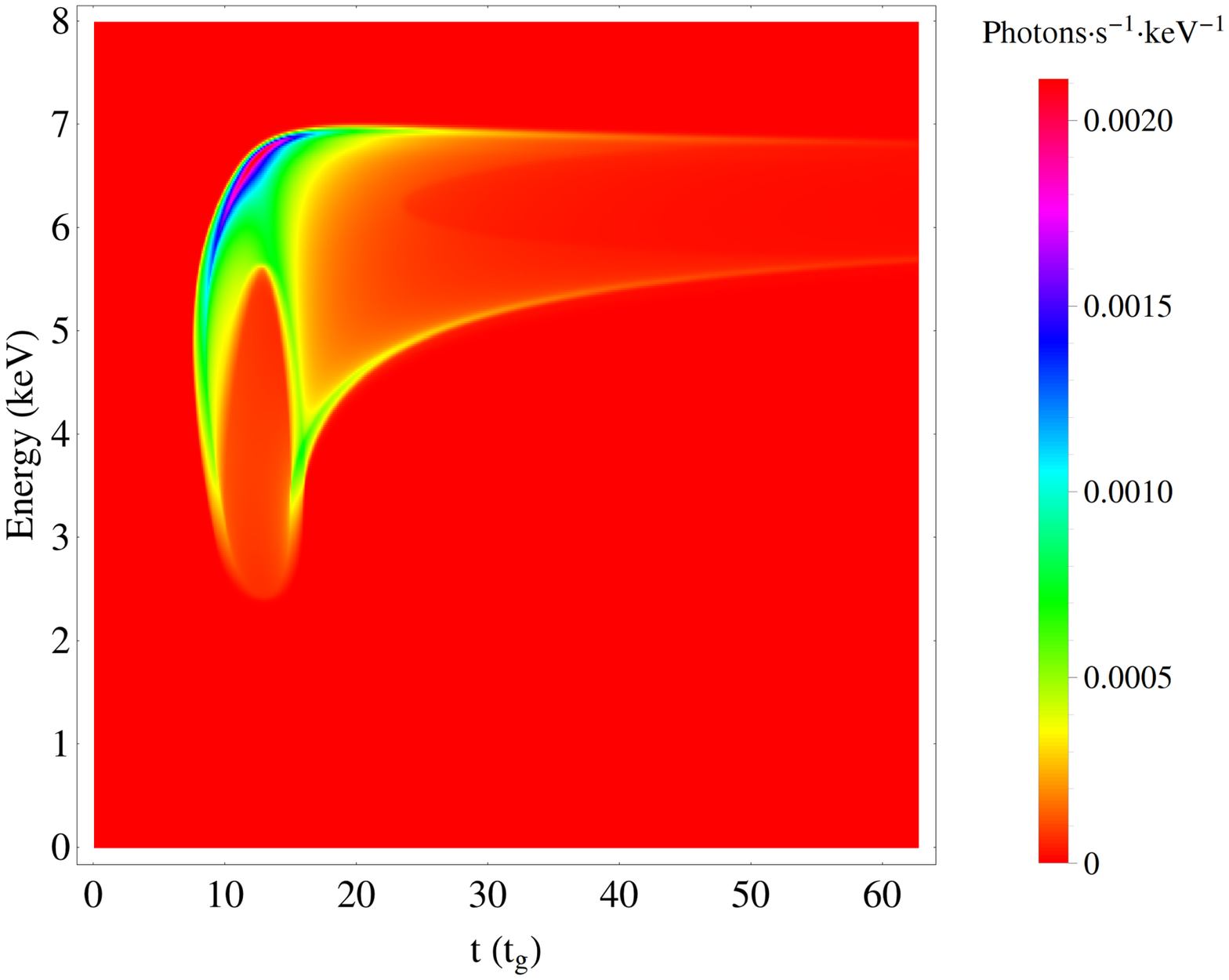}}}\\ 
\parbox{0.5\linewidth}{\hspace{18em}\includegraphics[width=2.85in]{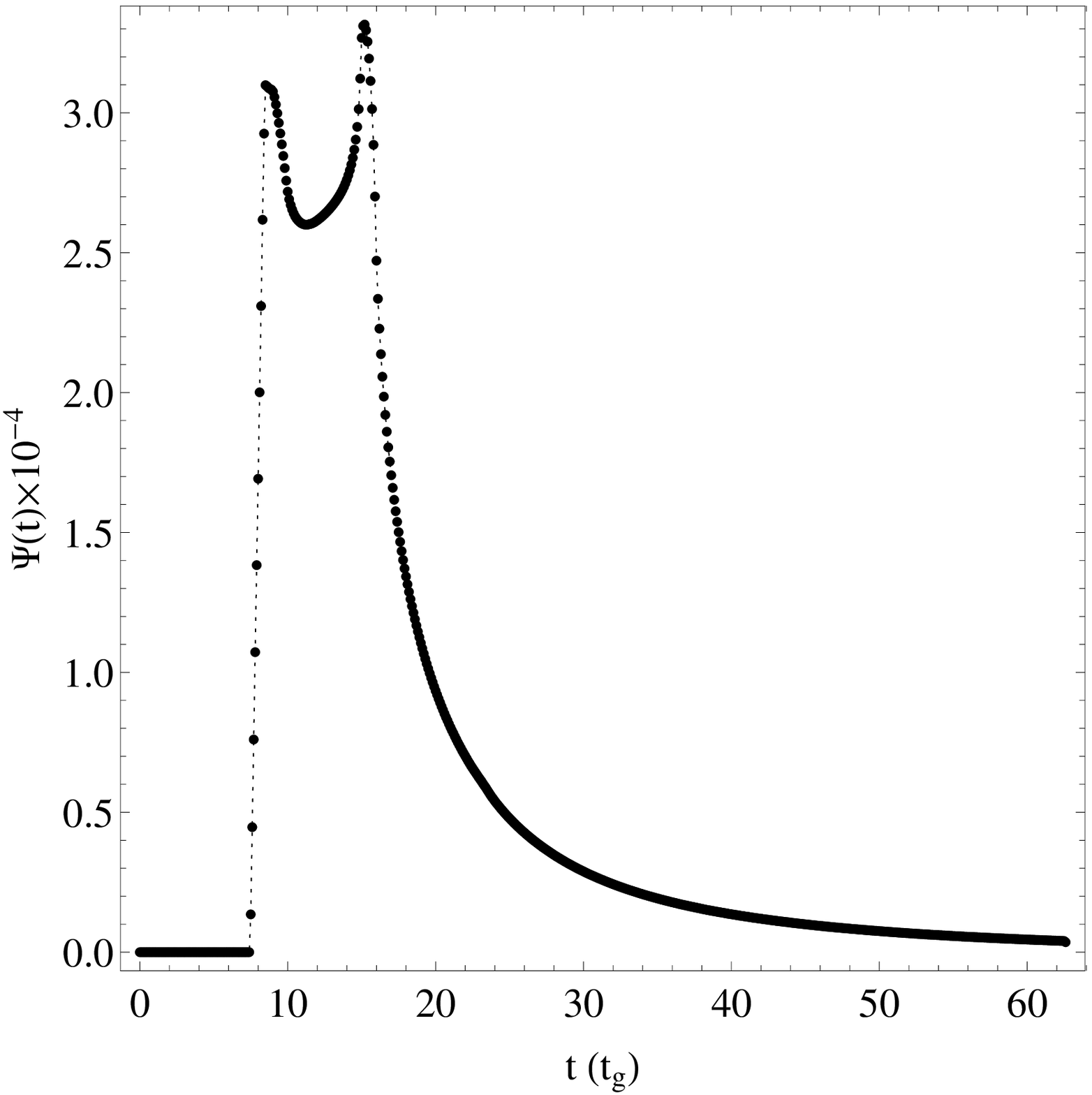}}
\caption{Estimation of the GRIRF for the lamp-post model with $\alpha=0.676$, $\theta=40\degr$ and $h=3.6$ \rg. Left-hand panel: The dynamic reflection spectrum with a time resolution of 0.1 \tg. --A colour version of this figure is available in the online version of the journal-- Right-hand panel: The GRIRF derived by adding up the \fa line flux, at a given time (black points), and the dotted line is the result of the cubic interpolation.}
\label{fig:dynSpec}
\end{figure*}
\begin{figure*}
\includegraphics[width=6.5in]{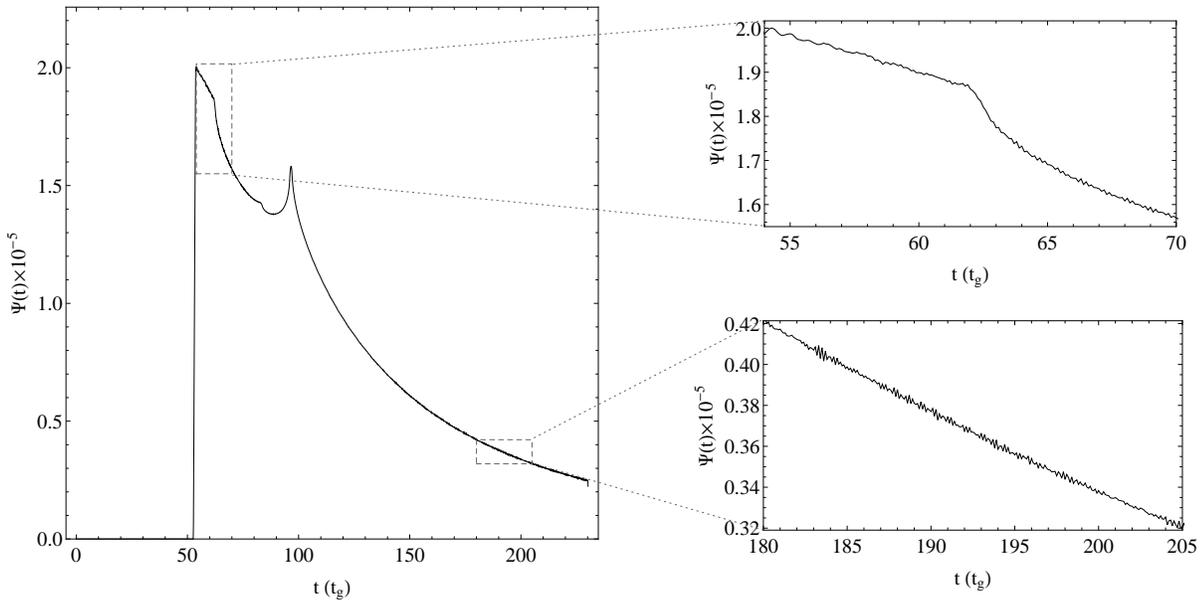}
\caption{Small scale oscillatory artefacts. The plot shows the linearly interpolated GRIRF for the lamp-post model with $\alpha=0$, $\theta=60\degr$ and $h=51.5$ \rg. The dashed rectangles zoom in two regions of the GRIRF where the oscillatory artefacts become prominent.}
\label{fig:oscillBehav}
\end{figure*}

To compute the \fa line flux we use the Monte Carlo multi-scattering code {\sc noar} \citep{dumont00}, which takes into account both the direct and inverse Compton scattering processes. We assume a neutral accretion disc and an iron abundance equal to the Solar value. The resultant flux at the observer is computed using all the general relativistic effects \citep{dovciak04}, without taking into account higher order images of the disc. Thus, we exclude photons emitted from the accretion disc (either from the top 
or the bottom surface) that go around the BH (any number of times), which reach the observer by travelling through the gap between the ISCO and the BH. Higher order images are more prominent for a Schwarzshild BH where the ISCO is the furthest out and the gap between the inner edge of the disc and the BH is the largest. Nevertheless, even in this case these photons do not contribute very much to the total flux \citep{beckwith04}.\par
An interesting point is that in the case of the lower-spin BHs, matter from the accretion disc falls down onto the BH rather quickly (on the orbital dynamical time-scale at the ISCO). Thus, it is expected that this free-falling material, filling the gap between the ISCO and the BH, will have much lower density, and consequently it could be heavily ionised contributing to the overall reflection spectrum \citep{reynolds97}. In this case higher order images would not be produced. The soft excess, of the contributed re-processed emission, is more prominent in the states with higher ionisation, but due to the much lower density of the matter in the plunging region, we do not expect a very large contribution to the overall emission and thus to the response functions of the system. Therefore, in our computations we do not take into account the ionised emission below ISCO in case of lower-spin BHs.\par
The GR effects include the light bending that causes the gravitational lensing, energy shift -- Doppler and gravitational-- and the relativistic time delays for both the photons travelling from the primary source to the disc, as well as the re-processed photons emitted from the disc travelling towards the observer. Due to the lamp-post geometry and relativistic effects, the illumination of the accretion disc, from the primary X-ray source, is uneven (i.e.\ depending on the radius). The reflected flux from the disc is proportional to the incident flux on the disc which is a power-law with its normalisation being a function of $\alpha$, $h$, $\theta$ and $\Gamma$ \citep[see equation 3 in][]{dovciak11}.\par
Due to the GR effects, the flux of the \fa line is spread over a wide range of energies, as a function of time, yielding for each lamp-post configuration a \textit{dynamic reflection spectrum}\footnote{Sometimes the dynamic reflection spectrum is called `2D transfer function' \citep{campana95,reynolds99} despite the fact that strictly speaking in signal processing the term `transfer function' refers to the Fourier domain i.e.\ ratio of the Fourier transformed input to output signals.}, estimated with a time resolution of 0.1 \tg. In the left-hand panel of Fig.~\ref{fig:dynSpec} we show an example of a \fa dynamic reflection spectrum for the case of $\alpha=0.676$, $\theta=40\degr$ and $h=3.6$ \rg. As we can see the flux of the \fa line is spread over a wide range of energies between 2.5--7 keV and fades-out as time goes on, indicating the decay of the `echo' as it is moving away from the centre of the accretion disc.\par
In order to derive the GRIRF, $\Psi(t)$, of the accretion disc for the \fa emission line we use the dynamic reflection spectrum and we add up for a given time all the \fa line fluxes in the energy band 0--8 keV. In the right-hand panel of Fig.~\ref{fig:dynSpec} we show the GRIRF which corresponds to the dynamic reflection spectrum of the left-hand panel. This plot in effect shows how the \fa line flux (in 0--8 keV) evolves with time, as seen by a distant observer. The actual sums are depicted by the black points and the dotted line is a cubic interpolation. Note that the cubic interpolation is needed because in some cases there are very small numerical errors during the estimation of the line flux in the dynamic spectra (of the order of $10^{-5}$) which can create oscillatory artefacts (Fig.\ref{fig:oscillBehav}) in the corresponding GRIRF (during the summation process). By performing a cubic interpolation (as opposed to a linear or quadratic) these artefacts are suppressed and the integration process which 
is used to derive the time-lag spectra (Section~\ref{ssect:TL_model}) is much faster i.e.\ there is no need to take globally a very small integration step.\par
As we can see from the right-hand panel of Fig.\ref{fig:dynSpec}, the general shape of the IRF consists of an abrupt rise followed by two peaks and a decay. The onset of the first peak corresponds to the time that the first hard X-ray photons hit the disc and from that point, a ring of an expanding echo is created on the disc around the BH. As the part of the echo approaches the BH it is deformed in such a way that eventually two echoes are created, inner and outer, which correspond to the second peak in the IRF. The inner ring is then moving towards the BH, the outer one continues with its radial expansion away from the centre of the disc and the IRF is gradually fading out. For low spin values (e.g.\ for the Schwarzschild BH) the inner ring is not created due to the existence of the gap in the accretion disc below the ISCO. The second peak in the transfer function in this case corresponds to the re-appearance of the part of the outer ring, behind the BH, after it was `hidden' below the ISCO.

\section{TIME-LAG SPECTRA ESTIMATION}
\label{sect:TL_spec}
\subsection{Model time-lag spectra}
\label{ssect:TL_model}

We can now use the GRIRFs, that we have computed for the \fa line photons, to construct the model time-lag spectra. Assume that the X-ray spectrum of AGN can be described by a power-law form $a(t)E^{-{\Gamma}}$, where $\Gamma$ is constant as a function of time and all the observed flux variability is due to the variations of the normalization $a(t)$ (primary continuum). Then, the source's hard X-ray emission in the 1.5--4 keV band, $h(t)$, which is dominated by the X-ray continuum source can be written as 
\eqb
h(t)=b a(t)
\label{eq:hard_emis}
\eqe
where $b=\int_{1.5\;{\rm keV}}^{4\;{\rm keV}}E^{-\Gamma}dE$ (in Section~\ref{sect:sum_disc} we discuss further the contamination of the continuum by the reflection component). Let us then assume that $s(t)$ is the variable source's soft X-ray emission in the 0.3--1 keV energy range where, in addition to the continuum, we also detect emission from the reprocessing component so that
\eqb
s(t)=k a(t)+f\int_{0}^{\infty}\Psi(t^\prime)a(t-t^\prime)dt^\prime
\label{eq:soft_emis}
\eqe
where $k=\int_{0.3\;{\rm keV}}^{1\;{\rm keV}}E^{-\Gamma}dE$ and $\Psi(t^\prime)$ are the GRIRFs for the \fa line photons over the 0--8 keV band that we discussed in the previous section (Section~\ref{ssect:relativ_respFunct}). In this way, we basically assume that the photons of the reflected component, which contribute to the observed 0.3--1 keV energy band, and the \fa line photons, at 6.4 keV, are produced in the same parts of the accretion disc, hence the shape of the soft band GRIRF is similar to the GRIRFs that we have estimated for the \fa line photons. However, even if this is the case, we do not expect the flux of the soft-band reflection spectrum to be the same as the \fa line flux (i.e.\ difference in the normalisation), hence the use of the constant $f$ in the above equation. As we explain in detail in Sections~\ref{sect:fits} and \ref{sect:sum_disc}, as well as in the Appendix~\ref{app:reflect_fracti}, during our analysis we set this value to 0.3.\par
In order to derive the time-lag spectrum between the $s(t)$ and $h(t)$ bands we need to estimate the cross-covariance function between the two time series. This function at a time-lag $\tau$ is defined as:
\eqb
r_{s,h}(\tau)=\operatorname{E}\left[\left(s(t)-\langle s(t)\rangle\right)\left(h(t+\tau)-\langle h(t)\rangle\right)\right]
\label{eq:cross_covari1}
\eqe
where $\operatorname{E}$ is the expectation operator, and the values in angle brackets denote mean values. It can be shown that in our case,
\eqb
r_{s,h}(\tau)=b k r_{a,a}(\tau)+f b\int_{0}^{\infty}\Psi(t^\prime)r_{a,a}(\tau+t^\prime)dt^\prime
\label{eq:cross_covari2}
\eqe
where $r_{a,a}(\tau)$ is the auto-covariance function of the continuum variability process,
\eqb
r_{a,a}=E\left[\left(a(t)-\langle a(t)\rangle\right)\left(a(t+\tau)-\langle a(t)\rangle\right)\right]
\label{eq:auto_covari}
\eqe
If we take the Fourier transform of both sides of the above equation, we obtain
\eqb
\mathscr{P}_{s,h}(\nu)=b k \mathscr{P}_{a,a}(\nu)[1+\frac{f}{k}\int_{0}^{\infty}\Psi(t^\prime)e^{-i 2\pi\nu t^\prime}dt^\prime]
\label{eq:cross_spec_cont}
\eqe
where $\mathscr{P}_{s,h}(\nu)$ is the cross-spectral density function between $h(t)$ (i.e.\ the input hard X-ray emission) and $s(t)$ (i.e.\ the output soft reflected emission) at frequency $\nu$, and $\mathscr{P}_{a,a}(\nu)$ is the power spectral density function of the continuum. The function $\mathscr{P}_{s,h}(\nu)$ is a complex function, and its complex argument defines its phase, at frequency $\nu$, i.e.\ the phase-lag between the time series $s(t)$ and $h(t)$. Since $\mathscr{P}_{a,a}(\nu)$ is a real function, equation~\ref{eq:cross_spec_cont} implies that the time-lag, $\tau(\nu)$, between the two time series at frequency $\nu$ is given by
\eqb
\tau_\nu=-\frac{\arg\left[1+(f/k)\int_{0}^{\infty}\Psi(t^\prime)e^{-i 2\pi\nu t^\prime} dt^\prime\right]}{2\pi\nu}
\label{eq:timeLagSpectra}
\eqe

We use equation~\ref{eq:timeLagSpectra} to estimate the model time-lag spectra for each one of the model GRIRF (in total 171, Section~\ref{sect:lampPost_irf}). For each $\Psi(t^\prime)$, the integral in this equation was estimated at 491 frequencies: $\nu=10^{k}$ Hz with $k=-5,-4.99,\ldots,-0.11,-0.1$. These frequencies cover the typical frequency range observed by \textit{XMM}-\textit{Newton}. Note that for the integration procedure we use adaptive integration method which identifies the problematic integration areas, which in our case are usually the regions of the two peaks, and concentrate the computational effort (i.e. sampling points) on them \citep{malcolm75,krommer98}. The resulting time-lag model spectra have an almost continuous profile, due to the very fine frequency resolution we have adopted. Therefore, this allows us to interpolate linearly among the various frequencies without adding additional structure into the resulting time-lag spectra.\par
These time-lag spectra (equation~\ref{eq:timeLagSpectra}) contain the integral of the GRIRF, $\Psi(\tau^\prime)$, in normalised time units (i.e.\ \tg). Thus, if one wants to work in conjunction with real data (e.g.\ for fitting purposes, as we are doing in Section~\ref{sect:res}) then one has to transform the time-lag spectra into real physical units (i.e.\ seconds) by dividing and multiplying the abscissas (frequencies, $\nu$) and the ordinates (time-lag estimates, $\tau_{\nu}$) respectively by $t_{{\rm g},M}$ (equation~\ref{eq:mass_scaling}).\par
\begin{figure*}
\includegraphics[width=3.2in]{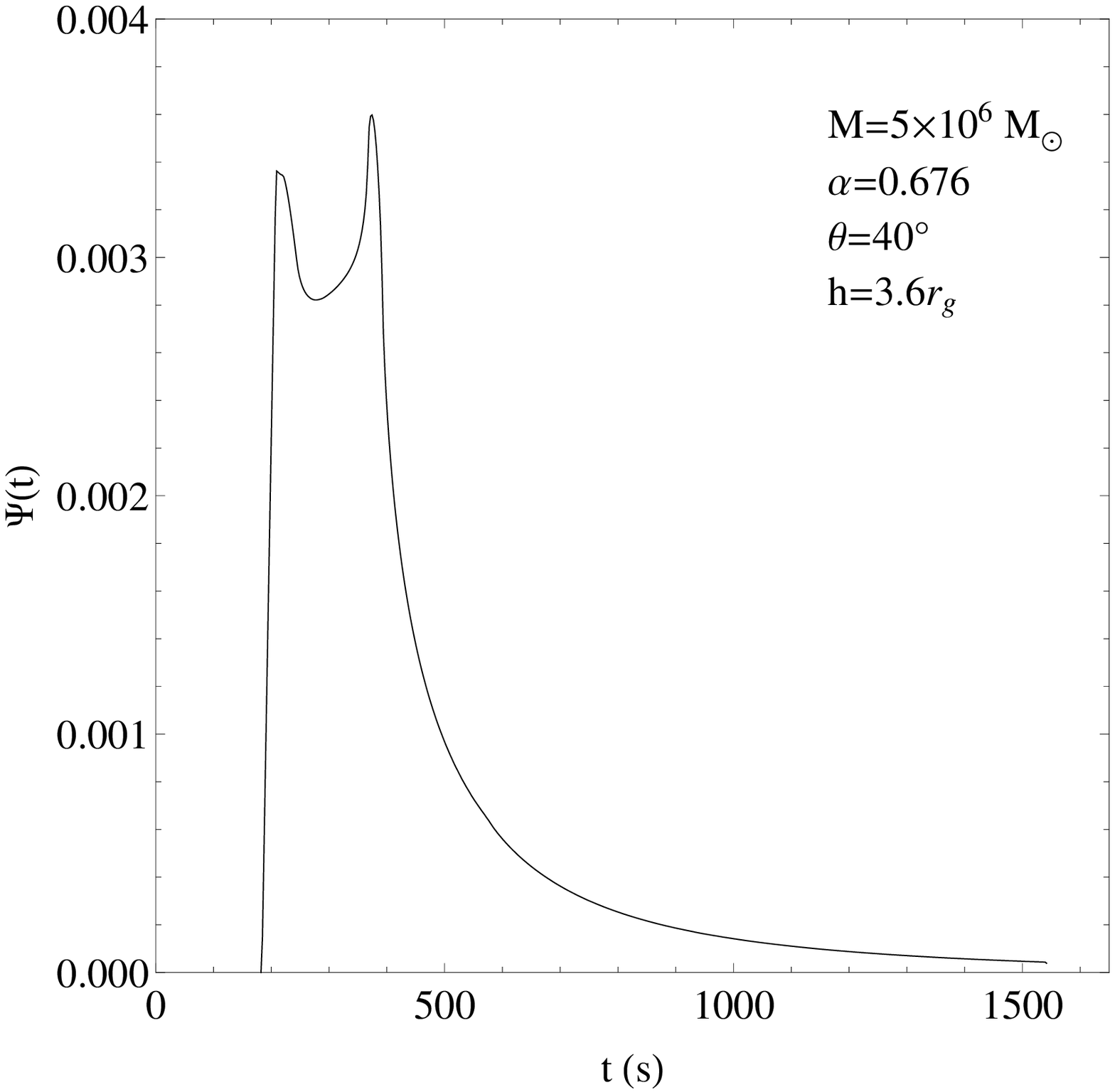}\hspace{3em}
\includegraphics[width=3.18in]{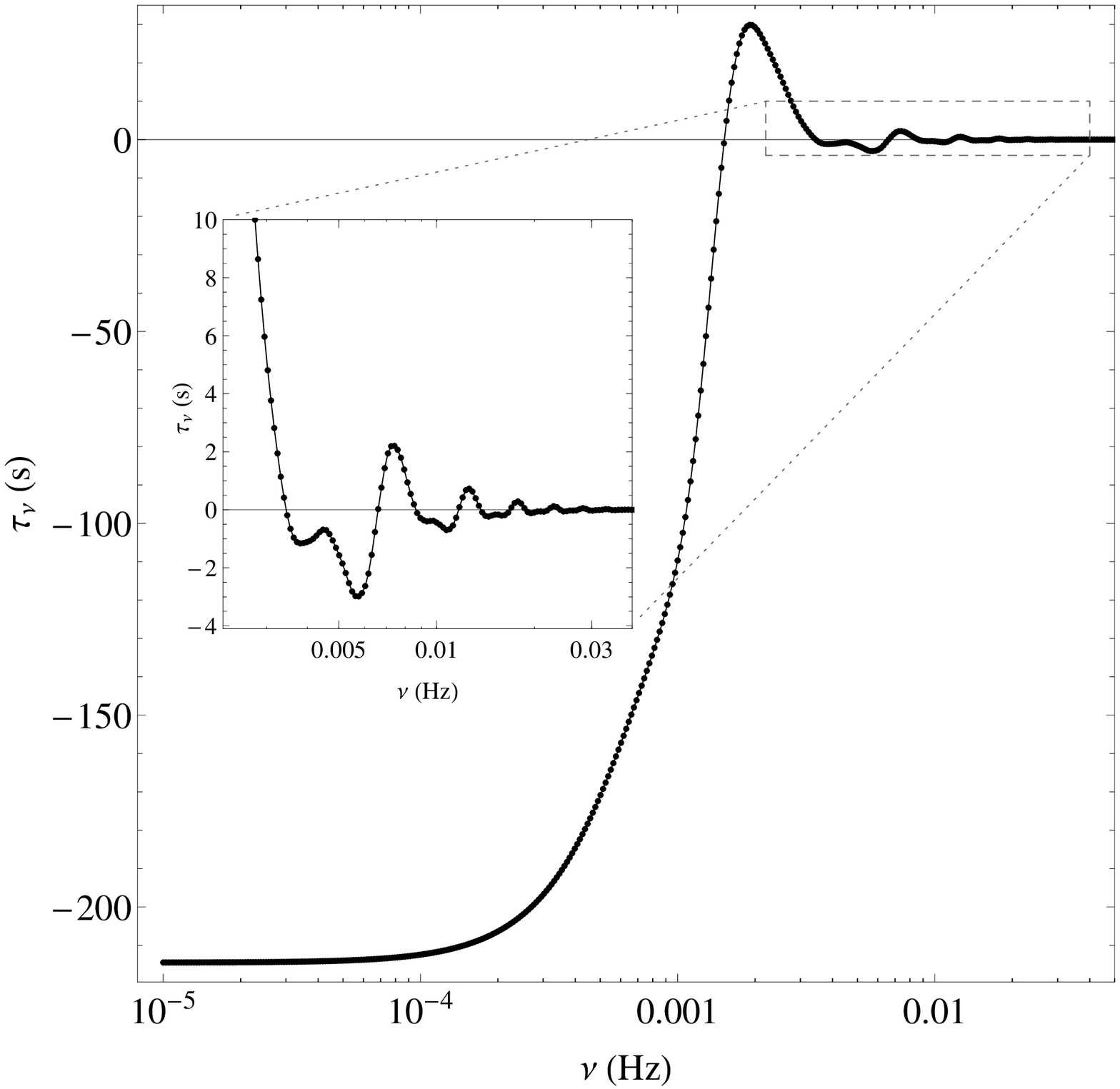}
\caption{Estimation of the time-lag spectrum for the lamp-post model with $\alpha=0.676$, $\theta=40\degr$ and $h=3.6$ \rg, for $M=5\times10^6$ \ms. Left-hand panel: The GRIRF model in physical units (a scaled version of Fig.~\ref{fig:dynSpec}, right-hand panel). Right-hand panel: The corresponding time-lag spectrum.}
\label{fig:tlSpec}
\end{figure*}

For demonstrative purposes in the left-hand panel of Fig.~\ref{fig:tlSpec} we show the GRIRF in physical time units for a BH mass of $5\times10^6$ \ms\ corresponding to the GRIRF shown in normalised units in the right-hand panel of Fig.~\ref{fig:dynSpec} ($\alpha=0.676$, $\theta=40\degr$ and $h=3.6$ \rg). The corresponding time-lag spectrum for this GR reflection scenario is in general a negative function (Fig.~\ref{fig:tlSpec}, right-hand panel). The first morphological characteristic of this time-lag spectrum is that at low frequencies, $10^{-5}-10^{-4}$ Hz (i.e.\ long time scales) it exhibits a negative tail that forms a constant plateau at -215 s. This plateau defines the most negative time delayed reflected emission from the disc (i.e.\ its ordinate) and the corresponding time scales that these delays occur (i.e.\ 10--100 ks). Then, the values of the time-lag spectrum increase as a function of frequency and they become positive at $1.5\times10^{-3}$ Hz, peaking around ($2\times10^{-3}$ Hz, 30 s). 
Finally, the time-lag spectra exhibit a damped oscillating behaviour (Fig.~\ref{fig:tlSpec}, inset) around zero.\par
In general, all the time-lag spectra exhibit these features but they appear at different frequencies depending on the BH mass, the height of the X-ray source and $r_{\rm in}$. Note, that whether the time-lag values become positive or remain always on a negative level depends only on the shape of corresponding GRIRF (as shown in equation~\ref{eq:timeLagSpectra}). In Appendix~\ref{ssect:modPar_effects} we explore the model parameter space for different configurations of the lamp-post model.

\begin{figure*}
\includegraphics[width=3.2in]{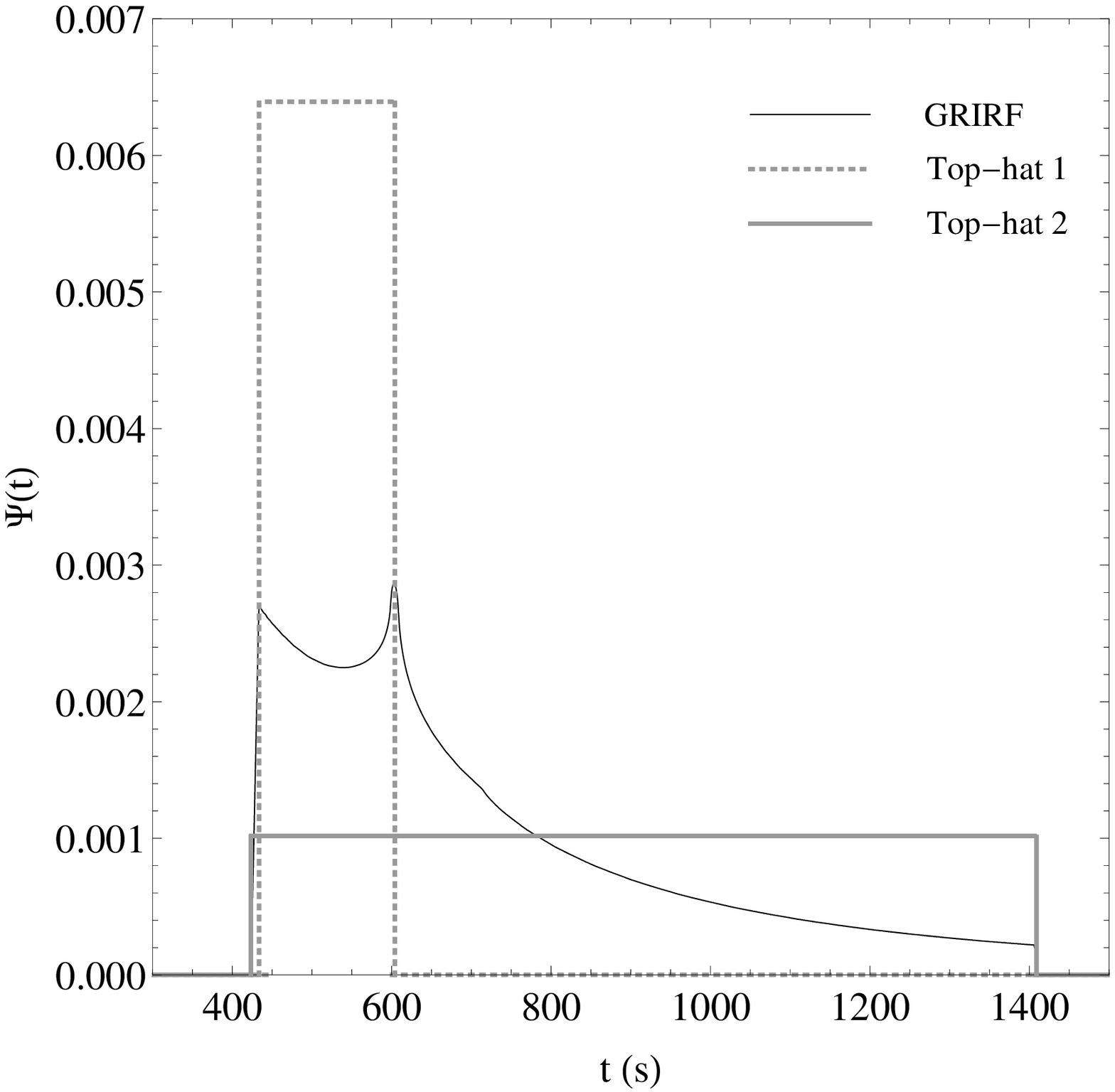}\hspace{3em}
\includegraphics[width=3.18in]{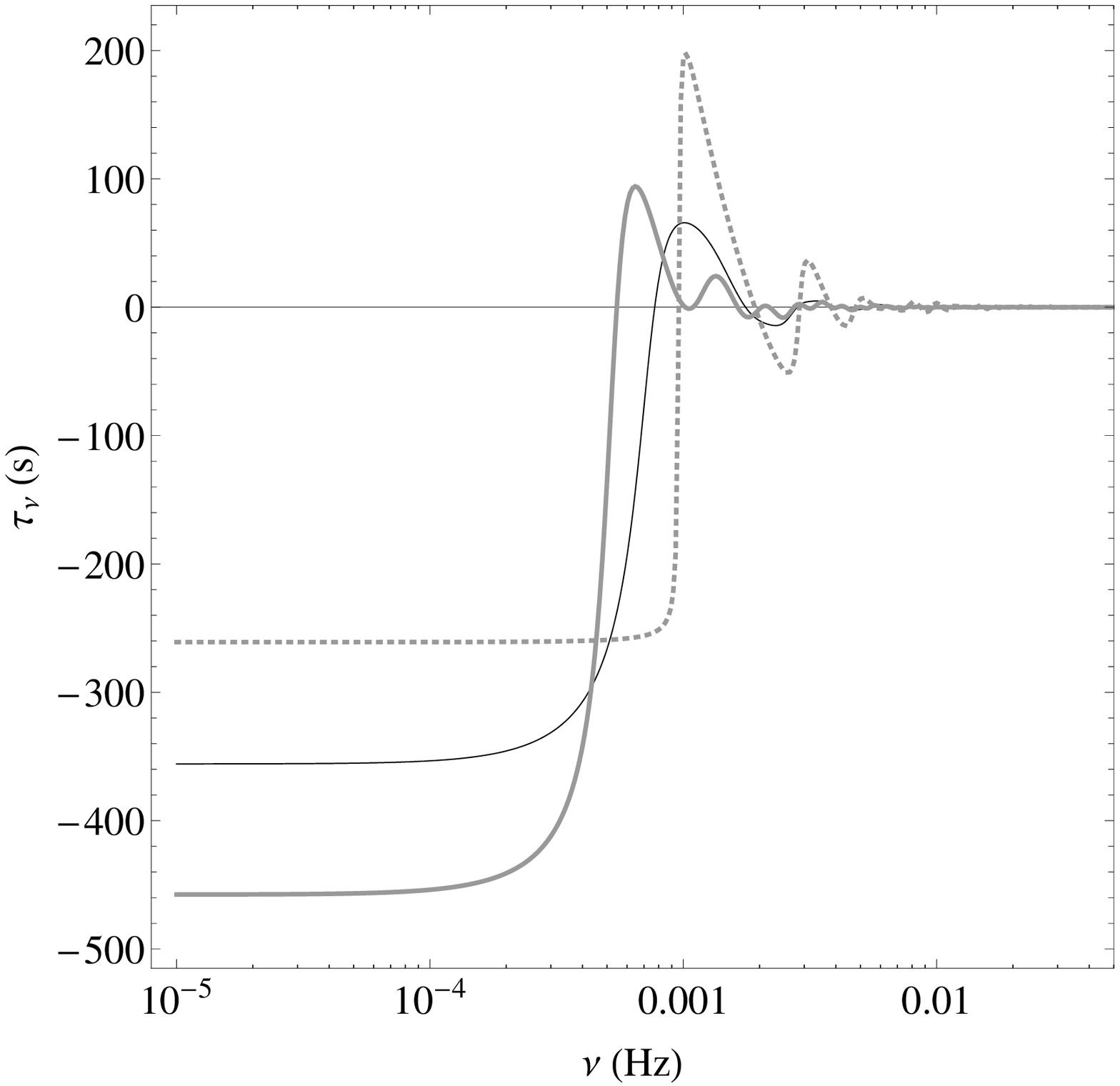}
\caption{GRIRF versus THIRF for the lamp-post model with $\alpha=1$, $\theta=40\degr$ and $h=26.5$ \rg, for $M=2\times10^6$ \ms. Left-hand panel: The GRIRF (black solid line) and the two top-hat parametrisation models each one depicting the two peaks (dotted thick grey line) and the overall shape (solid thick grey line) of the GRIRF, respectively. Right-hand panel: The corresponding time-lag spectra.}
\label{fig:GRIRFvsTopHat}
\end{figure*}
\begin{figure*}
\includegraphics[width=3.16in]{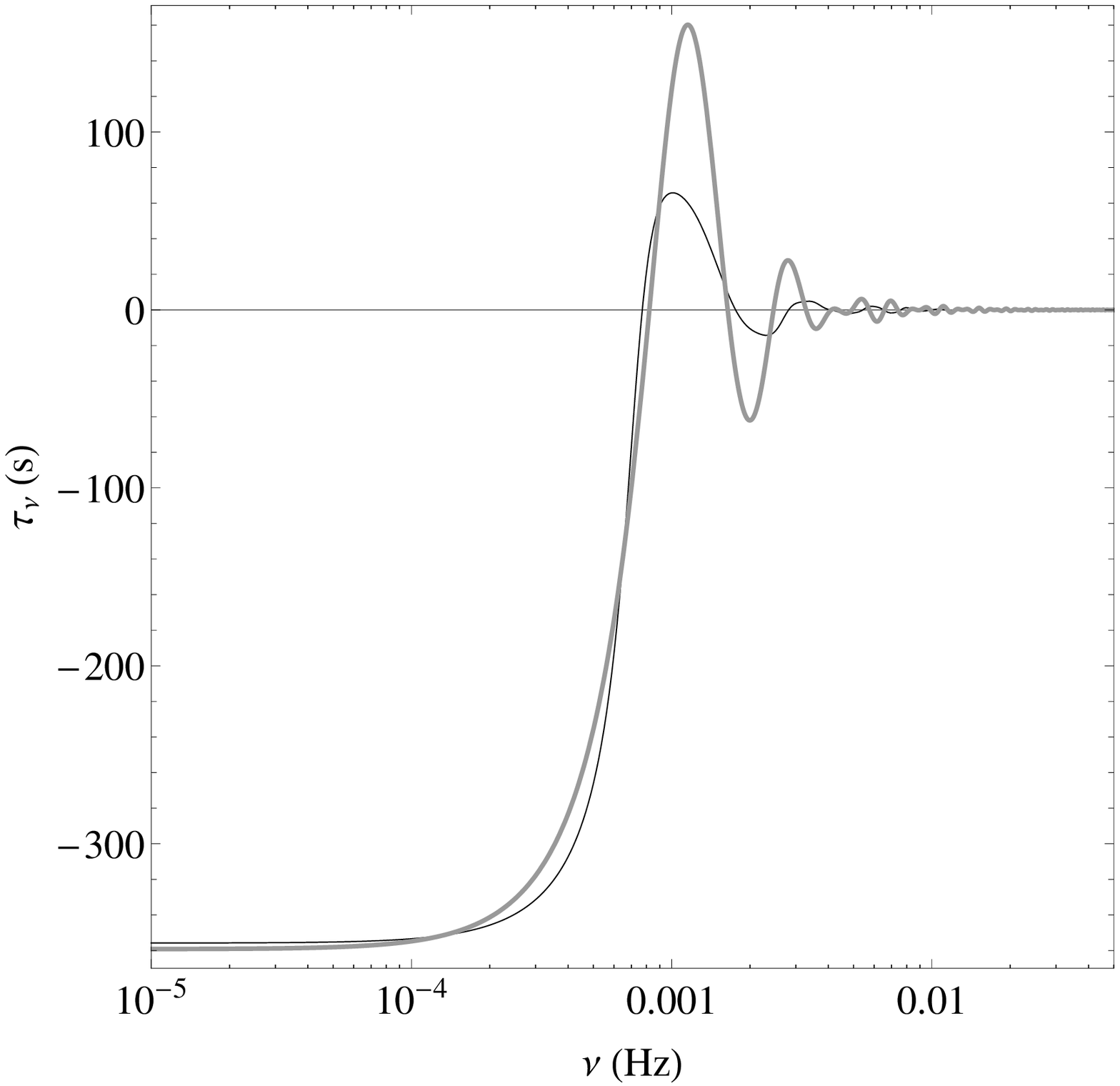}\hspace{3em}
\includegraphics[width=3.28in]{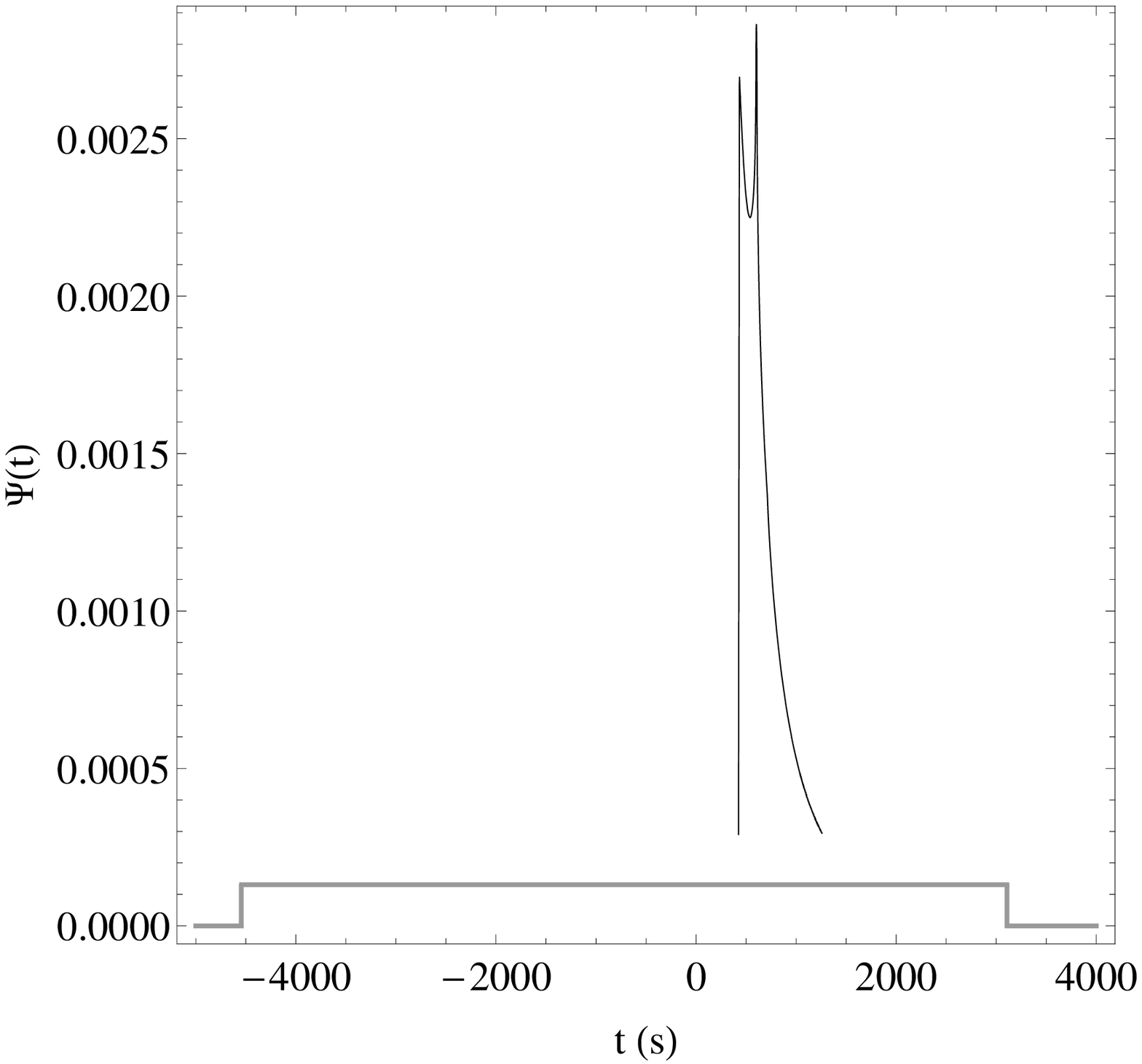}
\caption{Fitting the time-lag spectrum of the lamp-post model with $\alpha=1$, $\theta=40\degr$ and $h=26.5$ \rg, for $M=2\times10^6$ \ms with a time-lag spectrum coming from a THIRF. Left-hand panel: The best-fitting time-lag spectrum (thick grey line), for $\{\tau^\prime_1, \tau^\prime_2\}=\{-4545,3108\}$ s, and the lamp-post time-lag spectrum (black line, also shown in the right-hand panel of Fig.~\ref{fig:GRIRFvsTopHat}). Right-hand panel: The corresponding THIRF, derived from the best-fitting parameters, (thick grey line) together with the GRIRF of the lamp-post model (black line, also shown in the left-hand panel of Fig.~\ref{fig:GRIRFvsTopHat}).}
\label{fig:fitTopHat}
\end{figure*}

\subsubsection{General relativistic IRF vesrus top-hat IRF}
\label{sssect:comparisonTopHat}
In this section we compare the physically justified GRIRFs with the widely used (see for references Section~\ref{sect:intro}) THIRF parametrisation model. We employ the case of the lamp-post model with the following parameters: $\alpha=1$, $\theta=40\degr$ and $h=26.5$ \rg, for a for a BH mass $M=2\times10^6$ \ms. For this case, the GRIRF is shown in the left-hand panel of Fig.\ref{fig:GRIRFvsTopHat} with the black solid line. Then we consider two top-hat parametrisation scenarios (both of them normalised to unity) to represent the given GRIRF: one with width defined to be equal to the separation of the two peaks in the real GRIRF (Top-hat 1, Fig.\ref{fig:GRIRFvsTopHat}, left-hand panel, dotted thick grey line) and the other one starting at the same time as the real GRIRF and extending over the full time range covered by it (Top-hat 2, Fig.\ref{fig:GRIRFvsTopHat}, left-hand panel, solid thick grey line). As we can see from the right-hand panel of Fig.\ref{fig:GRIRFvsTopHat}, the corresponding time-lag 
spectra 
of the two THIRFs differ genuinely from that of the GRIRF. The position of the negative constant plateau, the position and the amplitude of the first positive peak as well as the behaviour is high frequencies (above $10^{-3}$ Hz) differ significantly from those in the time-lag spectrum of the GRIRF. Note that similar discrepancies occur for lower heights of the X-ray source.\par
The top-hat parametrisation yields always two numbers: the start- and the end-time of the rectangular pulse. In both scenarios that we have considered, the start-time is associated with the beginning of the reverberation phenomenon on the accretion disc (first peak of the GRIRF) i.e.\ the time that the first hard X-ray photons hit the disc, which is a physically real parameter of the system. The end-time could correspond either to the second peak of the GRIRF where the iso-delayed arcs (travelling opposite and around the BH) meet for the first time, (Top-hat 1) or to the overall time of the reverberation phenomenon (Top-hat 2) (see Section~\ref{ssect:relativ_respFunct} for the various peaks and times of the GRIRF). However, none of these scenarios yield a time-lag that matches that of GRIRF.\par
Finally, in order to retrieve the best-fitting THIRF model, that could correspond to the time-lag spectrum of the given lamp-post geometry, we fit to the time-lag spectral estimates (shown in Fig.\ref{fig:GRIRFvsTopHat}, right-hand panel, black line) equation~\ref{eq:timeLagSpectra}, using as $\Psi(\tau)$ a top-hat function leaving as free parameters $\tau^\prime_1$ and $\tau^\prime_2$ (i.e.\ start and end-times) under the condition $\tau^\prime_1<\tau^\prime_2$ \citep[e.g.][]{zoghbi11,emmanoulopoulos11b}. This is done by finding the pair $\{\tau^\prime_1,\tau^\prime_2\}$ that minimises the squared sum of the distance of between the two functions estimated over the 491 frequencies points of the time-lag spectrum (Section~\ref{ssect:TL_data}).\par
The best-fitting THIRF model is shown in the left-hand panel of Fig.\ref{fig:fitTopHat} with the thick grey line, together with the physically realistic GRIRF model (black line). The quality of the fit is poor (squared sum equals to 188288 s$^2$ for 489 degrees of freedom, --d.o.f.--), but more importantly, the best-fitting THIRF yields physically unrealistic times $\{\tau^\prime_1,\tau^\prime_2\}=\{-4545,3108\}$ s that can not be associated with any time-scale of the system\footnote{Constraining the problem only to the positive domain i.e.\ $0<\tau^\prime_1<\tau^\prime_2$ yields practically a zero width best-fitting THIRF with $\{\tau^\prime_1,\tau^\prime_2\}=\{1.1, 1.3\}\times10^{-5}$ s, having a very poor fit characterised by a squared sum of $2.12\times10^7$ s$^2$ for 489 d.o.f.}. As we can see from the right-hand panel of Fig.\ref{fig:fitTopHat}, the top-hat model, that corresponds to the best-fitting times (thick grey line), covers completely different time-scales from the GRIRF model (black line) and 
thus can not be associated with any genuine physical property of the system.

\subsection{Observed time-lag spectra}
\label{ssect:TL_data}
In order to estimate the time-lag spectra between two light curves we use we use the standard analysis method outlined in \citet{bendat86,nowak99}. In brief, consider for a given source a soft and a hard light curve, $s(t)$ and $h(t)$, obtained simultaneously, consisting of the same number of $N$ equidistant observations with a sampling period $t_{\rm bin}$ (these are discretized and finite length versions of equations~\ref{eq:soft_emis} and \ref{eq:hard_emis}, respectively). For a given Fourier frequency, $f_j=j/(N t_{\rm bin})$ for $j=0,1,\ldots,[N/2-1\;\rm{or}\;(N-1)/2]$ (for even or odd $N$) we estimate the cross-spectrum, $\mathscr{C}(f_j)$\footnote{This is a natural estimator of the continuous cross-spectrum equation~\ref{eq:cross_spec_cont}.}, \citep[e.g.][]{priestley81} between the two light curves in a phasor form as
\eqb
\mathscr{C}_{s,h}(f_j)=S^*(f_j)H(f_j)=\lvert S(f_j)\rvert\lvert H(f_j)\rvert e^{i\left(\phi_H(f_j)-\phi_S(f_j)\right)}
\label{eq:cross_spec}
\eqe
in which $S(f_j)$ and $H(f_j)$ are the discrete Fourier transforms\footnote{The discrete Fourier transform for the soft light curve, $s(t)$, at a given Fourier frequency, $f_j$, is defined as follows:
\eqb
DFT_s(j)=\sum_{k=1}^{N}s(t_k)e^{-2\pi i (k-1) j/N}
\label{eq:dft}
\eqe
for $j=0,1,\ldots,[N/2-1\;\rm{or}\;(N-1)/2]$ (for even or odd $N$). Note, that the exponential function contains as a running index $k-1$ instead of $k$, since the data start for $k=1$ and not $k=0$.} of $s(t)$ and $h(t)$, respectively, with phases $\phi_S(f_j)$ and $\phi_H(f_j)$ and amplitudes $\lvert S(f_j)\rvert$ and $\lvert H(f_j)\rvert$, respectively. The asterisk denotes complex conjugation.\par
Then, we average the complex cross-spectrum estimates, coming from all the observations, over a number of at least 10 consecutive frequency bins, yielding $m$ average cross-spectra estimates, $\left<\mathscr{C}_{s,h}(f_{{\rm bin},i})\right>$ at $m$ new (averaged) frequency bins $f_{{\rm bin},i}$ for $i=1,2,\ldots,m$. Finally, for each average cross spectrum we derive its complex argument i.e.\ its angle with the positive real axis, known also as \textit{phase}, $\phi(f_{{\rm bin},i})$ and we convert it to physical time 
units
\eqb
\tau(f_{{\rm bin},i})=\frac{\phi(f_{{\rm bin},i})}{2\pi f_{{\rm bin},i}}
\label{eq:tl_data}
\eqe
For each time-lag estimate we calculate the corresponding standard deviation, $\operatorname{std}\{\tau(f_{{\rm bin},i})\}$ via equations 16 and 17 in \citet{nowak99}.\par
At the same time from the cross-spectrum we estimate the coherence between $s(t)$ and $h(t)$ as a function of Fourier frequency \citep{vaughan97}
\eqb
\gamma^2_{s,h}(f_j)=\frac{\lvert\langle \mathscr{C}_{s,h}(f_j)\rangle\rvert^2}{\langle\lvert S(f_j)\rvert^2\rangle\langle\lvert H(f_j)\rvert^2\rangle}
\eqe
This quantity takes values between 0 and 1 and it is a measure of the linear correlation between the two light curves at a given Fourier frequency. A very important cautionary point is that small coherence values correspond to uncorrelated phases, $\phi_S(f_j)$ and $\phi_H(f_j)$, whose differences are actually depicted by the $\phi(f_{{\rm bin},i})$ (averaged over a range of frequencies). Thus, for uncorrelated phases, $\phi(f_{{\rm bin},i})$ has a rather uniform distribution in the range $(-\pi,\pi]$ (due to phase-wrapping) that averages always to zero. That means that for small coherence values we get a time-lag of 0 that has small uncertainties, due to the large number of averaging points, appearing statistical meaningful even if there is not a real correlation between the phases and hence no meaningful time delay.\par 
In all our analysis we estimate the time-lag spectra down to $(3-5)\times10^{-3}$ Hz, but for the fitting procedure we consider only the time-lag estimates for which the coherence is greater that 0.15 corresponding to a physically meaningful phase correlation.

\subsubsection{Selection of energy bands}
\label{sssect:sele_energyBands}
As we discussed in Section~\ref{sect:obs} for all the sources we extracted the light curves between 0.3--1 (soft band) and 1.5--4 (hard band) keV energy bands and these are the ones that we use for the extraction of the time-lag spectra. These bands depict quite accurately the general behaviour of the reflection component (i.e.\ soft excess) and that of the X-ray source (i.e.\ continuum) despite the fact that for each source separately the exact limits could be shifted slightly towards higher or lower energies.\par
In this paper we actually model the response of the 6.4 keV \fa line as a proxy for the soft reprocessed emission (Section~\ref{ssect:relativ_respFunct}). Given the possible uncertainties between the response of the low energy reprocessed emission and that of the 6.4 keV line, and also the fact that any hard band will contain a small contribution from reprocessed emission, there is little point in worrying too much about the exact choices of soft and hard energy bands. Thus, the selected energy bands offer us a simple homogeneous description of both the soft and the hard behaviour for the ensemble of sources.\par

\section{THE FITTING PROCEDURE}
\label{sect:fits}
In order to fit the observed time-lag spectra, $\tau(f_{{\rm bin},i})$ binned into $m$ frequency bins (Section~\ref{ssect:TL_data}), we require two time-lag spectral model components. The first one corresponds to the GR reflected component, estimated from equation~\ref{eq:timeLagSpectra} for a given GRIRF. This component, carries all the physical and geometrical information about the lamp-post model and gives rise predominantly to negative lags, particularly at the low frequency range. In total, we have 171 of these negative time-lag spectral models, $\tau_\nu(M,\alpha,\theta,h)$, (each one corresponding to a different set of $\{\alpha,\theta,h\}$ lamp-post model parameters for a given $M$) estimated for the ensemble of GRIRFs (Section~\ref{ssect:TL_model}). In principle, the constant $f$ (appearing in equation~\ref{eq:timeLagSpectra}) should be left as a free model parameter. However, given the complexity of the model fitting (as explained below), this would result in a prohibitively large number of model 
spectra to compare to a limited number of observed points for each object. For this reason, we fix $f$ to 0.3, so that the total emission flux of the \fa line, over all energies, is 30 per cent that of the input X-ray continuum spectral flux. This is in rough agreement to the observed flux ratio between the reprocessing component flux over the continuum flux at soft energies \citep[e.g.][]{crummy06}. The effect of $f$, in the resulting time-lag spectra, is discussed in Section~\ref{sect:sum_disc} and in the Appendix~\ref{app:reflect_fracti}. The second component consists of a simple power-law, $PL_\nu(A,s)=A\nu^{-s}$, providing us with positive time-lags. Thus, we deal with a five-dimensional model parameter space consisting of model parameter vectors of the form, $\mathbf{v}=\{M,\alpha,\theta,h,A,s\}$.\par
In this framework, the overall time-lag spectral model at a given frequency, $\nu$, is given by the sum of the two components
\eqb
TL_\nu(\mathbf{v})=\tau_\nu(M,\alpha,\theta,h)+PL_\nu(A,s)
\label{eq:overall_tlSpec}
\eqe
\begin{figure*}
\includegraphics[width=2.3in]{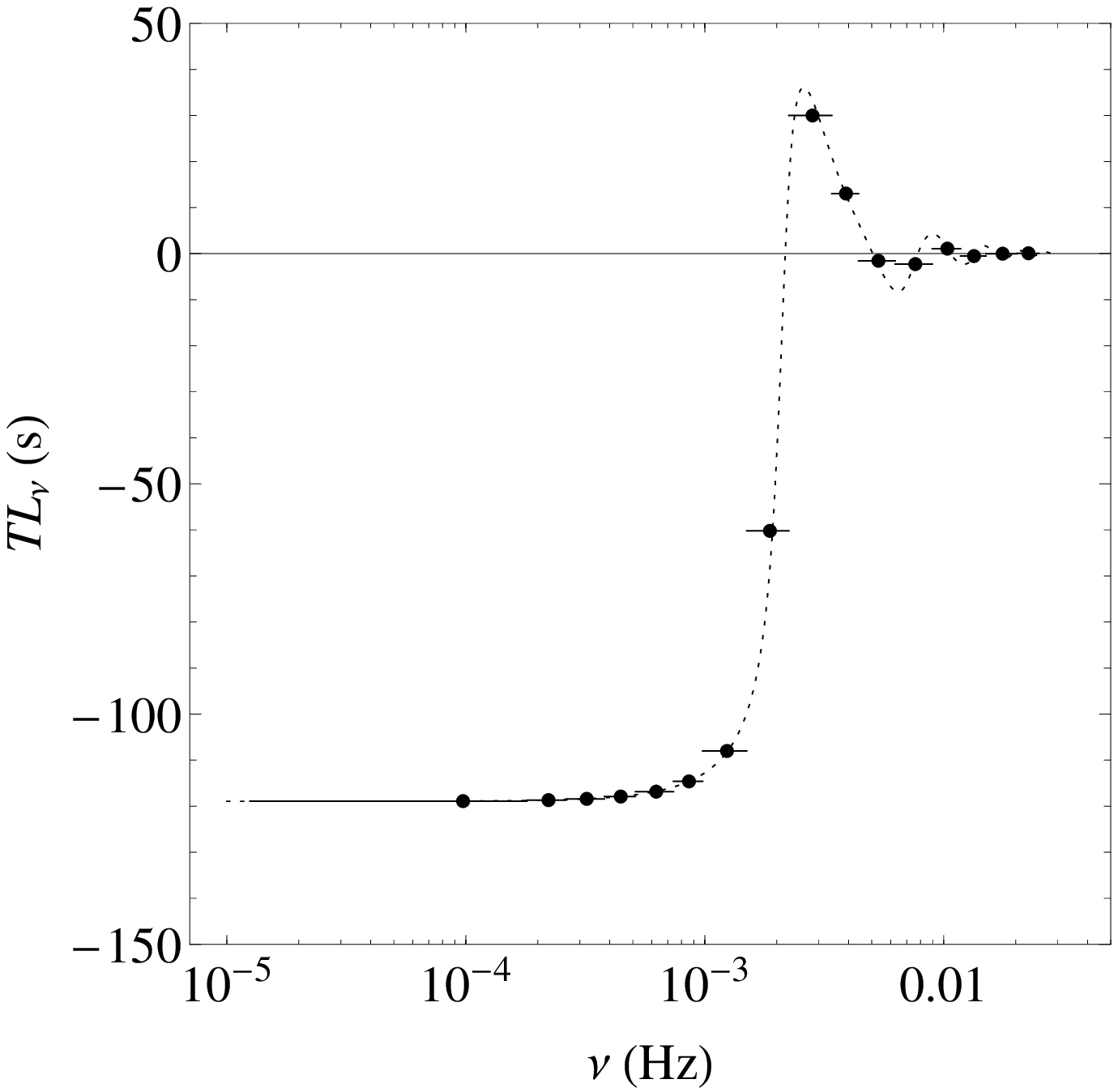}
\includegraphics[width=2.3in]{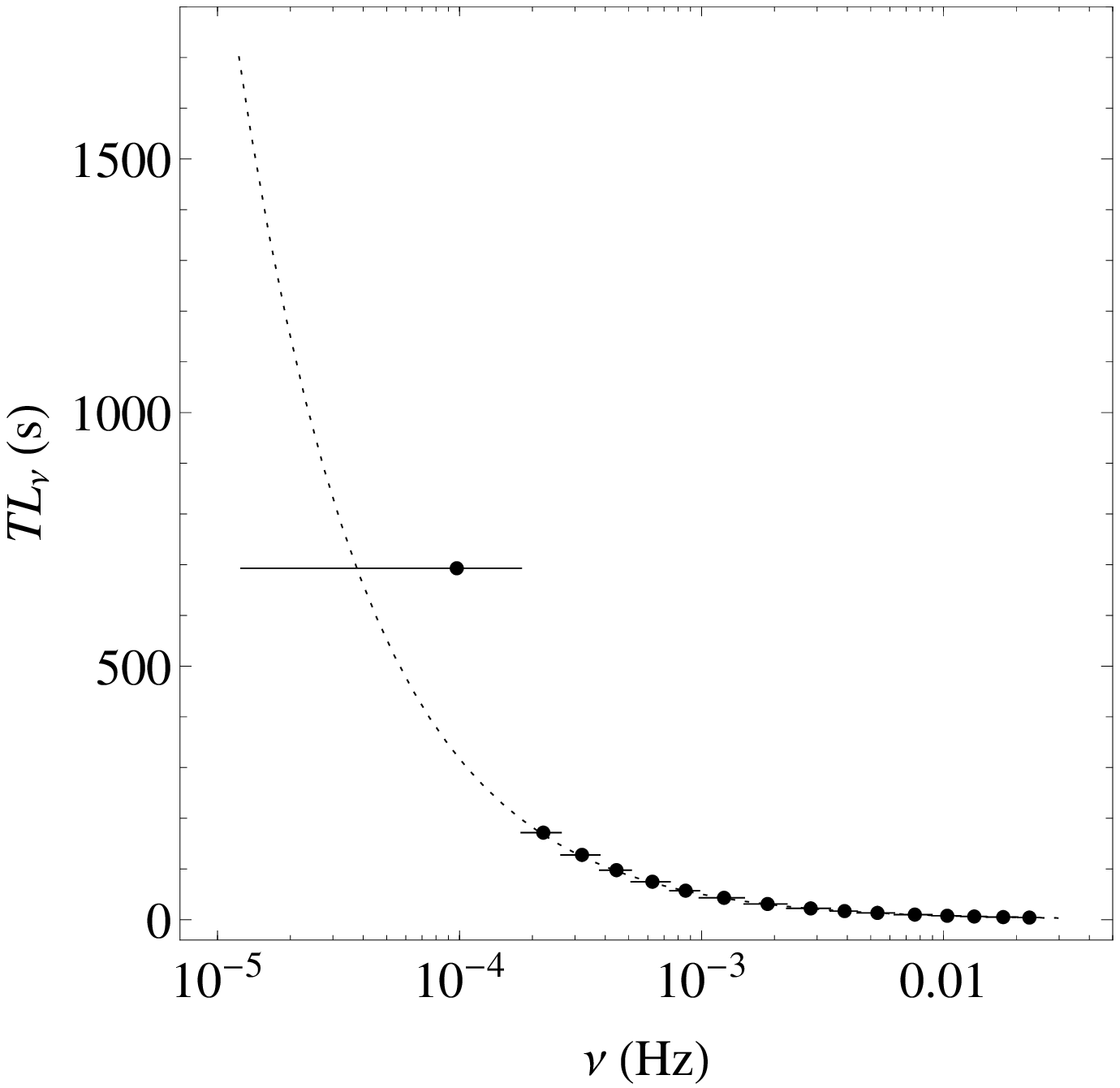}
\includegraphics[width=2.3in]{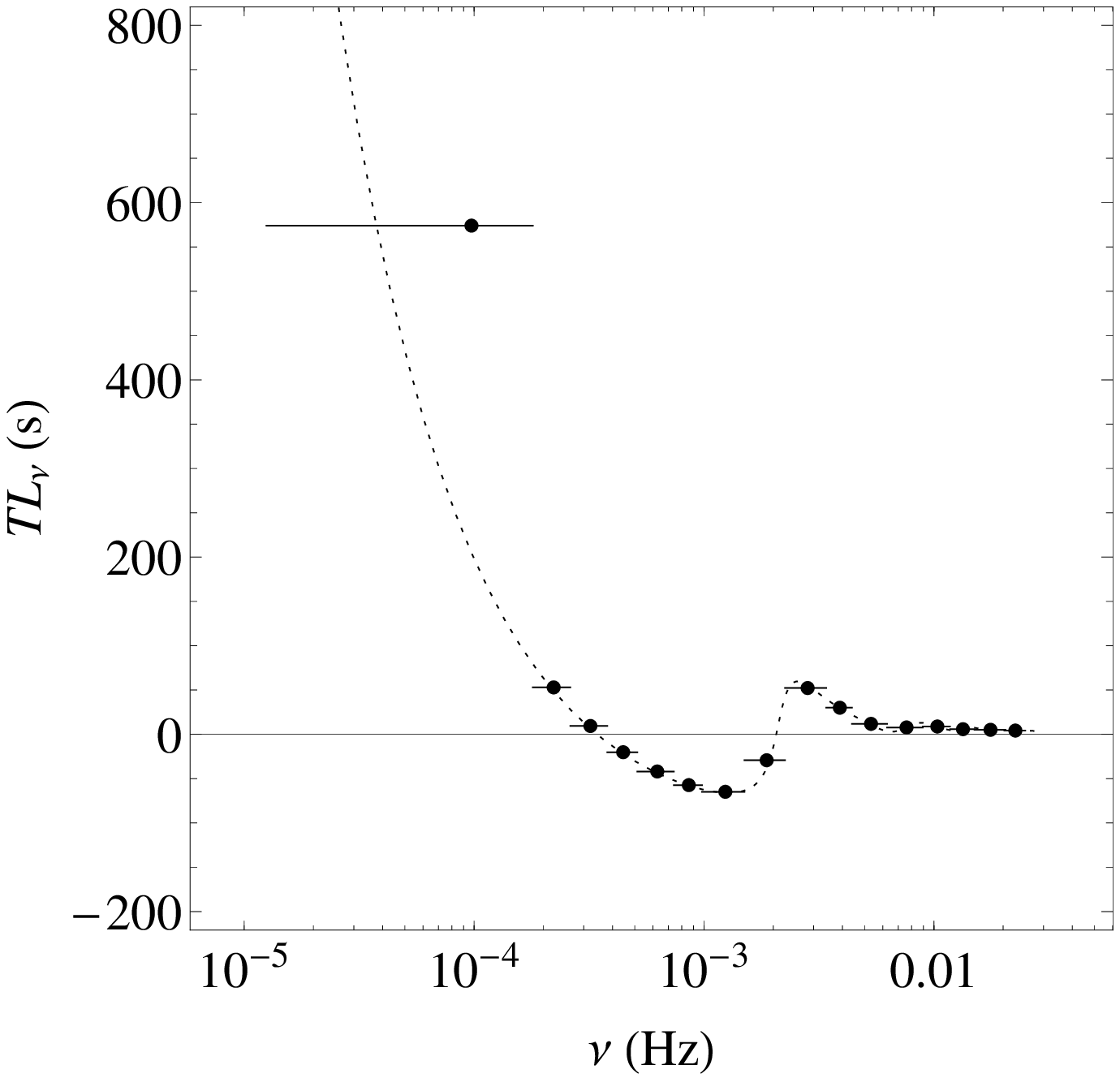}
\caption{Discretization and averaging effects for the lamp-post case model with $\alpha=0.676$, $\theta=20\degr$, $h=33.1$ \rg, for M=$2\times10^6$\ms\; and power-law parameters $A=0.2$ and $s=0.8$. The continuous models are shown with the dotted lines and the corresponding discrete estimates (averaged over the frequency ranges indicated by the horizontal lines) with the filled circles. Left-hand panel: The GR reflected component. Middle panel: The power-law component. Left-hand panel: The overall time-lag model.}
\label{fig:continTLfitModel}
\end{figure*}
Then, we transform each one of the 171 GR reflected model components into physical time units (using equation~\ref{eq:mass_scaling}) for an ensemble of 12 BH masses: $\left(0.01, 0.05,0.1,0.5,1,2,5,10,50,100,200,500\right)\times10^6$ \ms, yielding a grid of 2052 cells, each one corresponding to a given set of parameter values $\{M,\alpha,\theta,h\}$. For each grid-cell we treat the model (i.e.\ equation~\ref{eq:overall_tlSpec}) in exactly the same way as the observed time-lag spectra i.e.\ we discretize it and average it over exactly the same number of frequency bins. This yields for each frequency, $f_{{\rm bin},i}$, an average value of the overall time-lag model, $\left<TL_{f_{{\rm bin},i}}(\mathbf{v})\right>$ for $i=1,2,\ldots,m$.\par
For demonstrative purposes, in Fig.\ref{fig:continTLfitModel} we show the effects of the discretization and averaging on the continuous time-lag model $TL_\nu(2\times10^6\;{\rm M}_{\rm \sun},0.676,20\degr,33.1\;r_{\rm g},0.2\;\rm{s},0.8)$ and on its two components. For the discretization we assume an observational data set of 80 ks binned in 20 s, yielding 2000 model estimates at Fourier frequencies between $(1.25\times10^{-5}-2.5\times10^{-2})$ Hz. Then, these binned model estimates are averaged over 16 frequency bins whose overall range is shown with the horizontal line. The left-hand panel of Fig.\ref{fig:continTLfitModel} shows the GR reflected component, $\tau_\nu(2\times10^6\;{\rm M}_{\rm \sun},0.676,20\degr,33.1\;r_{\rm g})$, exhibiting in the continuous version a great deal of `oscillatory-structures' towards high frequencies $(4\times10^{-3}-2\times10^{-2})$ Hz which are suppressed during the averaging process. The middle-panel of Fig.\ref{fig:continTLfitModel} shows the power-law model component, 
$PL_\nu(0.2\;\rm{s},0.8)$, in which, once again, the discrete estimates differ from the actual continuous power-law model estimates, at the corresponding frequencies, due to the averaging process. The most prominent differences are at the lowest frequency bins, particularly the first one at $10^{-4}$ Hz. Finally, the right-hand panel of Fig.\ref{fig:continTLfitModel} shows the overall time-lag model, i.e.\ the sum of the previous two model components. The discrete averaged version of the overall time-lag model carries all the previously mentioned differences with respect to its continuous version.\par
Then for each averaged discretized overall time-lag spectral model, within each grid-cell, $k$, we estimate the squared differences between the model and the data using the following $\chi^2$ indicator
\eqb
\chi^2_k(\mathbf{v})=\sum_{i=1}^{m}\frac{\left(\left<TL_{f_{{\rm bin},i}}(\mathbf{v})\right>-\tau(f_{{\rm bin},i})\right)^2}{\operatorname{std}\{\tau(f_{{\rm bin},i})\}^2}
\label{eq:chi_indic}
\eqe
and for each grid-cell we minimise this quantity with respect to the two power-law model parameters, $\{A,s\}$. Finally, we end up with an ensemble of 2052 global minimum values of $\chi^2_k(\mathbf{v})$ from which the smallest one corresponds to the set of the best-fitting grid model parameter values $\mathbf{v}_{\rm gbf}=\{M_{\rm gbf},\alpha_{\rm gbf},\theta_{\rm gbf},h_{\rm gb},A_{\rm gbf},s_{\rm gbf}\}$. For the minimisation we use the the classical Levenberg-Marquardt method \citep{bevington92}.\par
After localising the $\mathbf{v}_{\rm gbf}$, we estimate the 68.3 per cent confidence bands for each best-fitting model parameter in the usual way i.e.\ by varying its value over a given range and deriving each time for the rest model parameters their best-fitting value that yield a $\Delta\chi^2=1$ from the minimum $\chi^2_k(\mathbf{\mathbf{v}_{\rm gbf}})$. During the error estimation process we cubically interpolate the $\chi^2_k(\mathbf{v})$ space around its BF best-fitting grid parameter values in the following way: two BH mass cells above and below $M_{\rm gbf}$, two height cells above and below $h_{\rm gbf}$ and all the spin parameters and viewing angles (each one consisting of three cells). This yields a continuous version of $\chi^2_k(\mathbf{v})$ around $\mathbf{v}_{\rm gbf}$, $\chi^2(\mathbf{v})$. Thus, during the error estimation we use the $\chi^2_k(\mathbf{v})$ space around the best-fitting grid model parameters covering $5\times5\times3\times3=225$ grid-cells, leaving the power-law model 
parameters $\{
A,s\}$ to vary freely. Since $\mathbf{v}_{\rm gbf}$ has been derived from the grid-cells (i.e.\ fixed values), during the error estimation process new values of $\chi^2(\mathbf{v})$ are emerging from the intermediate grid-cell values, $\chi^2(\mathbf{v_{\rm bf}})$, within the selected best-fitting parameter region. The quantity $\chi^2(\mathbf{v_{\rm bf}})$ indicates the final best-fitting parameter values that we keep from our fitting process, $\mathbf{v}_{\rm bf}=\{M_{\rm bf},\alpha_{\rm bf},\theta_{\rm bf},h_{\rm bf},A_{\rm bf},s_{\rm bf}\}$.\par
Despite the fact that we can not visualize the interpolated $\chi^2(\mathbf{v})$ space due to its high dimensionality (6-dimensions) in Appendix~\ref{ssect:interpolTLspectra} (Fig.~\ref{fig:interpolSpin}, \ref{fig:interpolAngle}, \ref{fig:interpolHeight}) we show the interpolated versions of the reflected components, $\tau_\nu(M,\alpha,\theta,h)$, which are the actual discrete components that we interpolate in equation~\ref{eq:chi_indic} which appear via $TL_{f_{{\rm bin},i}}(\mathbf{v})$ (equation~\ref{eq:overall_tlSpec}).\par
Some important points of the overall procedure:
\begin{itemize}
 \item The best-fitting values derived from the grid parameter space, $\mathbf{v}_{\rm gbf}$, are, in fact, very close to the best-fitting values derived from the interpolated parameter space, $\mathbf{v}_{\rm bf}$. All the $\mathbf{v}_{\rm gbf}$ lie within the 86.6 per cent confidence intervals (1.5 standard deviations) of the best-fitting values of $\mathbf{v}_{\rm bf}$. Note that the only reason that we perform the interpolation in the first place is for the derivation of the 68.3 per cent confidence bands and the $\mathbf{v}_{\rm bf}$ is a natural product of this process. 
 \item We leave the power-law parameters free since their values fix the normalisation of the overall model. Note that for a given set of physical lamp-post model parameter the negative reflected component, $\tau_\nu(M,\alpha,\theta,h)$ is absolutely fixed. Thus both the normalisation and the index, $\{A,s\}$, of the positive power-law component define the level of the final overall time-lag spectral model.
 \item The cubic interpolation, that we use for the derivation of the final best-fitting model parameters and their corresponding uncertainties, smooths the $\chi^2_k(\mathbf{v})$ space adequately in order the estimation of the various gradients (needed for the orientation of the minimisation algorithm) to become easier. Linear or quadratic interpolation yields edges in the parameter space, higher order interpolations (i.e.\ greater than three) create artefacts in the parameter space. Note that spline interpolation yields equivalent results as the cubic interpolation.
 \item In principle, the BH mass can be left as a free fitting parameter, through equation~\ref{eq:mass_scaling}, together with the power-law parameters $A$ and $s$. However, such an addition slows significantly the minimization process since one has to minimise equation~\ref{eq:chi_indic} simultaneously with another equation describing the corresponding scaling in the abscissae (i.e.\ frequency domain). The second equation would consist of the squared differences in the abscissae between the model and the data. For the needs of this paper and the quality of our time-lag spectra the grid approach for the BH masses is very robust.
\end{itemize}

\section{RESULTS}
\label{sect:res}
\begin{figure*}
\includegraphics[width=2.2in]{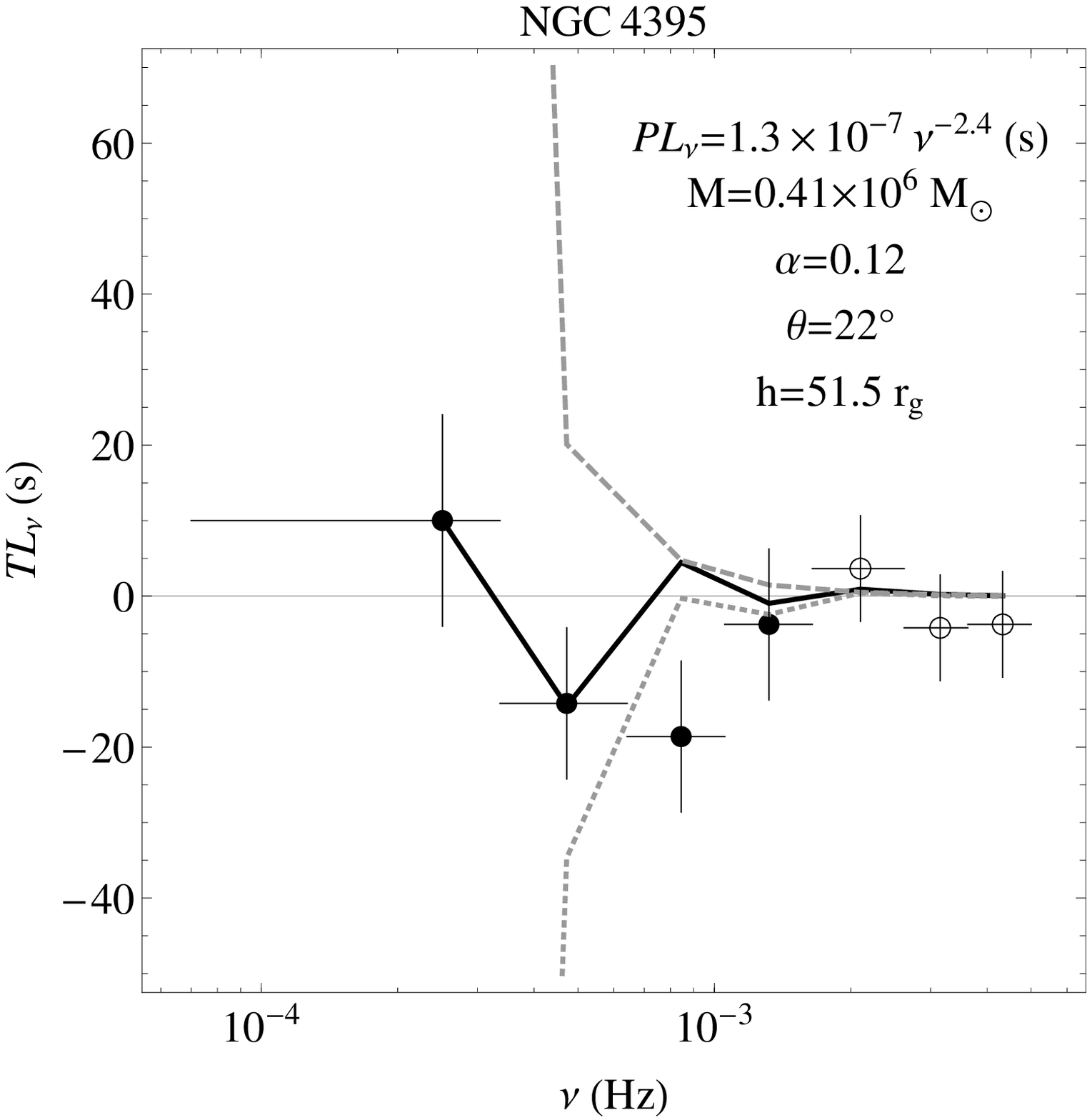}
\includegraphics[width=2.2in]{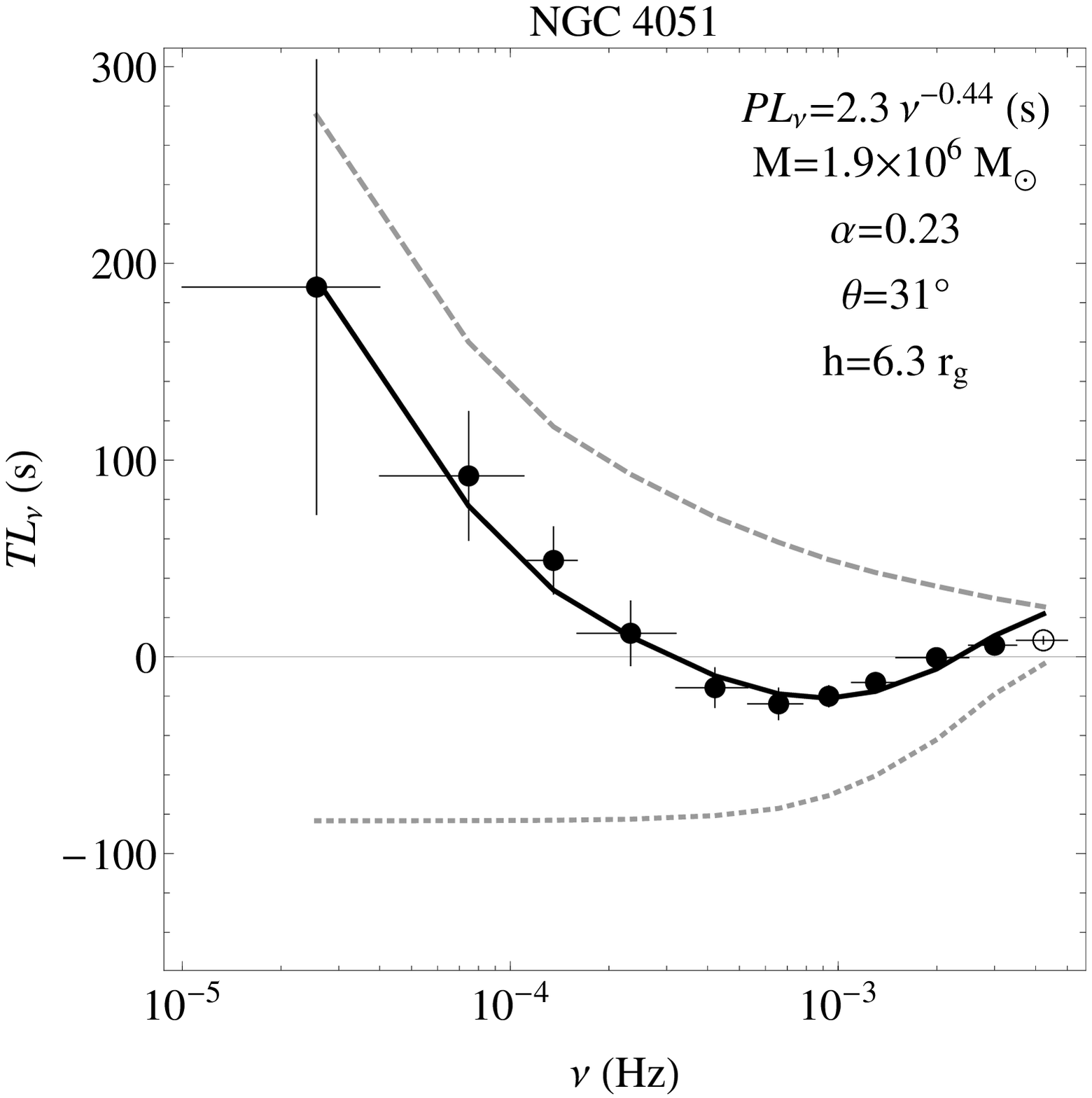}
\includegraphics[width=2.2in]{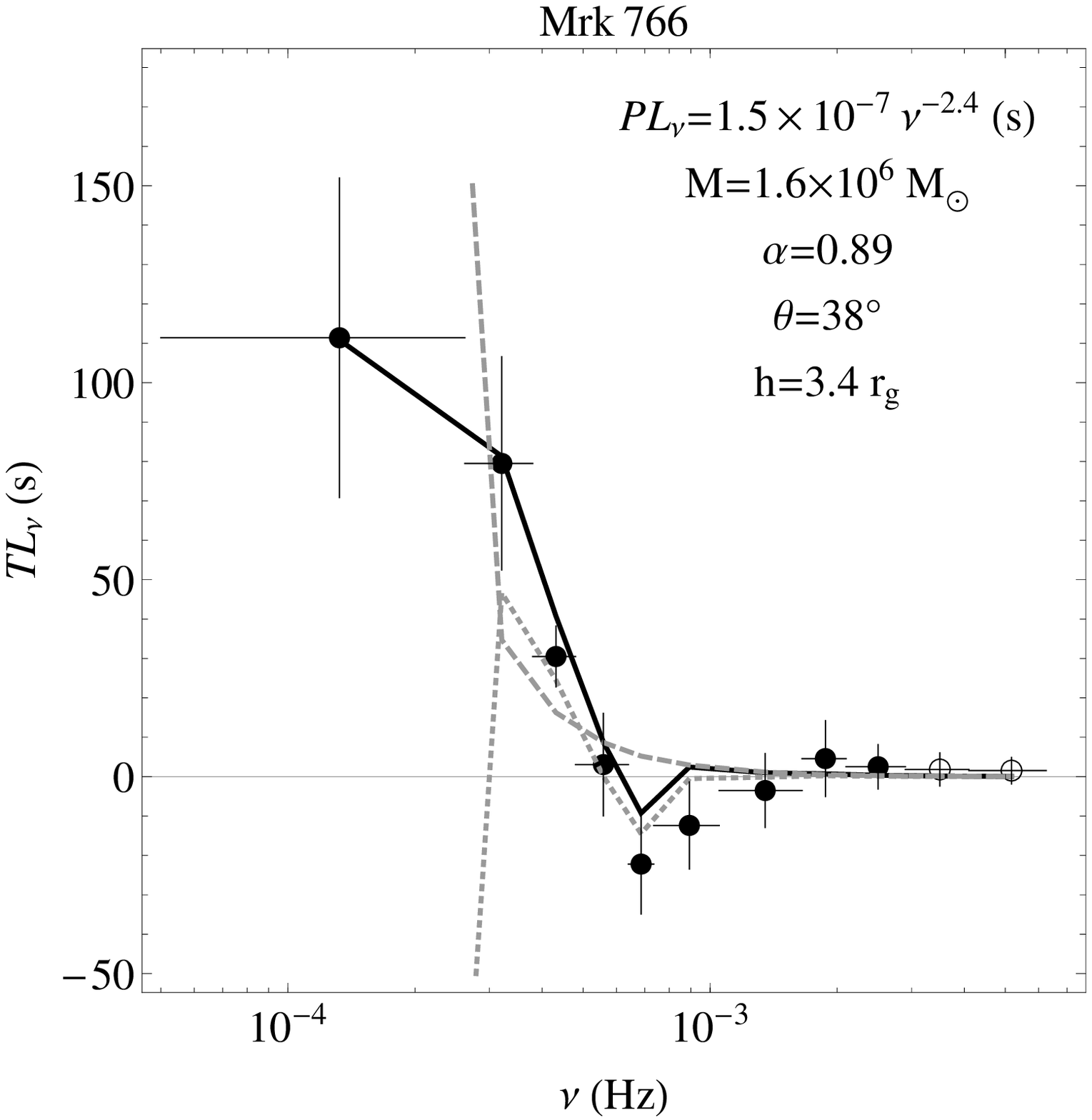}\\
\includegraphics[width=2.2in]{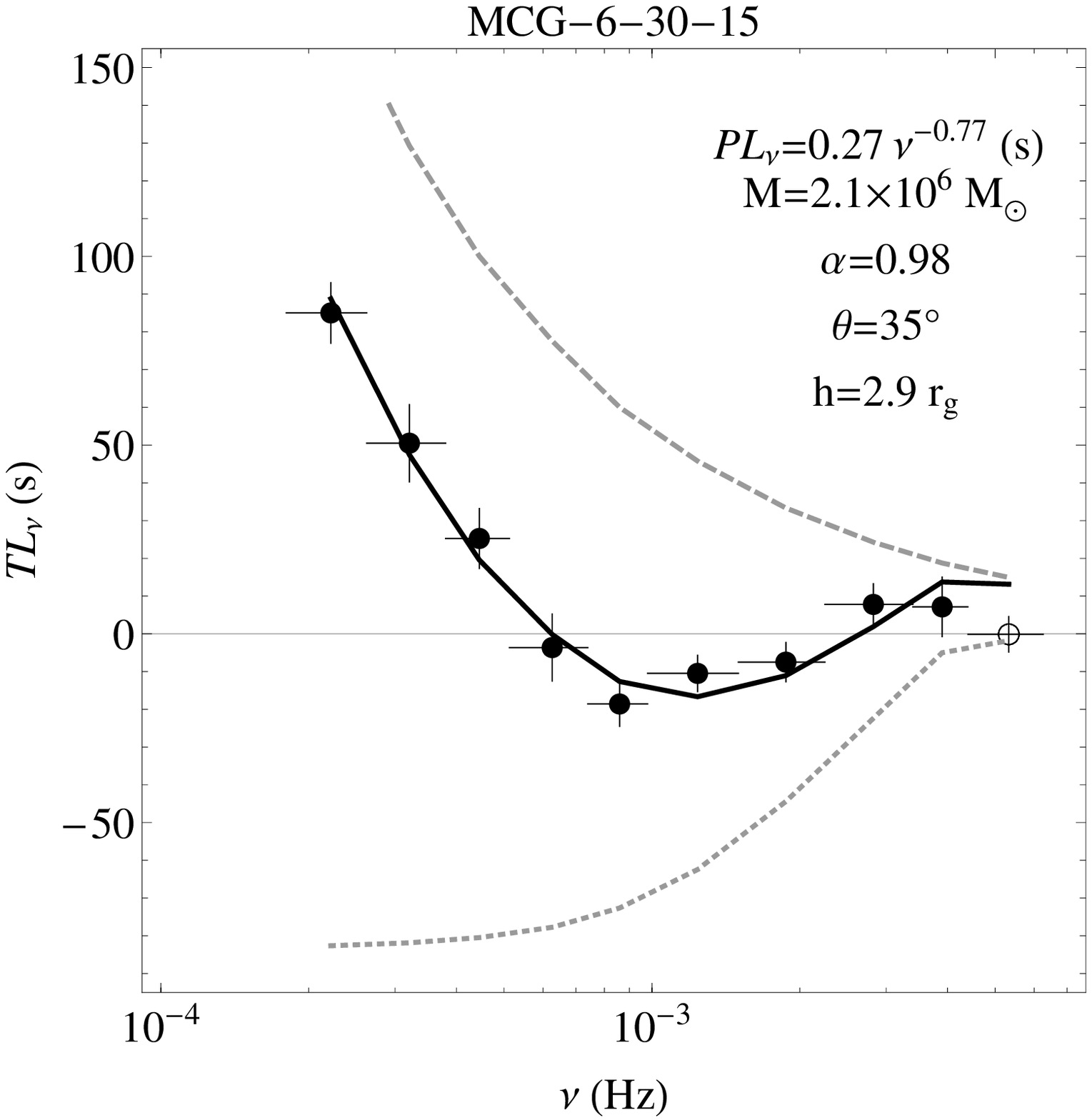}
\includegraphics[width=2.2in]{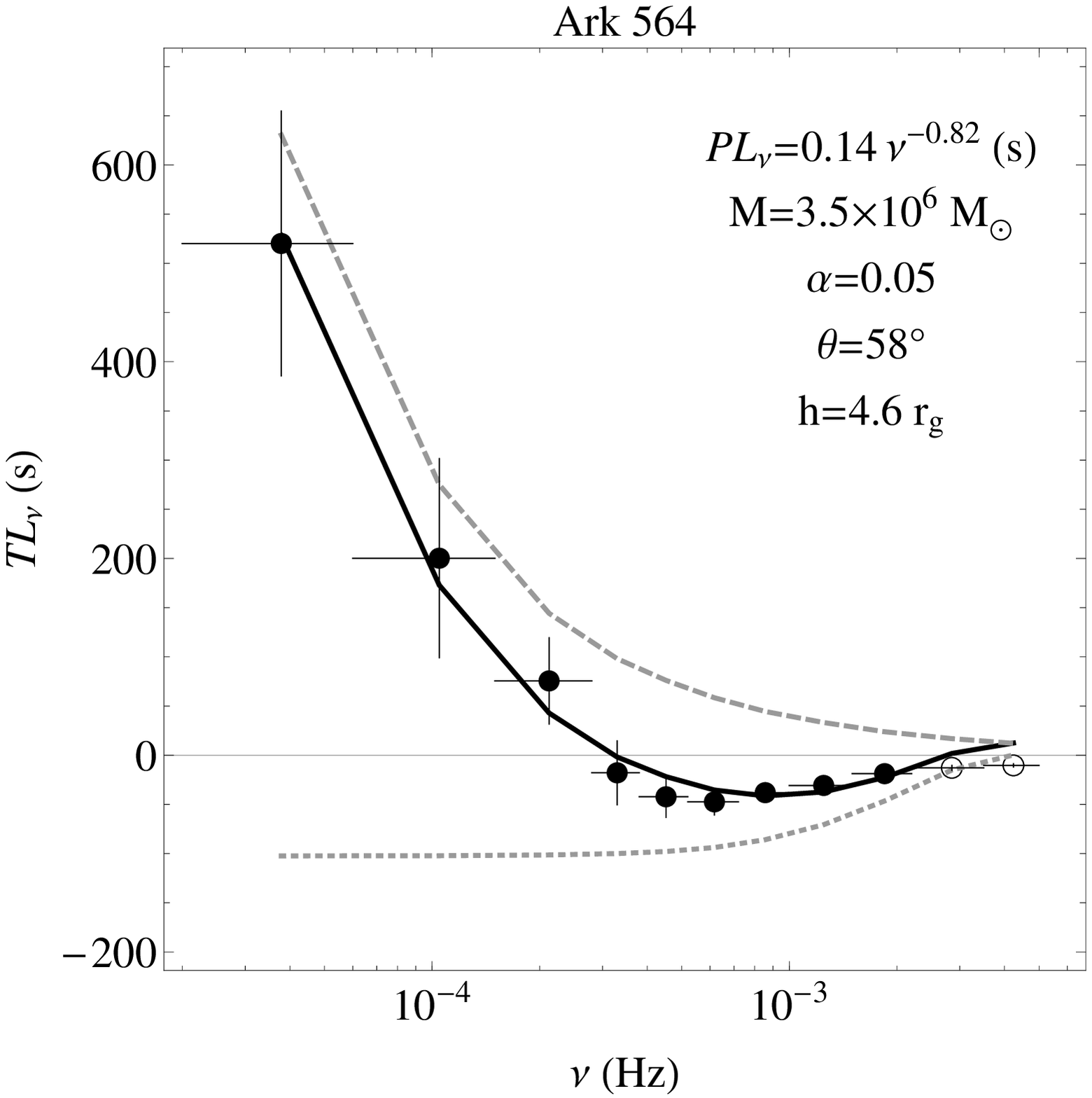}
\includegraphics[width=2.2in]{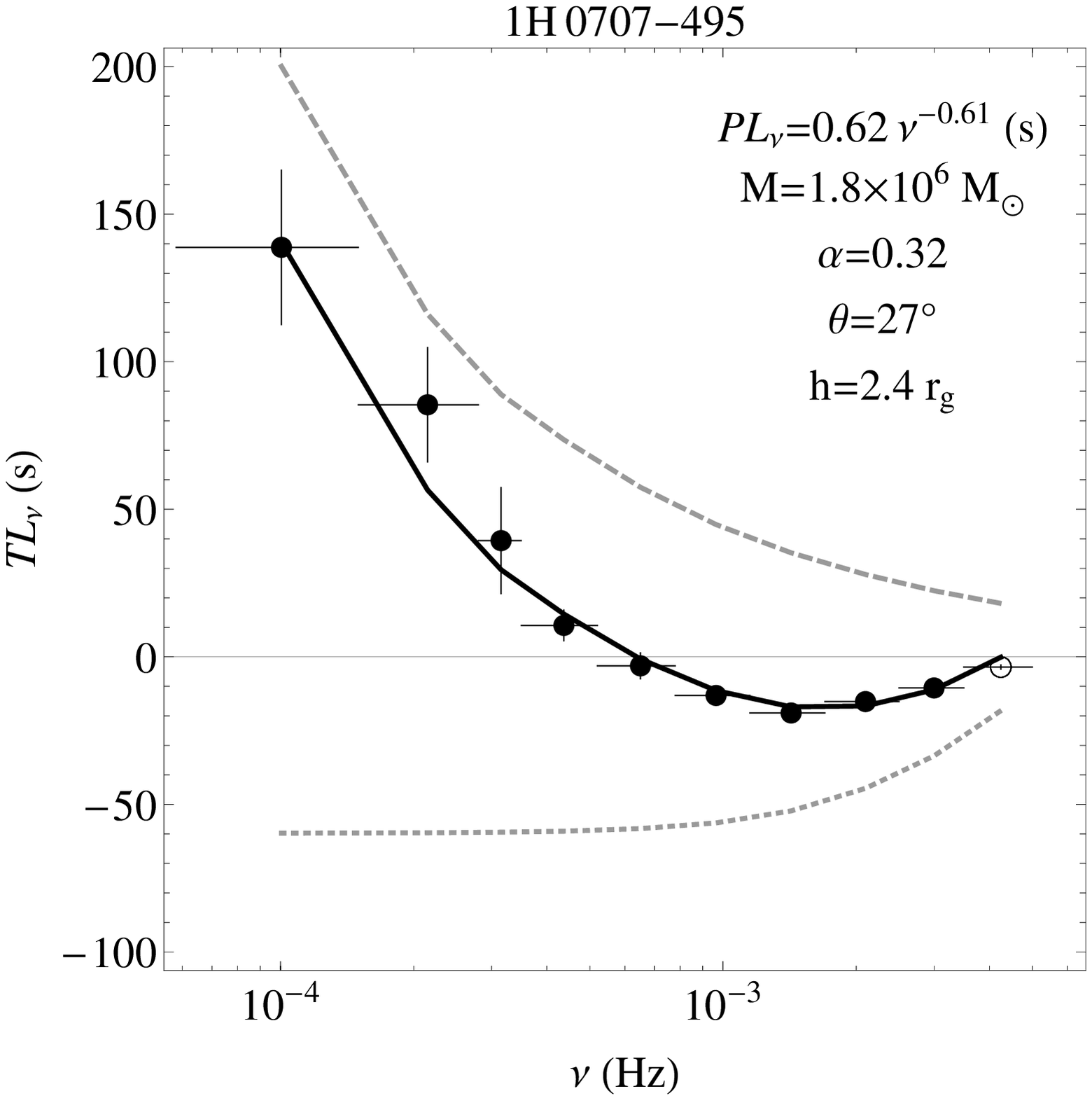}\\
\includegraphics[width=2.2in]{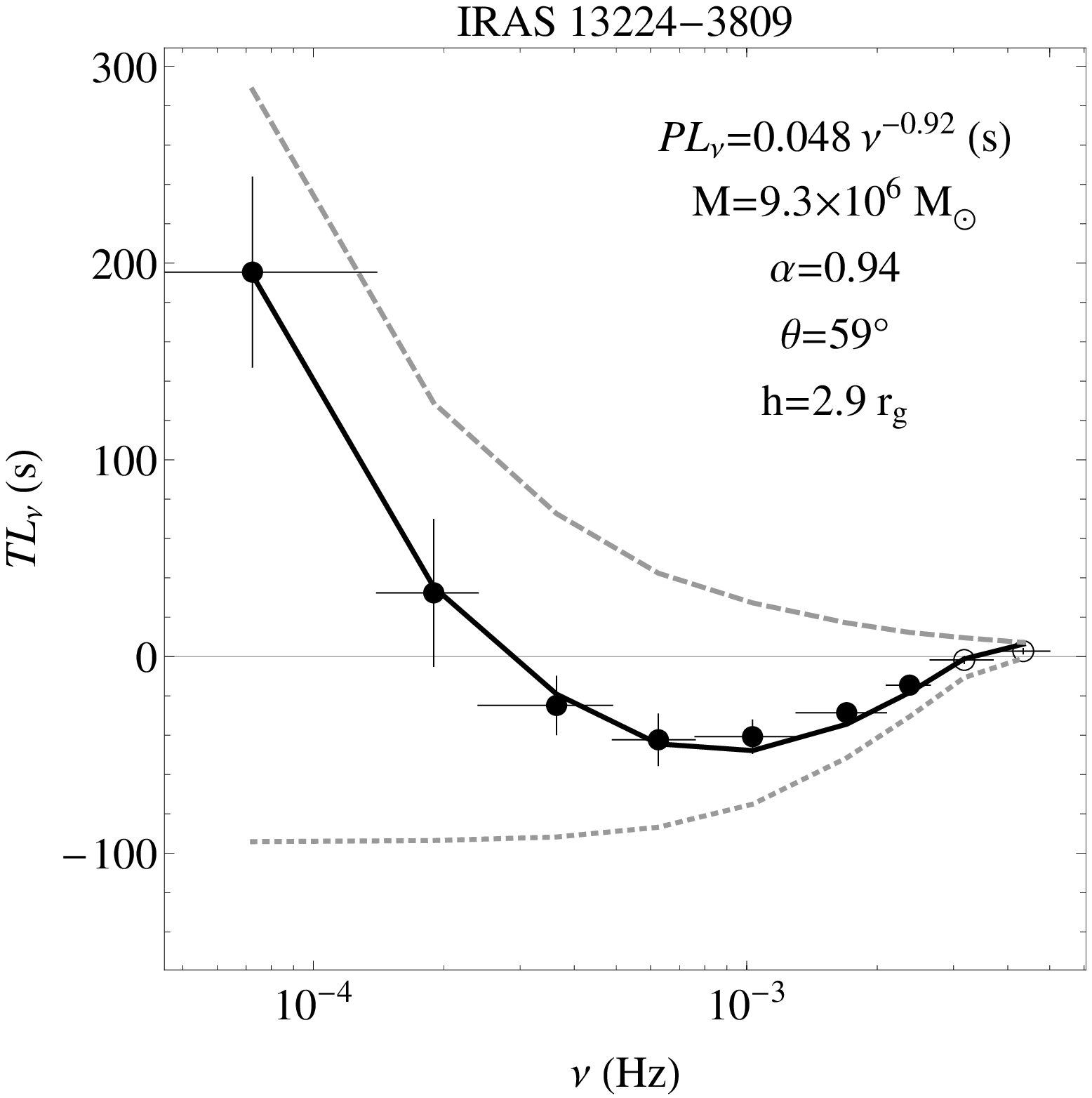}
\includegraphics[width=2.2in]{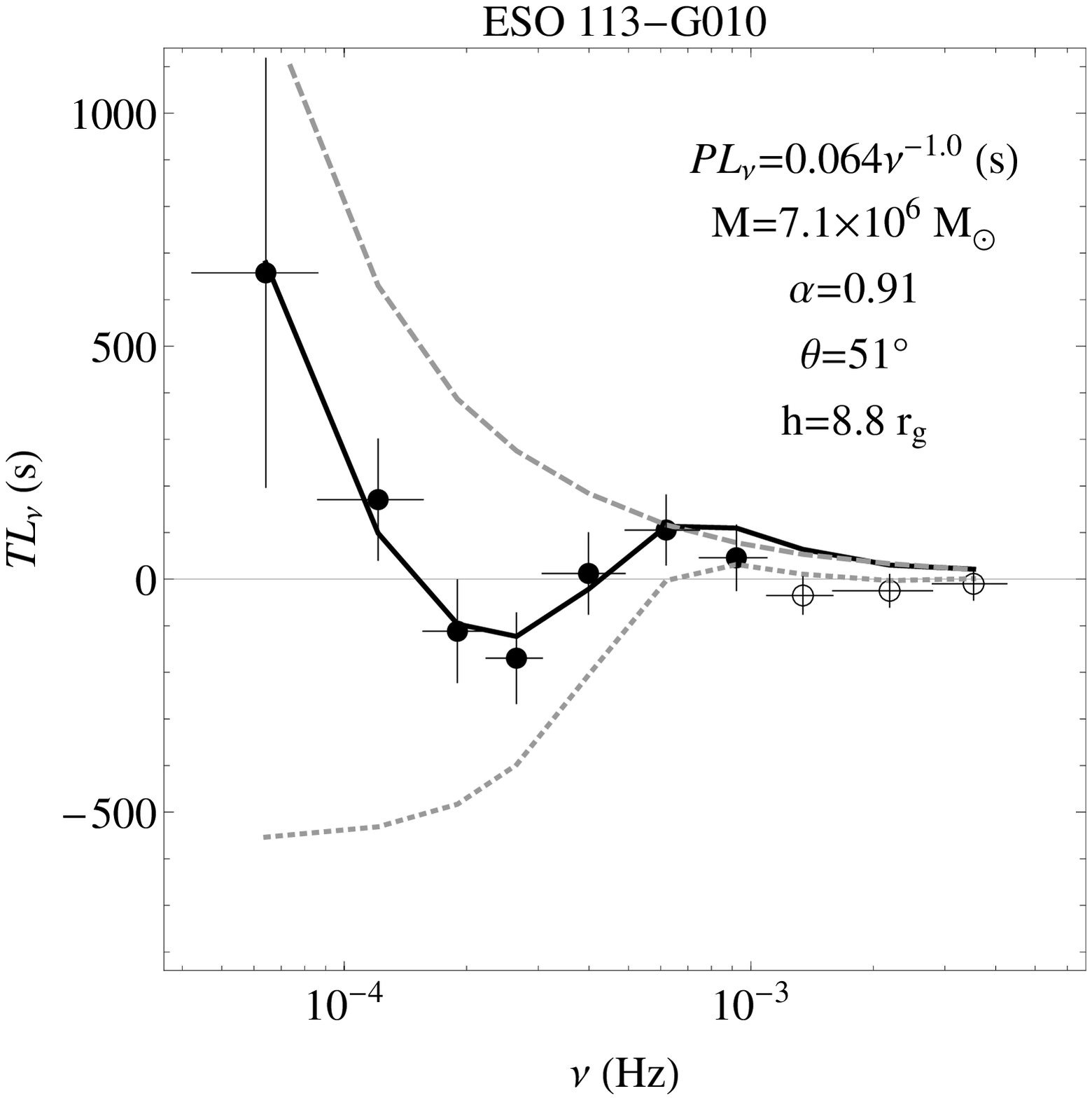}
\includegraphics[width=2.2in]{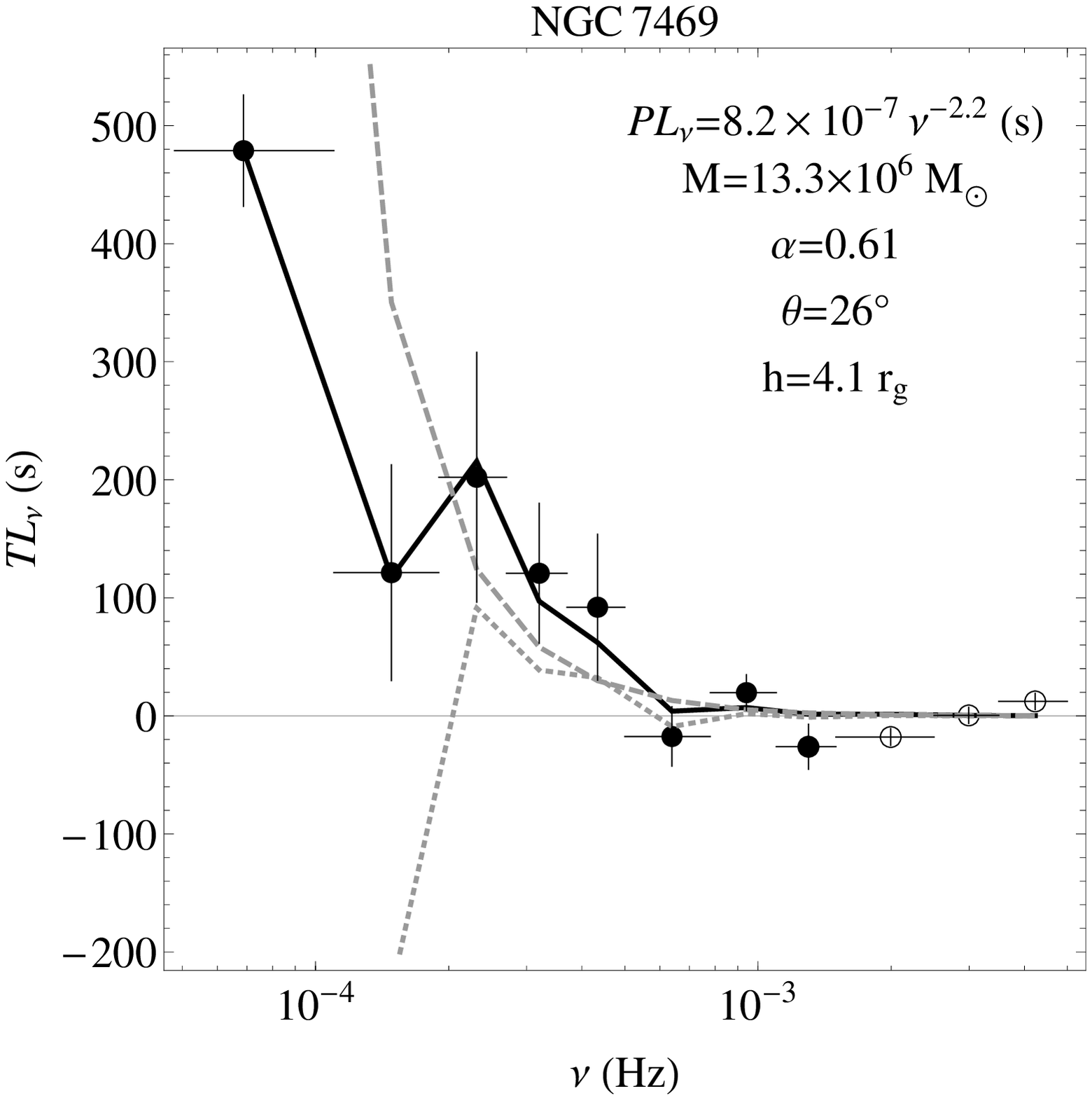}\\
\includegraphics[width=2.2in]{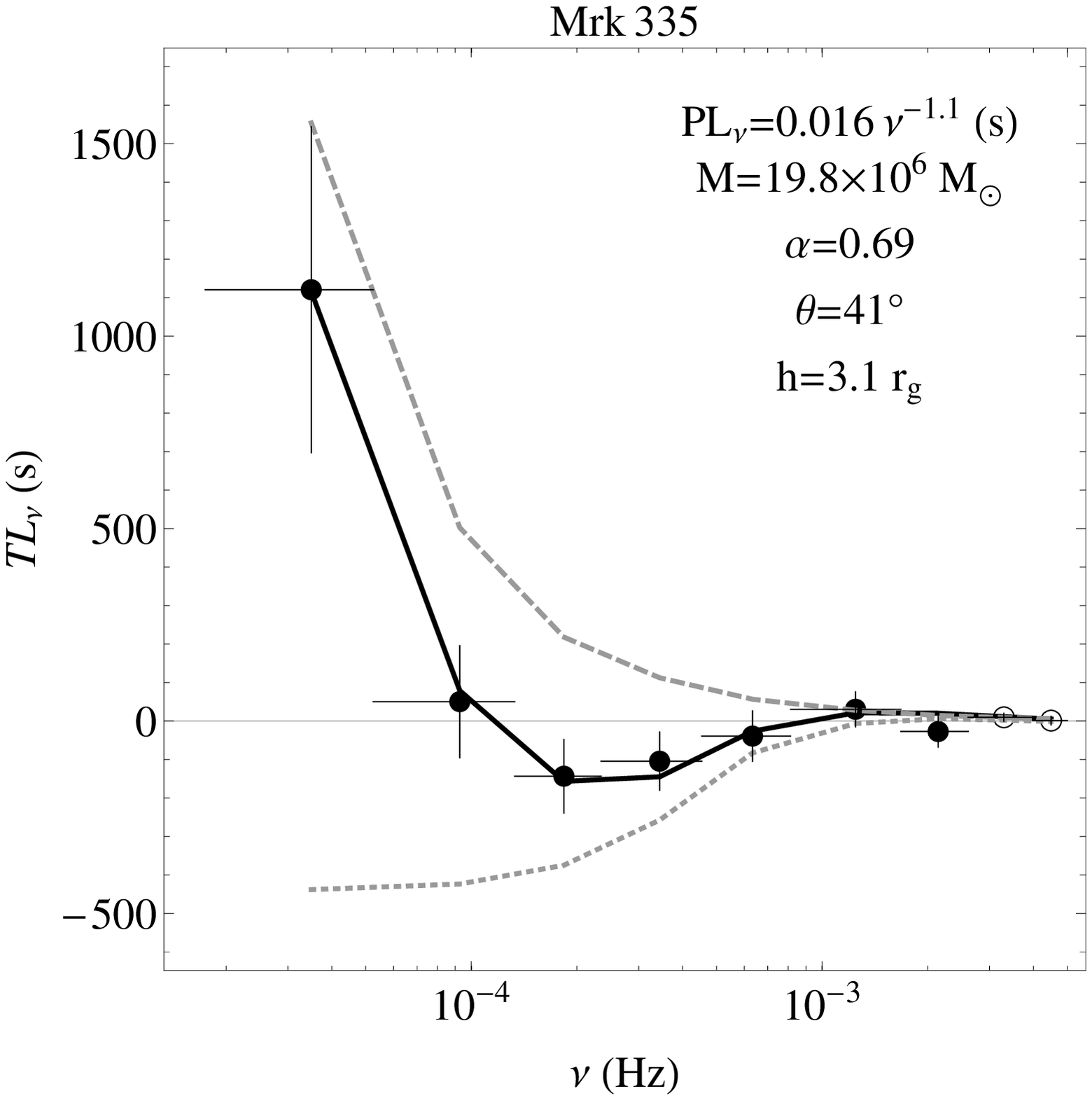}
\includegraphics[width=2.2in]{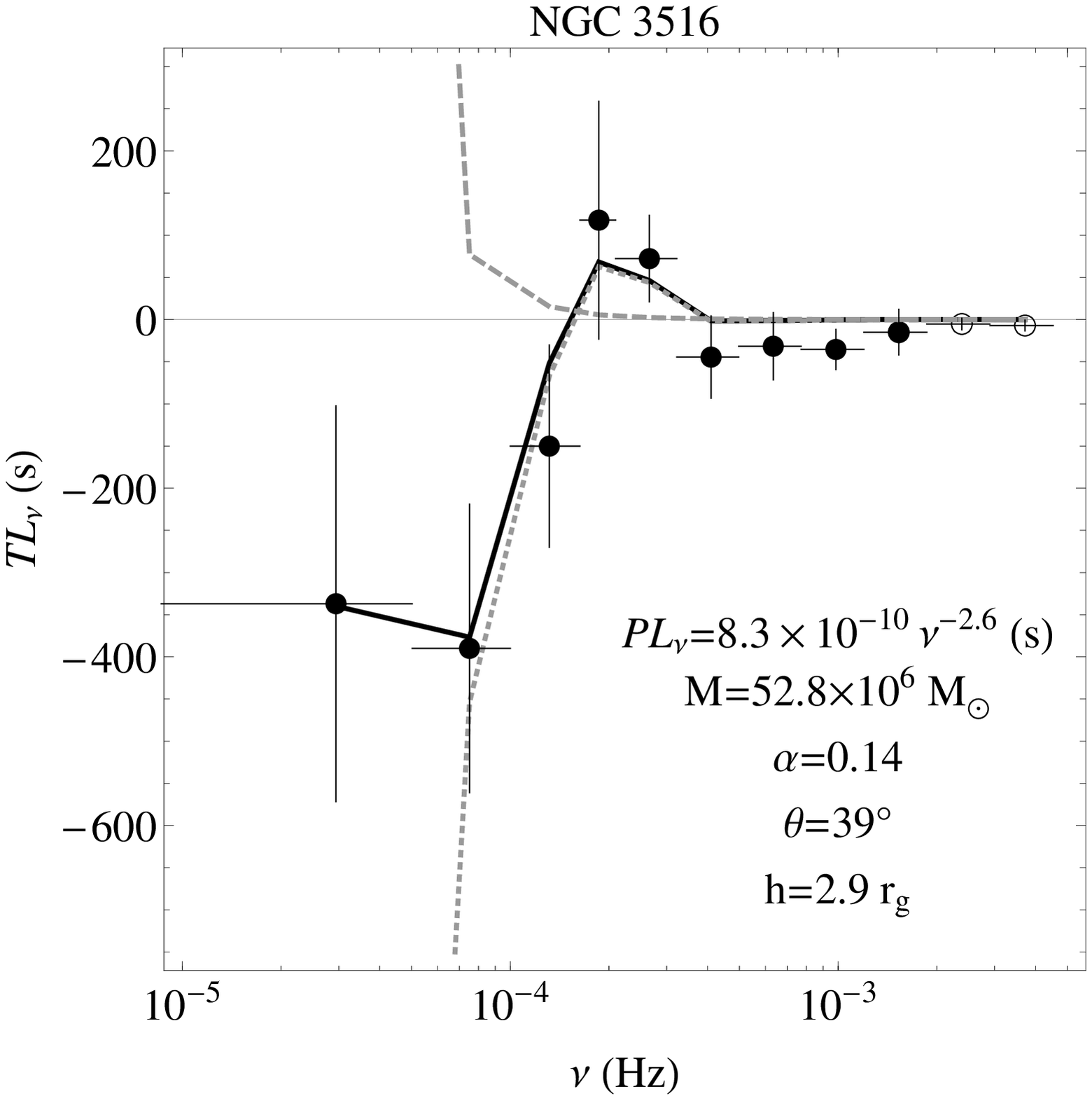}
\includegraphics[width=2.2in]{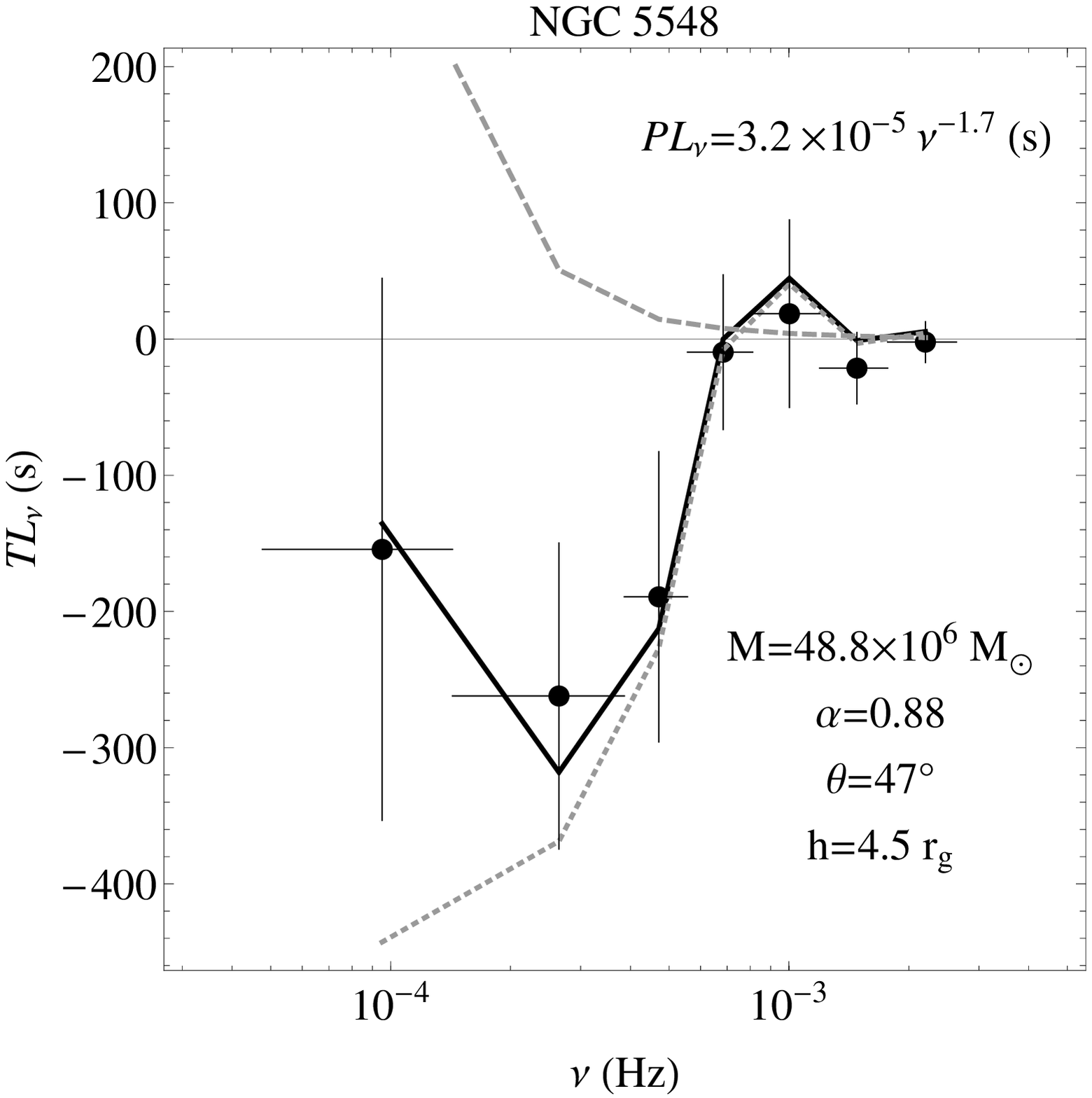}
\caption{The best-fitting time-lag spectral models (black line) and the time-lag estimates with coherence greater than 0.15 (filled circles), --open circles correspond to lower coherence values-- The two components of the best-fitting time-lag model are the GR reflected component (grey-dotted line) and the power-law (grey-dashed line).}
\label{fig:TLSpecFit}
\end{figure*}

The observed time-lag spectra for all the 12 AGN together the best-fitting overall time-lag spectral models are shown in Fig.~\ref{fig:TLSpecFit}. The filled circles correspond to the time-lag estimates with a coherence greater than 0.15 (Section~\ref{ssect:TL_data}) and the open circles to those time-lag estimates with coherence smaller than 0.15 (i.e.\ those excluded from the fitting procedure). All the time-lag spectra agree very well, within the estimated errors, with those reported by \citet{demarco13} with the only exception being that of NGC\;7469. This difference could be caused by the slightly different selection of beginning and end times or/and by the different version of the Current Calibration Files (CCFs) which are updated on a regular basis. The best-fitting overall time-lag spectral models are shown with the black line. The panels also include the constituents components of the best-fitting time-lag spectral model; the power-law and the GR reflected component with the grey-dashed and the grey-
dotted lines, respectively. The best-fitting parameters, together with the 68.3 per cent confidence intervals, are given in Table~\ref{tab:bfTL}.

\begin{table*}
\begin{minipage}{180mm}
\caption{The best-fitting time-lag spectra models consisting of the GR reflected component and the power-law (PL). The first column, (1), is the name of the AGN source, the next four columns, (2,3,4 and 5), are the model parameters of the GR lamp-post component (reflected component). The next two columns, (6 and 7), are the model parameters of the power-law. The last column, (8), is the $\chi^2$ indicator coming from the interpolated region around $\chi^2_k(\mathbf{v})$. The quoted errors correspond to the 68.3 per cent confidence intervals around the best-fitting model parameters with `\text{---}' indicating the cases where the uncertainty could not be estimated due to the finite extension of the interpolated grid.}
\label{tab:bfTL}
\begin{tabular}{@{}cccccccc}
\hline
(1) & (2) & (3) & (4) & (5) & (6) & (7) & (8) \\
AGN name & BH mass & BH spin & Viewing angle & Height & PL normalisation & PL index & $\chi^2(\mathbf{v_{\rm bf}})$/d.o.f. \\ 
 & $M$  ($\times 10^6$ \ms)   & $\alpha$   & $\theta\degr$ & $h$ (\rg) & $A$ (s) & $s$ & \\
\hline
NGC\;4395 & $0.41^{+0.19}_{-0.29}$ & $0.12^{+0.42}_{-\text{---}}$ & $22^{+18}_{-\text{---}}$ & $51.5^{+21.2}_{-36.8}$ & $\left(1.3^{+0.8}_{-0.6}\right)\times10^{-7}$ & $2.4^{+0.4}_{-1.3}$ & 6.24\\
NGC\;4051 & $1.9^{+1.3}_{-1.1}$ & $0.23^{+0.31}_{-\text{---}}$ & $31^{+6}_{-8}$ & $6.3^{+2.2}_{-3.8}$ & $2.3^{+0.3}_{-0.2}$ & $0.44^{+0.15}_{-0.08}$ & 1.89\\
Mrk\;766 & $1.6^{+1.4}_{-1.2}$ & $0.89^{+\text{---}}_{-0.33}$ & $38^{+7}_{-9}$ & $3.4^{+2.8}_{-1.8}$ & $\left(1.5^{+1.1}_{-0.8}\right)\times10^{-7}$ & $2.4^{+0.3}_{-1.1}$ & 1.78\\
MCG--6-30-15 & $2.1^{+0.9}_{-1.6}$ & $0.98^{+\text{---}}_{-0.26}$ & $35^{+11}_{-8}$ & $2.9^{+0.4}_{-0.7}$ & $0.27^{+0.4}_{-0.3}$ & $0.77^{+0.2}_{-0.1}$ & 1.59 \\
Ark\;564 & $3.5^{+1.3}_{-1.8}$ & $0.05^{+0.45}_{-\text{---}}$ & $58^{+\text{---}}_{-12}$ & $4.6^{+0.9}_{-0.7}$ & $0.14^{+0.11}_{-0.07}$ & $0.82^{+0.34}_{-0.29}$ & 1.77\\
1H\;0707-495 & $1.8^{+1.7}_{-1.2}$ & $0.32^{+0.24}_{-0.22}$ & $27^{+9}_{-5}$ & $2.4^{+0.6}_{-0.3}$ & $0.62\pm0.23$ & $0.61^{+0.3}_{-0.1}$ & 2.70 \\
IRAS\;13224-3809 & $9.3^{+3.4}_{-2.9}$ & $0.94^{+\text{---}}_{-0.28}$ & $59^{+\text{---}}_{-11}$ & $2.9^{+0.8}_{-\text{---}}$ & $\left(4.8^{+0.5}_{-0.8}\right)\times10^{-2}$ & $0.92^{+0.32}_{-0.29}$ & 1.93\\
ESO\;113-G010 & $7.1^{+3.8}_{-4.2}$ & $0.91^{+\text{---}}_{-0.22}$ & $51^{+8}_{-7}$ & $8.8^{+0.9}_{-2.3}$ & $\left(6.4^{+1.4}_{-0.9}\right)\times10^{-2}$ & $1.0^{+0.4}_{-0.8}$ & 1.56 \\
NGC\;7469 & $13.3^{+7.1}_{-4.6}$ & $0.61^{+0.22}_{-0.29}$ & $26^{+9}_{-\text{---}}$ & $4.1^{+0.3}_{-0.8}$ & $\left(8.2^{+2.1}_{-1.5}\right)\times10^{-7}$ & $2.2^{+0.4}_{-0.9}$ & 1.97 \\
Mrk\;335 & $19.8^{+11.8}_{-10.5}$ & $0.69^{+0.28}_{-0.31}$ & $41^{+8}_{-9}$ & $3.1^{+0.5}_{-0.6}$ & $1.6\pm0.3\times10^{-2}$ & $1.1^{+0.2}_{-0.4}$ & 1.81\\
NGC\;3516 & $52.8^{+15.2}_{-14.2}$ & $0.14^{+0.35}_{-\text{---}}$ & $39^{+12}_{-9}$ & $2.9^{+1.3}_{-\text{---}}$ & $\left(8.3^{+2.4}_{-1.9}\right)\times10^{-10}$  & $2.6\pm0.5$ & 1.38\\
NGC\;5548 & $48.8^{+13.2}_{-12.6}$ & $0.88^{+\text{---}}_{-0.42}$ & $47^{+13}_{-11}$ & $4.5^{+0.9}_{-1.1}$ & $\left(3.2^{+1.9}_{-1.8}\right)\times10^{-5}$ & $1.7^{+0.5}_{-0.3}$ &1.34\\
\end{tabular}
\end{minipage}
\end{table*}

\subsection{The best-fitting BH masses}
\label{ssect:bhmass}
\begin{figure}
\includegraphics[width=3.3in]{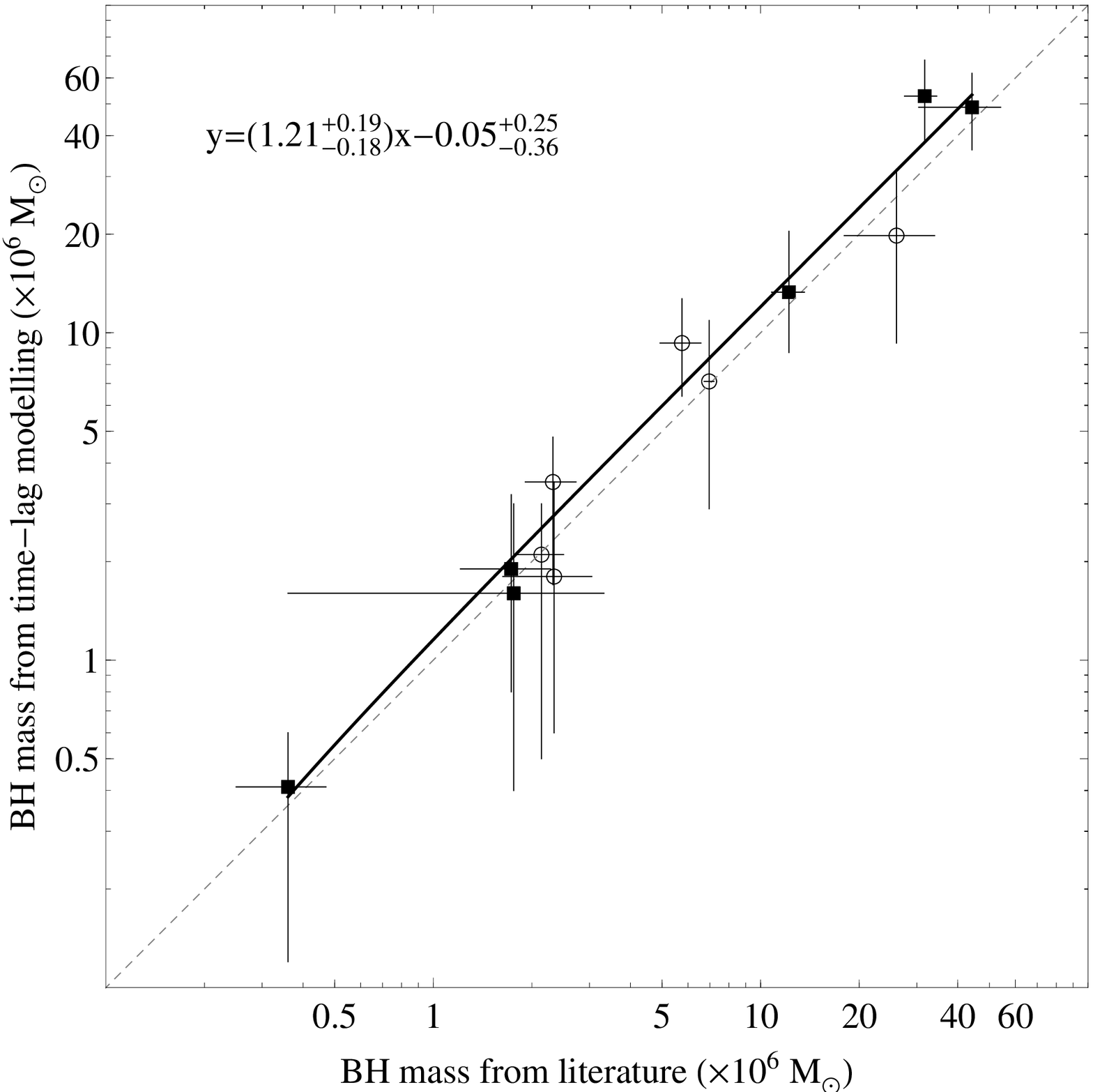}
\caption{Best-fitting BH masses versus literature values. The filled squares correspond to the literature values estimated via the reverberation technique (given with an (r) in the first column of Table~\ref{tab:obs}) and the open circles to all the other BH mass estimates. The black solid line corresponds to the best-fitting linear model and the dashed line depicts the direct proportionality between the two quantities, $y=x$.}
\label{fig:BHanalogy}
\end{figure}

In order to check the validity of our best-fitting results, in Fig.~\ref{fig:BHanalogy} we plot the derived best-fitting BH masses (Table~\ref{tab:bfTL}, second column) versus those originating from the existing literature (Table~\ref{tab:obs}, first column). The literature values estimated using the reverberation technique are shown with the filled squares and for all the other cases with open circles.\par
In order to quantify the obvious correlation between the two quantities we use two simple methods. Initially we compute the Kendall’s $\tau$ rank correlation coefficient \citep{press92a}, using only the actual data values plotted in Fig.~\ref{fig:BHanalogy}, ignoring their uncertainties. This gives a value of $\tau$ equal to 0.818 which corresponds to a very low probability for the null hypothesis, $H_0$, i.e.\ the two quantities are not associated, of $2.13\times10^{-4}$, indicating that there is a strong correlation between the corresponding BH mass estimates. Then, we fit to the data a simple linear model of the form $y = \kappa x + \lambda$, considering the uncertainties in both coordinates \citep{press92a}. For each point a mean error is estimated for its abscissa and ordinate by averaging the corresponding upper and lower error estimate in each direction respectively. The best-fitting model (Fig.~\ref{fig:BHanalogy}, black solid line) has a slope of $\kappa=1.21^{+0.19}_{-0.18}$ and an intercept 
of $\lambda=-0.05^{+0.25}_{-0.36}$ and is derived from a $\chi^2$ merit function of 2.98 for 10 d.o.f. yielding a very low probability for $H_0$ (i.e.\ obtaining by chance a value of $\chi^2$ smaller or equal to 2.98 from an uncorrelated data of the same length) of 0.018. In the same plot we show with the grey-dashed line the direct proportionality between the two quantities, i.e.\ $y=x$. As we can see the overall agreement is very good, and within the quoted best-fitting parameter uncertainties the two lines are consistent with each other.\par
This result validates that the derived BH mass estimates are entirely consistent with the estimates derived from other independent methods, hence supporting the validity of our modelling process and thus giving confidence in our estimation of the other best-fitting lamp-post model parameters, i.e.\ spin, inclination and height.

\subsection{The lamp-post plane and parameter correlations}
In this section, we examine if there are any significant correlations between the best-fitting model parameter values. Such correlations could be of physical origin, but on the other hand, spurious dependencies could also be introduced in the presence of model degeneracies.\par
\subsubsection{The lamp-post plane}
\label{sssect:lpp}

\begin{figure*}
\includegraphics[width=5in]{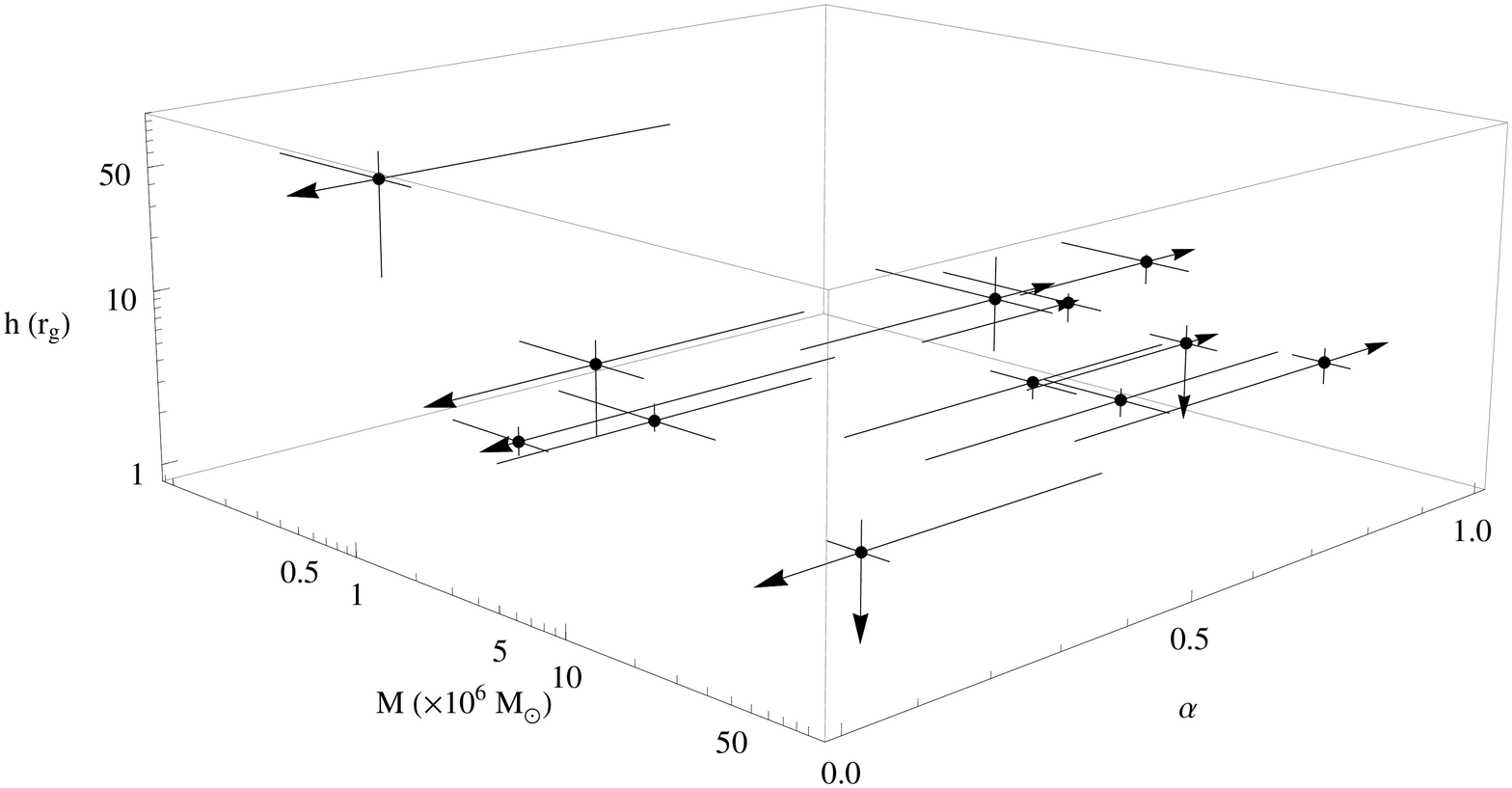}\\
\parbox{0.5\linewidth}{\vspace{-10em}\hspace*{-15.5em}\scalebox{0.45}{\includegraphics[trim = 2.3mm 229mm 200.4mm 190mm clip width=2cm]{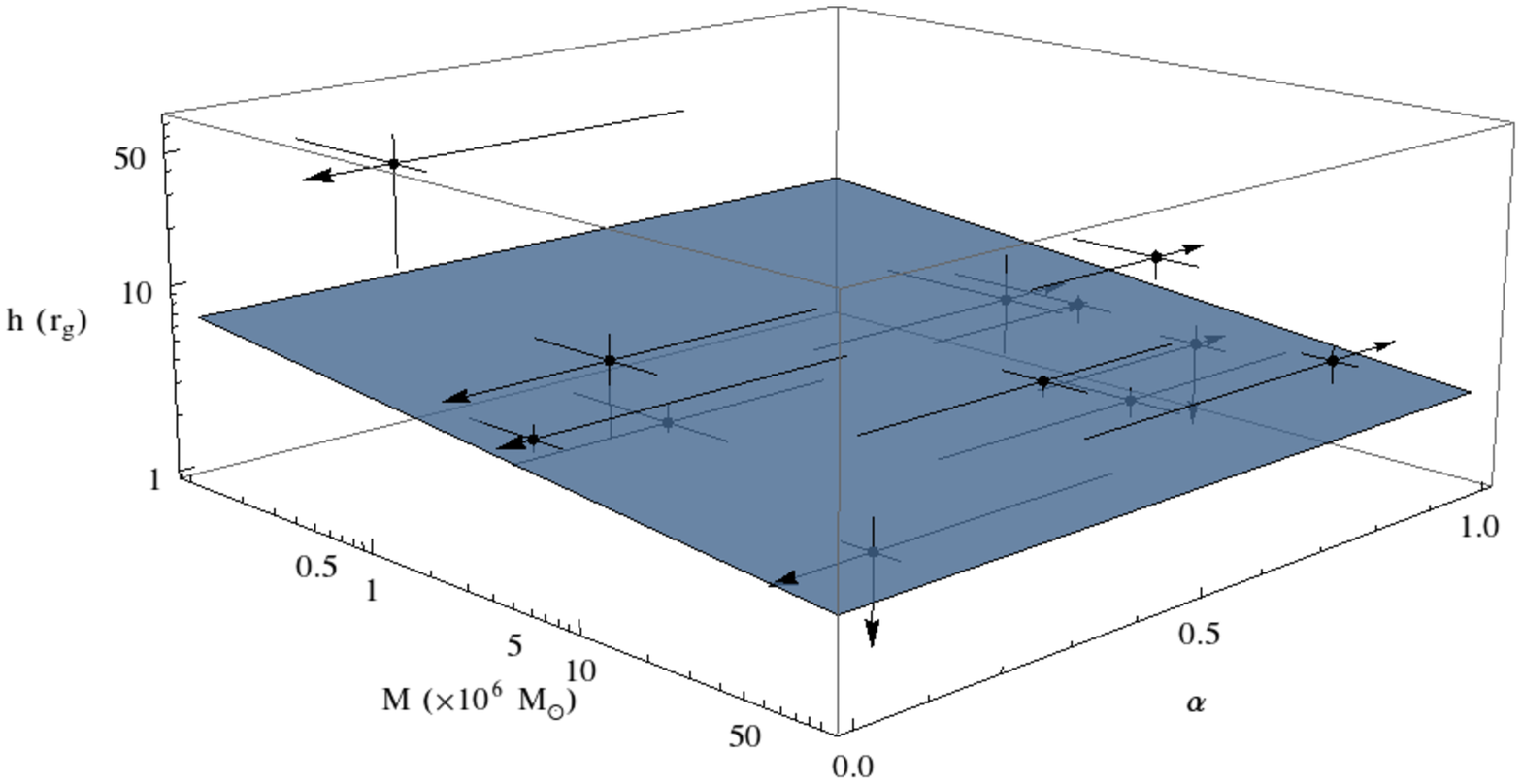}}}\\
\parbox{0.5\linewidth}{\vspace{-12em}\hspace*{14.2em}\scalebox{0.45}{\includegraphics[trim = 2.3mm 229mm 200.4mm 190mm clip width=2cm]{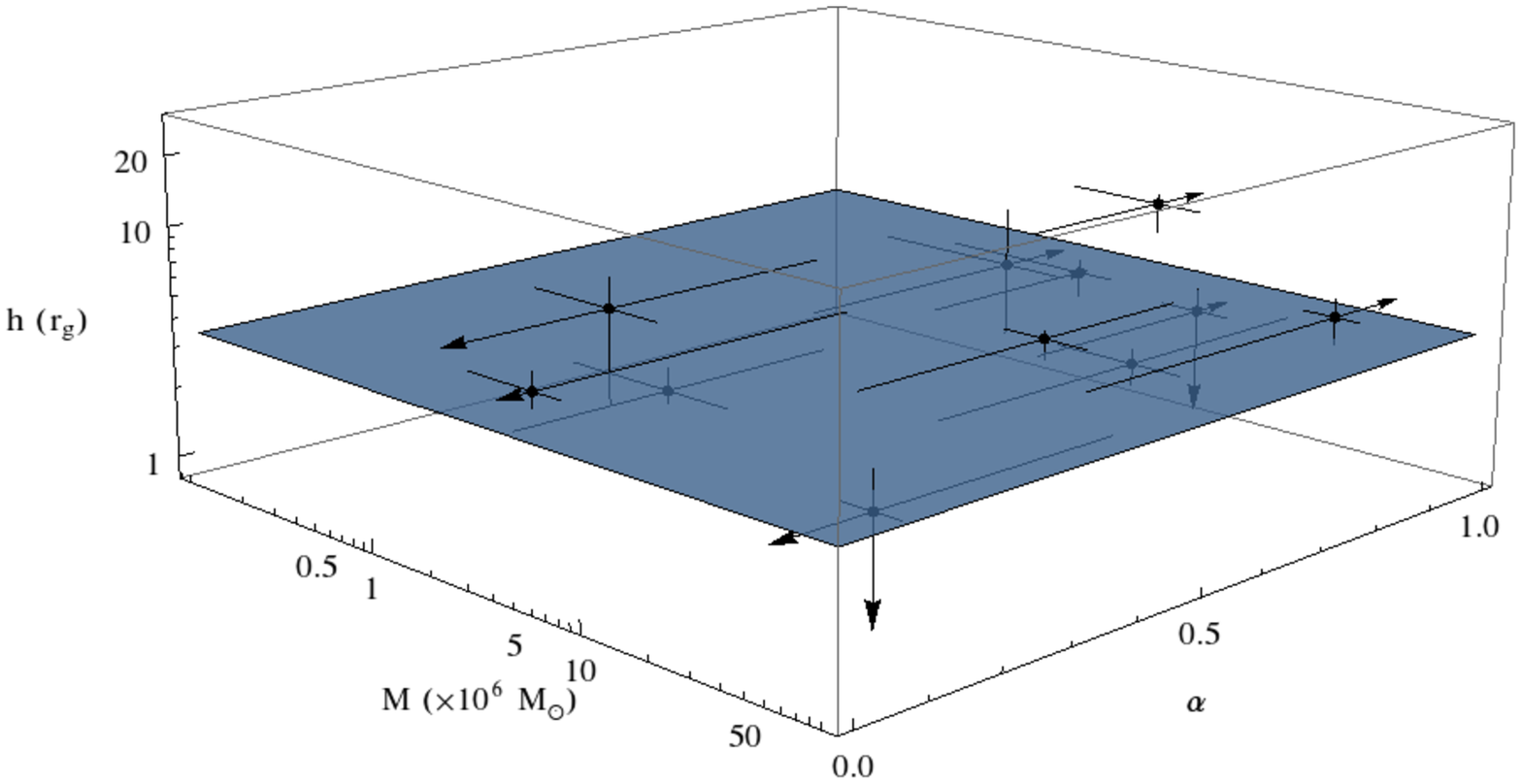}}}\\[12.2em]
\caption{The lamp-post parameter space. Top panel: The best-fitting values for the 12 AGN as  given in Table~\ref{tab:bfTL}. Bottom left-hand panel: The best-fitting LPP including all the AGN, given by equation~\ref{eq:lppBF1}. Bottom right-hand panel: The best-fitting LPP excluding NGC\;4395, given by equation~\ref{eq:lppBF2}.}
\label{fig:lpp}
\end{figure*}

Initially, we study the lamp-post parameter space by defining the lamp-post plane (LPP) consisting of the physical quantities of, $M$, $\alpha$ and $h$ ($\theta$ is excluded since it is an observational property). This approach enable us to unveil potential dependencies among the various model parameters that could lead to model degeneracies. In the top panel of Fig.~\ref{fig:lpp} we show the lamp-post parameter space consisting of the 12 AGN as given in Table~\ref{tab:bfTL}. The arrows indicate the measurement error estimates that exceed the interpolated grid limits (values with `\text{---}' in Table~\ref{tab:bfTL}).\par
Assuming that the LPP can be described from the equation:
\eqb
k_M \log_{10}\left[M\over{10^6 \rm{M}_{\rm \sun}}\right]+k_\alpha\alpha+k_h\log_{10}\left[h\over{r_{\rm g}}\right]+k_0=0
\label{eq:lpp}
\eqe
the distance of a point $\left\{M_i,\alpha_i,h_i\right\}$ from this plane is \\
\hspace*{-2em}\parbox{1\linewidth}{
\eqb
&&\mathscr{D}_i=\frac{\left|k_M \log_{10}\left[M_i\over{10^6 \rm{M}_{\rm \sun}}\right]+k_\alpha\alpha_i+k_h\log_{10}\left[h_i\over{r_{\rm g}}\right]+k_0\right|}{\sqrt{k_M^2+k_\alpha^2+k_h^2}}\nonumber\\
\hspace{0em}
\label{eq:lppDistance}
\eqe
}
\\
In order to derive the LPP best-fitting parameters $k_M$, $k_\alpha$ and $k_h$ one can estimate the total distance of all the points from the LPP and minimize it. Since the measurements are characterised by large errors in all the three directions we perform a Monte Carlo simulation. Generalising the approach employed in Section~\ref{ssect:bhmass}, we estimate for each point a mean error along each direction by averaging the corresponding upper and lower error estimates. Thus, for each point we form a three-dimension multivariate normal distribution whose probability density function is characterised by a diagonal covariance matrix, $\mathbf{\Sigma}$. This enable us to draw for each point an ensemble of 10000 random numbers that form, around it, a three dimensional ellipsoid whose axes are in the directions of the eigenvectors of $\mathbf{\Sigma}$ and the length of the i\textsuperscript{th} longest axis is proportional to $\sqrt{\lambda_i}$ where $\lambda_i$ is the eigenvalue associated with the i\
textsuperscript{th} eigenvector of $\mathbf{\Sigma}$. Finally, we find the best-fitting parameters $\left\{k_M,k_\alpha,k_h\right\}$ that minimize the total distance for each set of points and we end up with 10000 best-fitting parameters whose distribution depicts the most possible value (i.e.\ maximum) and the corresponding uncertainties.\par
By considering all the sources, the best-fitting LPP (Fig.~\ref{fig:lpp}, bottom left-hand panel) is given by the following relation
\eqb
&&0.095^{+0.124}_{-0.123}\log_{10}\left[M\over{10^6 \rm{M}_{\rm \sun}}\right]-0.018^{+0.091}_{-0.110}\alpha+\nonumber\\
&&0.746^{+0.103}_{-0.084}\log_{10}\left[h\over{r_{\rm g}}\right]-0.526^{+0.116}_{-0.136}=0
\label{eq:lppBF1}
\eqe
By ignoring the first AGN in the sample, NGC\;4395, whose best-fitting estimates come from a very poor fit, the best-fitting LPP (Fig.~\ref{fig:lpp}, bottom right-hand panel) is given by:
\eqb
&&-0.001^{+0.088}_{-0.082}\log_{10}\left[M\over{10^6 \rm{M}_{\rm \sun}}\right]-0.020^{+0.086}_{-0.095}\alpha+\nonumber\\
&&0.756^{+0.090}_{-0.082}\log_{10}\left[h\over{r_{\rm g}}\right]-0.411^{+0.096}_{-0.090}=0
\label{eq:lppBF2}
\eqe
Both equations are consistent with each other and the best-fitting LPP parameters $k_M$ and $k_\alpha$ are consistent with 0 dictating that there is no statistical coupling among the various model parameters, and thus disfavouring the existence of parameter degeneracies in our model. Both equations are consistent with a LPP perpendicular to the height axis (i.e.\ parallel to the $M$, $\alpha$ plane) intersecting with it at around h=$4.28^{+0.51}_{-0.47}$\rg\ (the average from the two planes).

\subsubsection{Parameter correlations}
\label{sssect:correlations}
In order to investigate even more the existence of potential correlations and actually visualize their nature, in the panels of Fig.~\ref{fig:lampPostCorrel} we plot the three lamp-post parameters, $\alpha$, $\theta$ and $h$ versus $M$. As before, the arrows indicate the measurement error estimates that exceed the interpolated grid limits (values with `\text{---}' in Table~\ref{tab:bfTL}).\par
In agreement with the results of the analysis above, there is absolutely no evidence of a correlation for any of the lamp-post parameters with the BH mass. Within each panel of Fig.~\ref{fig:lampPostCorrel} we also quote the values of the Kendall’s $\tau$ rank correlation coefficient and the corresponding probability, $p$, for $H_0$ (i.e.\ the two quantities are not associated). The latter values are quite large, indicating that there is no ground to reject the $H_0$ for any of the three cases.\par

\subsection{Best-fitting parameters}
\label{ssect:bfparam}
\begin{figure*}
\includegraphics[width=2.2in]{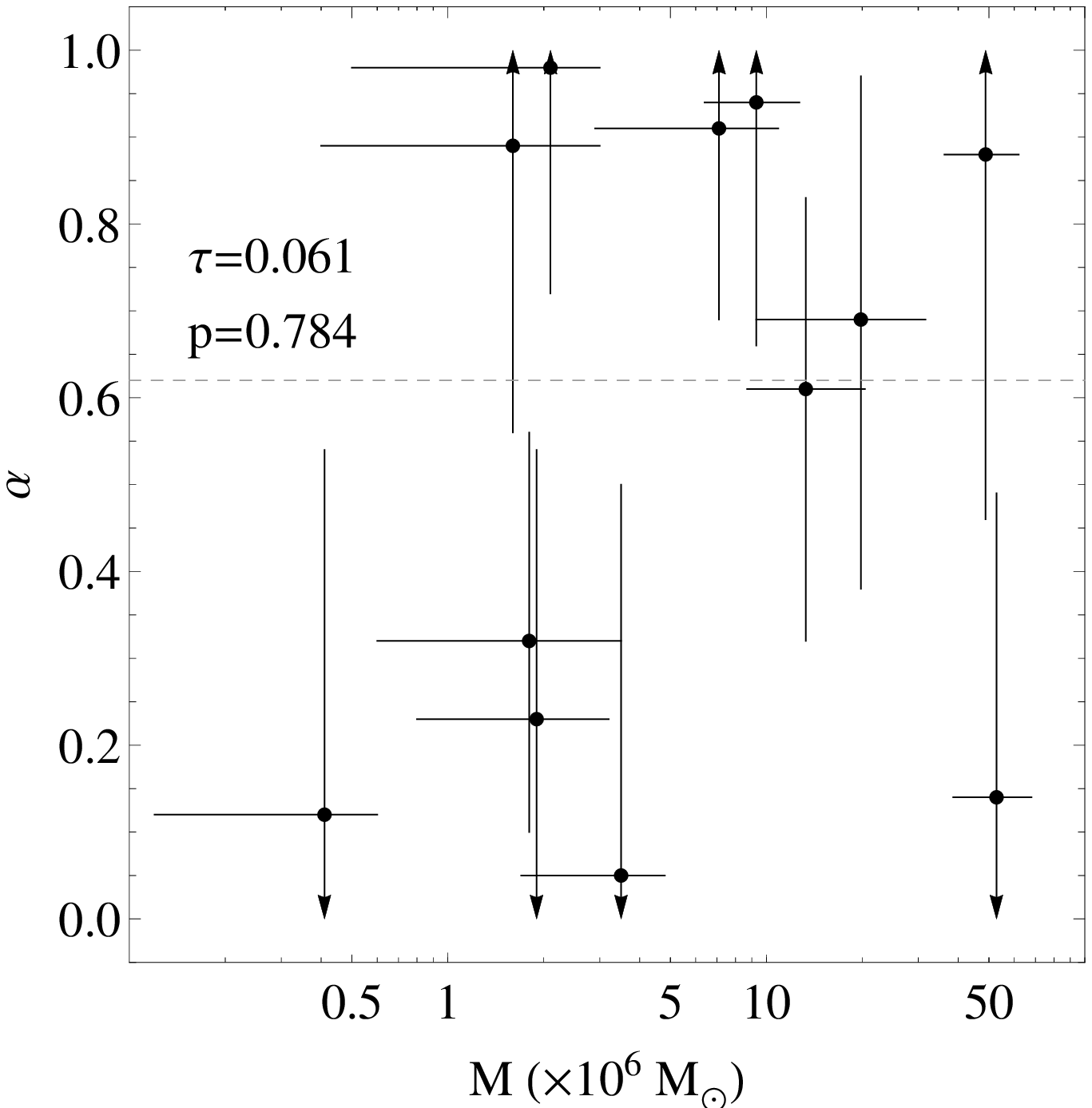}
\includegraphics[width=2.2in]{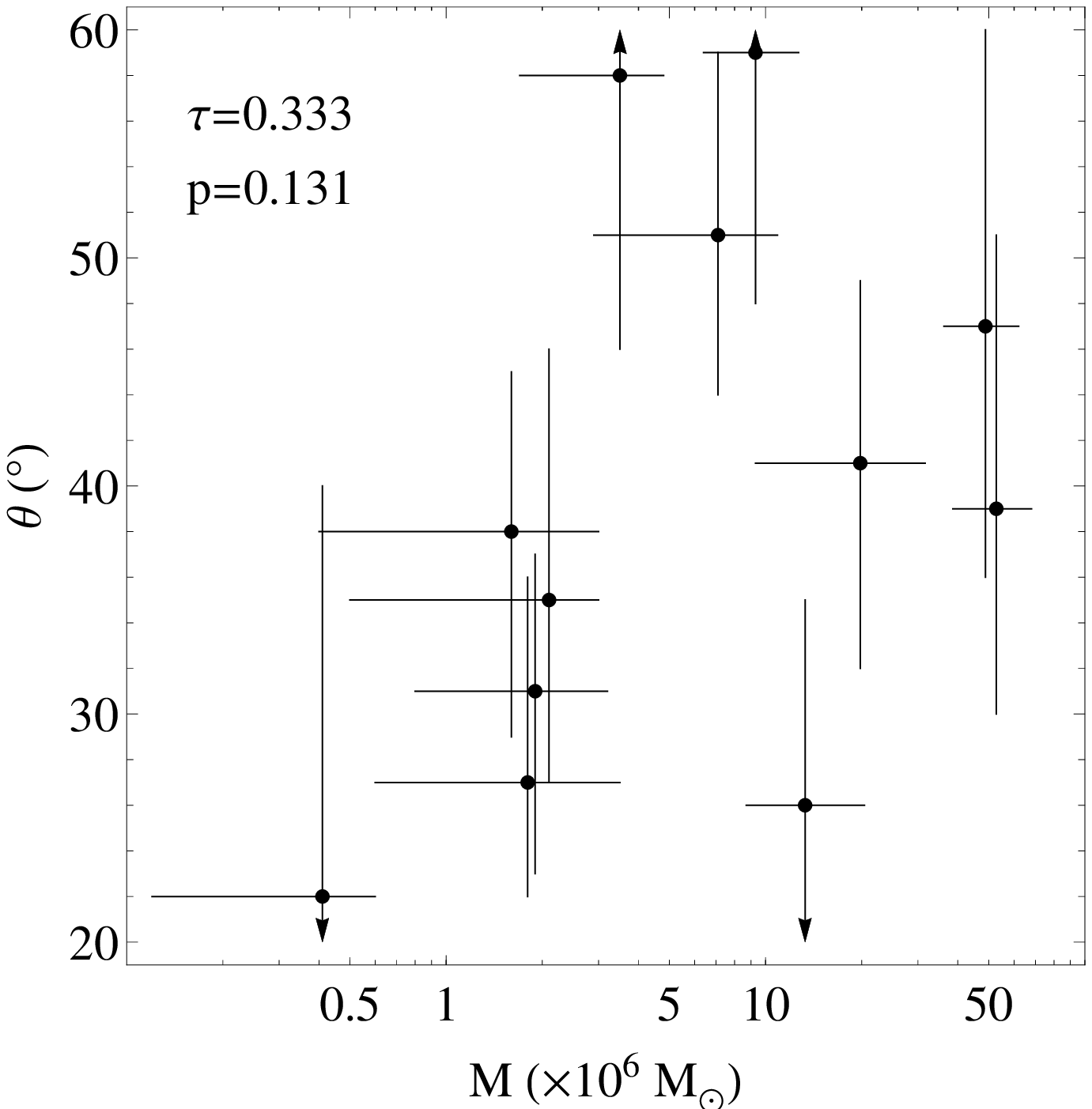}
\includegraphics[width=2.25in]{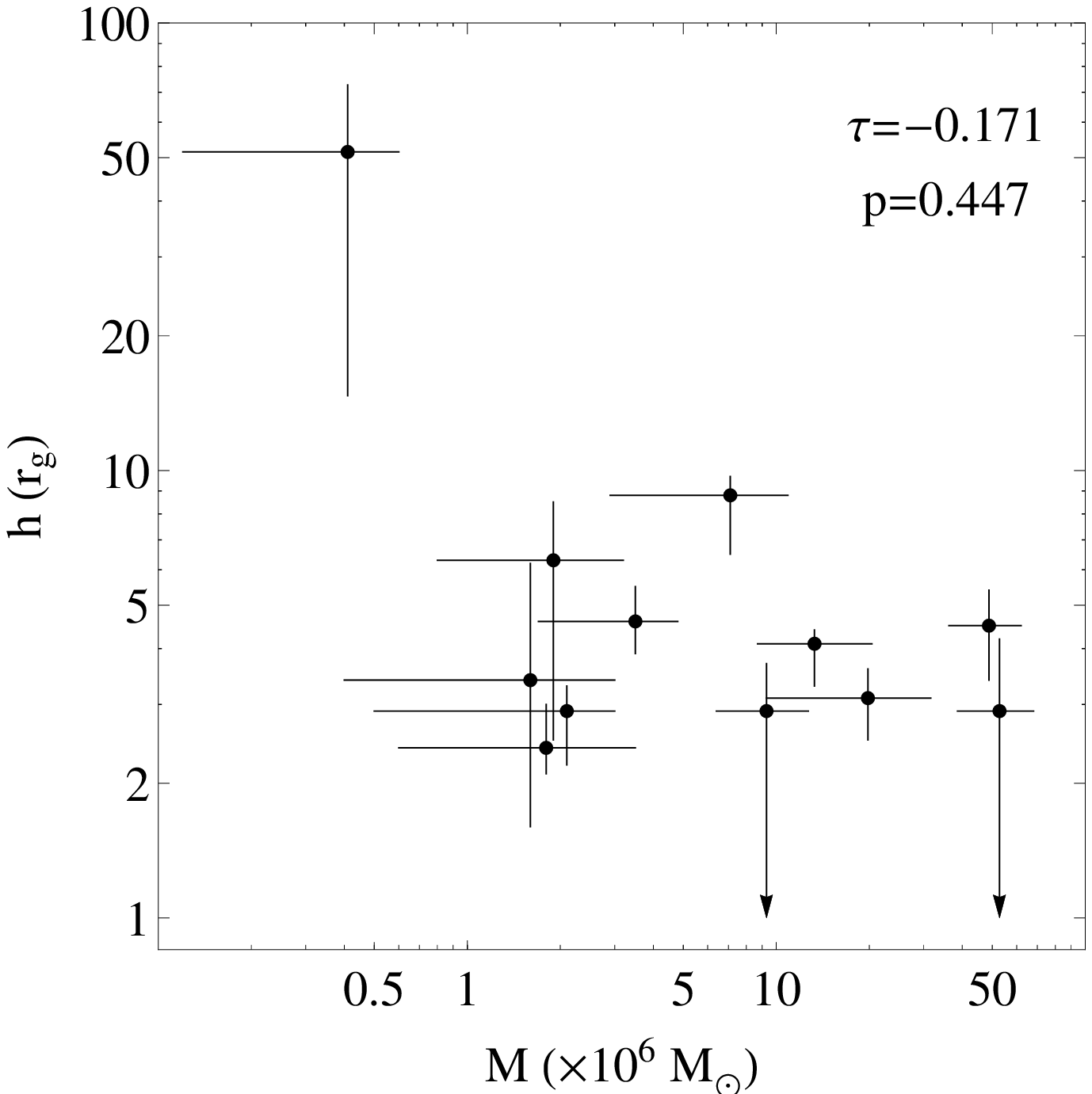}
\caption{Correlation plots of the various best-fitting lamp-post parameters versus the best-fitting BH mass, showing also the Kendall’s $\tau$ rank correlation coefficients and the corresponding probability, $p$, for $H_0$. Left-hand panel: The BH spin parameters. The grey-dashed line corresponds to the mean value of the ensemble of points, $0.62\pm0.09$. Middle panel: The viewing angles. Right-hand panel: The heights of the X-ray sources.}
\label{fig:lampPostCorrel}
\end{figure*}

In this section we analyse each lamp-post best-fitting parameter, coming from all the sources, individually.    
\begin{itemize}
\item \underline{Spin parameters}: The effects of the spin parameter on the corresponding time-lag spectra are rather small (see Fig.~\ref{fig:spin_param} in Appendix~\ref{ssect:spin_param}) and given the quality of our X-ray data sets, it is difficult to statistically distinguish between different spin-parameters, for each source individually. This is the reason for the rather large uncertainties on the model best-fitting spin parameters.\par
Nevertheless, the left panel in Fig.~\ref{fig:lampPostCorrel} suggests that there may exist two groups of objects within our sample: those with high (i.e.\ $\alpha\sim1$)  and those with low (i.e. $\alpha\sim0$) BH spin parameters. To investigate a possible clustering scenario for the spin parameters, we apply an unsupervised partitioning algorithm to the derived spin parameters, using the Canberra distance \citep{everitt11}. Indeed, we do find that the spin parameters form two clusters of points; those defined above and below the mean value of the ensemble, which is $0.62\pm0.09$, indicated by the grey-dashed line. This is indicative of two populations of objects, as we mentioned above: objects with a low and high spin. The low-spin population is characterised by a mean value of $0.35^{+0.16}_{-\text{---}}$ and the high-spin population is characterised by a mean value of $0.84^{+\text{---}}_{-0.11}$.\par
In order to derive the probability of how significantly different these mean values are from that of the ensemble mean we perform a simple Student's ${\itl t}$-test. For the low-spin population, the value of ${\itl t}$-statistic is -4.55 for 5 d.o.f., yielding a small probability value for the null hypothesis, $H_0$, i.e.\ the population mean is 0.62, of $6.1\times10^{-3}$. Similarly, for the high-spin population the probability of $H_0$ is low, $1.4\times10^{-3}$, corresponding to a ${\itl t}$-statistic value of 6.37 for 5 d.o.f.\par
We therefore find significant evidence for the presence of two groups of objects: those with high BH spin parameters (larger than 0.75) and those with a low BH spin parameters (smaller than 0.5). We note that this result should be treated with some caution, as the initial selection of points was done by our clustering algorithm, whose results are hard to judge due to the small number of objects in the original sample.\par
Nevertheless, what we can state with high confidence is that the $\alpha$ best-fitting values for the high- and low-spin populations differ significantly from 0 and 1, respectively. Namely, for five sources:  Mrk\;766, MCG--6-30-15, IRAS\;13224-3809, ESO\;113-G010 and NGC\;5548, the derived spin parameter values differ significantly from 0, with $H_0$ (i.e.\ the quantity is consistent with zero) probabilities of 0.0034, $8.19\times10^{-5}$, $3.94\times10^{-4}$, $1.76\times10^{-5}$ and 0.018, respectively. On the other hand, the spin parameters for five AGN: NGC\;4395, NGC\;4051, Ark\;564, 1H\;0707-495 and NGC\;3516, differ significantly from 1, having $H_0$ (i.e.\ the quantity is consistent with one) probabilities of 0.018, 0.006, 0.017, 0.002 and 0.007, respectively. Thus, not all AGN have a maximally/minimally spinning BHs, and more importantly the actual value of the BH does not correlate with the BH mass i.e.\ Schwarzschild and Kerr BHs can exist in AGN with the same mass.
\item \underline{Angles}: The viewing angles are distributed uniformly between 20 to $60\degr$, with a mean value of $39.5^{+4.9}_{-4.3}$\degr. Within the unification models for AGN, this result would imply an average opening angle (i.e.\ twice the viewing angle) for the putative molecular torus of around $80^{+9.8}_{-8.6}\degr$. This is in agreement with recent results which find a Type II to Type I AGN ratio of the order of 1.75 \citep{rush93,burlon11,toba13}. This ratio implies a TypeI/(TypeI+TypeII) ratio of around 0.35, and hence an opening angle of the torus of around 100\degr. We consider this agreement as another indication that our approach yields reasonable values for the best-fit model parameters. However, note that as in the case of the spin parameter, the effects on the viewing angles to the corresponding time-lag spectra are also rather small (see Fig.~\ref{fig:angle_param} in Appendix~\ref{ssect:angle_param}).
\item \underline{Heights}: The `lamp' height is the best constrained fitting parameter (i.e.\ has the smallest error). As we can see from Fig.~\ref{fig:height_param} in Appendix~\ref{ssect:height_param} this is due to the fact the X-ray source height affects significantly the form of the corresponding time-lag spectra, since the onset of the initial negative plateau, as well as the strength and the position of peaks is greatly modified by different heights of the X-ray source. The particularly low best-fitting values for the height parameter indicate that, within the lamp-post geometry model, the X-ray source is situated very close to the BH. The mean average value of the height is $3.72^{+0.56}_{-0.52}$ \rg\ including the first estimate of NGC\;4395 (first point) coming from a very poor fit. Thus, the small heights indicate that for all AGN (within the lamp-post model scenario) the X-ray source is situated very close the BH, hence should also have a rather small size. This, is a solid result which does not 
suffer from model degeneracies as we showed in the previous section (Section~\ref{sssect:lpp}). 
\end{itemize}

\subsection{Mass scaling relation}
\label{ssect:mass_scale}
\citet{demarco13} have shown that for each source, from an ensemble of 15 AGN, the most negative time-lag value and the corresponding frequency scales linearly, in logarithmic space, with its BH mass. As we find from our physical modelling, the principal parameters affecting the form of the negative component of the time-lag spectra is the BH mass as well as the height of the X-ray source. As we show in the previous section (Section~\ref{ssect:bfparam}), in almost all the AGN the X-ray source lies above the accretion disc at the same small height, around 3--4 \rg\ above the BH. Thus, the \citet{demarco13} relation holds firstly because the source height appears to be comparable in all AGN, but also because the BH mass range is sufficiently small that, within the frequency range sampled by \textit{XMM}-\textit{Newton}, we measure the same part of the time-lag spectrum in almost all AGN.\par
\begin{figure}
\includegraphics[width=3.3in]{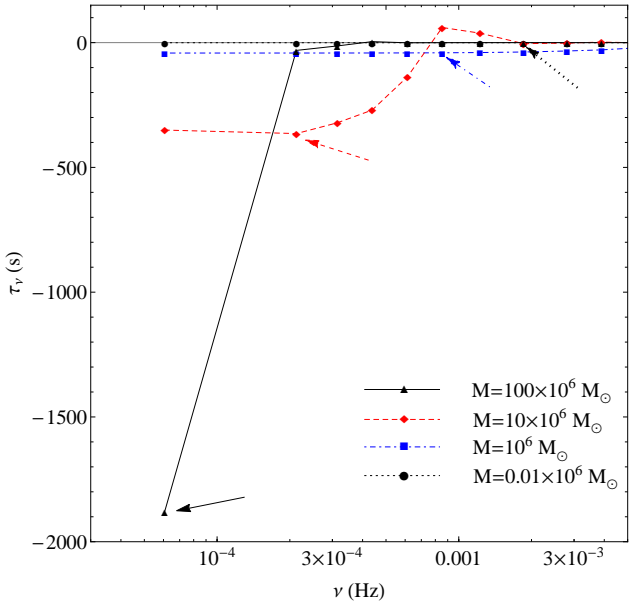}
\caption{The mass scaling relation. The discretized negative time-lag spectral component for the lamp-post models with $\alpha=1$, $\theta=40$\degr and $h=3.6$ \rg\ and BH masses: $M=\left(0.01,1,10,100\right)\times10^{6}$ \ms. The arrows indicate the most negative time-lag point for each BH mass. --A colour version of this figure is available in the online version of the journal--}
\label{fig:massScale}
\end{figure}

To clarify this issue, in Fig.~\ref{fig:massScale} we plot the negative model components of the time-lag spectra (discretized and binned in order to mimic the observations --see Section~\ref{sect:fits} and Fig.~\ref{fig:continTLfitModel}--) for objects with BH masses in the range between $10^{4}-10^{8}$ \ms. The arrows, in the same figure, indicate the most-negative time-lag measurements in each case. Clearly, within the observed frequency range, the larger the BH mass, the larger the most-negative time delay and the lower the corresponding frequency. The frequency range over which the time-lag spectra have been estimated corresponds to the frequency range (i.e.\ between $10^{-5}$ Hz and few times $10^{-3}$ Hz) that is currently `available' with the present day \textit{XMM}-\textit{Newton} observations. As we can see from Fig.~\ref{fig:massScale}, within this frequency range,  the amplitudes of the time-lag spectra indeed become smaller with increasing frequency. However, if we had longer observations (e.g.\ 
going down to $10^{-8}$ Hz) 
then the time-lag values for the highest BH mass would be larger at frequencies below those presently observable. Alternatively, if the observation length was shorter, so than the lowest observed frequency for the same BH mass was {\it higher} than $\sim 2\times 10^{-4}$ Hz, then the `most-negative' time-lag estimate would certainly be less than the maximum negative time-lag for this object. We therefore caution the readers that \textit{the most negative time-lag value and the corresponding frequency versus BH mass scaling relations} hold for objects with a BH mass not larger than $10^{8}$ \ms, as long as the X-ray source is situated on the same height for all objects, and the time-lags have been estimated in a {\it same} frequency range which is the same as the one used by \citet{demarco13}.

\section{SUMMARY AND DISCUSSION}
\label{sect:sum_disc}
We have modelled the time-lag spectra of 12 AGN, using with the highest-quality \textit{XMM}-\textit{Newton} observations, in terms of signal-to-noise ratio and observational length. We have employed physically realistic GRIRFs for the case of the lamp-post geometrical model in which the X-ray source lies above the BH. This is the first time that such a fully GR approach is employed to model in a statistically robust way the time-lags spectra of a large number of local AGN. Our results imply that such a model can adequately describe the observed X-ray time delays and thus it is a realistic representation of the innermost regions of most AGN.\par
We find that the observed time-lag spectra of all objects are fully consistent with the hypothesis of X-ray reverberation from the innermost part of the accretion disc. The main results from our analysis are summarised below.
\begin{itemize}
\item The best-fitting BH-masses are in very good accordance with those derived from other, independent, methods, e.g.\ optical reverberation mapping.
\item The best-fitting viewing angles imply an opening angle of around $80\degr$ for the putative molecular torus, which is fully consistent with current estimates for the ratio of Type II over Type I Seyfert galaxies in the local Universe.
\item There is no correlation between the BH mass and any of the lamp-post model parameters, i.e.\ spin, viewing angle and height (Fig.~\ref{fig:lampPostCorrel}).
\item There is a tentative evidence for bimodality in the distribution of the best-fitting spin parameters above and below $\alpha=0.62$: one group with a low BH-spin (with a mean of 0.35) and another group with a high BH spin (with a mean of 0.84), respectively. In any case, our results indicate, with a high significance, that the BH spin for member of the former group is not consistent with 1, while the BH spin for the objects in the latter group is not consistent with 0. But we do not find any correlation between BH mass and BH spin: our results suggest the existence of both Schwarzschild and Kerr BHs, at any given BH mass.
\item The average X-ray source height is 3.7 \rg, with little dispersion. As these heights are an approximation to the X-ray source size, they imply a very small source size.
\end{itemize}
Our modelling is limited by a few assumptions, which can potentially affect the validity of our results. To start with, as we have mentioned in Section~\ref{ssect:relativ_respFunct}, strictly speaking, our GRIRFs do not correspond to the actual response of the soft band but rather to the response of the neutral fluorescent \fa line, at 6.4 keV (in the rest frame of the accretion disc). However, if both the \fa line and the soft band photons are generated in the same part of the accretion disc, there should not be dramatic differences between the two GRIRFs.\par
Perhaps, more important may be our assumption that the contribution of the reflection spectrum, in the soft X-ray band is equal to 30 per cent of the total observed flux for all objects (see also Appendix~\ref{app:reflect_fracti}). If the contribution of the continuum emission in the soft band were insignificant (in which case $k$ in equation~\ref{eq:soft_emis} is zero by definition), then the value of $f$ would not affect our results at all, since the time-lag spectrum, at a given frequency, would be determined only by the response function, $\Psi(t)$ (equation~\ref{eq:timeLagSpectra}). In the opposite case, when $k\neq0$ and $f<<1$, the cross-correlation between the soft and hard band is basically equal to the autocorrelation of the continuum and thus the time-lags will be equal to those of the continuum (if any e.g.\ positive time lags). Intermediate values of $f$ predict time-lag values between those of the continuum, and those due to the reprocessing component only. As we show in Appendix~\ref{app:
reflect_fracti} the resulting time-lag spectra depend quite a lot on the value of $f$. However, as we explained at the beginning of Section~\ref{sect:fits}, we could not allow $f$ to be a variable parameter during the model fitting, due to the limited number of points in the observed time-lag spectra. However, the value of $f$ which we choose i.e.\ 0.3, is the best current estimate based on detailed X-ray spectral analysis \citep{crummy06}.\par
For similar reasons, to neglect the presence of a reflection component in the 2-–4 keV energy band, which we considered as a proxy for the direct continuum emission, will also affect our results. This issue has been discussed by \citet{wilkins13}, who show that the presence of a reflection component in the continuum hard X-ray band does affect the strength of the reflection component in the resulting time-lag spectra. However, full investigation of this issue would have been computationally prohibitive in our case, where we try to model the time-lags spectra of a large number of AGN. Nevertheless, we believe that the good agreement between the BH mass estimates, from the present work, and the estimates resulting from other independent methods, as well as the reasonable estimates that we get for the viewing angles for all the objects in the sample, imply that our results should be a valid representation of the true values of the other model parameters i.e.\ BH spin and X-ray source height and size as well. 
This could possibly also indicate that, based on the current data quality, we are not able to discern significant differences in the fitting parameters between the actual response of the soft band and that of the 
neutral \fa line.\par 
We note in passing that, given the good agreement between the BH estimates presented in this work, and others from the literature, the time-lag spectrum modelling with physically motivated models, is yet another way of measuring the BH mass in these systems. The uncertainty of our BH mass estimates is rather large, but this is not due to the method itself, but rather to the quality of the time-lags spectra, even though we have used the longest X-ray data sets currently available, obtained by \textit{XMM}-\textit{Newton}.
In the case of objects with a BH mass larger than few times $10^7$ \ms, the main limitation arises from the duration of the available X-ray light curves. For these objects, all the largest amplitudes time-lag spectral features (e.g.\ the the most negative time-lag estimate) are shifted towards frequencies which are lower than the lowest accessible frequencies, defined by the length the light curves, which consist, in the best cases, of around 120 ks, or few times $10^{-5}$ Hz in the frequency domain. As a result, the only time-lag spectral features available to fit are of small amplitudes, resulting in larger BH mass fitting uncertainties (i.e.\ the $\chi^2$ regions around the best fitting-values are characterised by a smaller curvature and thus larger errors).\par
From the other hand, for objects with smaller BH mass, below $10^6$ \ms, the largest time-lag amplitude variations are shifted towards the high Fourier frequencies above few times $10^{-3}$ Hz. In these frequencies, due to the Poisson noise, the very low coherence of our light curves prohibits any meaningful  physical modeling. Thus, in these cases we need to sample thoroughly the small time-scales in order to determine accurately the various lamp-post model parameters; something which is beyond the sensitivity limit of current X-ray observatories.\par
Both of these observational problems will be solved by future X-ray observatories. Athena, due for launch in 2028, will be placed at the L2 point and so will allow continuous uninterrupted observations for as long as it is needed, in order to sample completely the low frequency part of the time-lag spectra of the high BH mass AGN. Athena will also have almost an order of magnitude greater sensitivity than \textit{XMM}-\textit{Newton}, allowing detailed measurements of the time-lags in the high frequency domain. Moreover, if approved, LOFT, although not allowing long continuous observations, will improve sensitivity at high frequencies due to its superior effective area.\par
A major result from our analysis is that the height of the X-ray source must be very close to the BH, around 3.7 \rg, in all AGN. In the lamp-post geometry layout this X-ray source should somehow correspond to the centroid of a hemispherical X-ray corona lying above the accretion disc (a similar X-ray source exists below the disc) where the bulk of the hard X-rays are produced. The derived compactness of the X-ray region and its proximity to the central BH is in accordance with recent X-ray microlensing studies \citep{chartas09,chartas12}, who also find values for the X-ray source size as small as 6--10 \rg. The lack of correlation between the height and either the mass and/or the BH spin indicates that the dimensions of the X-ray region must by tuned by some other physical parameters, perhaps the (currently unknown) physical mechanism which generates this X-ray emission, which should be the same in all objects. Note that if there was any height-luminosity dependence, it would have been wiped out since 
we are using luminosity-averaged time-lag spectra. In a future work we will address this issue by modelling the time-lag spectra during high and low source states.\par
Another major result from our work is that the BH spin parameter, $\alpha$, does {\it not} appear to be the same in all objects. We find that five sources in our sample, namely Mrk\;766, MCG--6-30-15, IRAS\;13224-3809, ESO\;113-G010 and NGC\;5548, possibly host a rotating BH, since the best-fitting $\alpha$ for these objects differs significantly from zero. On the other hand, the best-fitting spin parameters differ significantly from unity in the case of NGC\;4051, Ark\;564, 1H\;0707-495 and NGC\;3516 suggesting that they may host a non-rotating BH. \citet{reynolds13} considered a sample of 20 AGN and, by compiling X-ray spectral fitting from the literature, found high spin parameter values ($\alpha>0.8$) for $M=\left(2\times 10^6-3\times10^7\right)$ \ms, with a hint of lower values ($\alpha\simeq0.5$) for $M<2\times 10^6$ and $M>5\times10^7$ \ms, favouring a prolonged ordered (coherent) accretion scenario that spins up the BH to high values. Our results are in agreement with theirs, although we do not 
observe a correlation between BH spin and mass. This may be due to the rather small size of our sample. On the other hand, our results i.e.\ high and low spin parameters and the lack of correlation between $\alpha$ and $M$, favour a galaxy mergers scenario, in which the BH growth occurs in gas-poor galaxies, resulting to a distribution of spins that has little dependence on BH mass \citep{volonteri13}.\par
Our work shows that derivation of the geometry of the X-ray source and surrounding reprocessing material (here modelled as an accretion disc) in the vicinity of BHs is possible from studies of the Fourier-resolved time-lags, between soft and hard X-ray bands. In future papers, we will study, via GR ray tracing, X-ray photons of all energies as they are reflected and reprocessed by the accretion disc. Not only will we then be able to properly trace the behaviour of soft photons, after they are emitted from the disc, but we
will also be able to determine the contribution of reflected photons to the observed 1.5--4 keV hard X-ray band. At the same time, we will model separately the high and low X-ray flux states of all AGN, in order to trace possible luminosity dependencies with respect to the X-ray source height. With future X-ray observatories such as Athena and LOFT it will be possible to improve significantly on our present time-lag spectral estimates by extending them to a much wider mass range, decreasing at the same time their uncertainties.

\section*{ACKNOWLEDGMENTS}
First of all, we would like to thank the anonymous referee for his/her very useful suggestions and comments. DE and IMM acknowledge the Science and Technology Facilities Council (STFC) for support under grant ST/G003084/1. The research leading to these results has received funding from the European Union Seventh Framework Programme (FP7/2007--2013) under grant agreement \textnumero 312789. This research has made use of NASA's Astrophysics Data System Bibliographic Services. 

\bibliographystyle{mnauthors}

\appendix
\section[]{THE REFLECTION FRACTION}
\label{app:reflect_fracti}
As we discussed in Section~\ref{sect:fits}, due to the complexity of the lamp-post models and the number of free parameters, during our fitting procedure we fix the parameter $f$ (of equation~\ref{eq:timeLagSpectra}) to the value 0.3 i.e.\ the contribution of the reflection spectrum in the soft band is equal to 30 per cent of the total observed flux.\par
\begin{figure}
\includegraphics[width=2.58in]{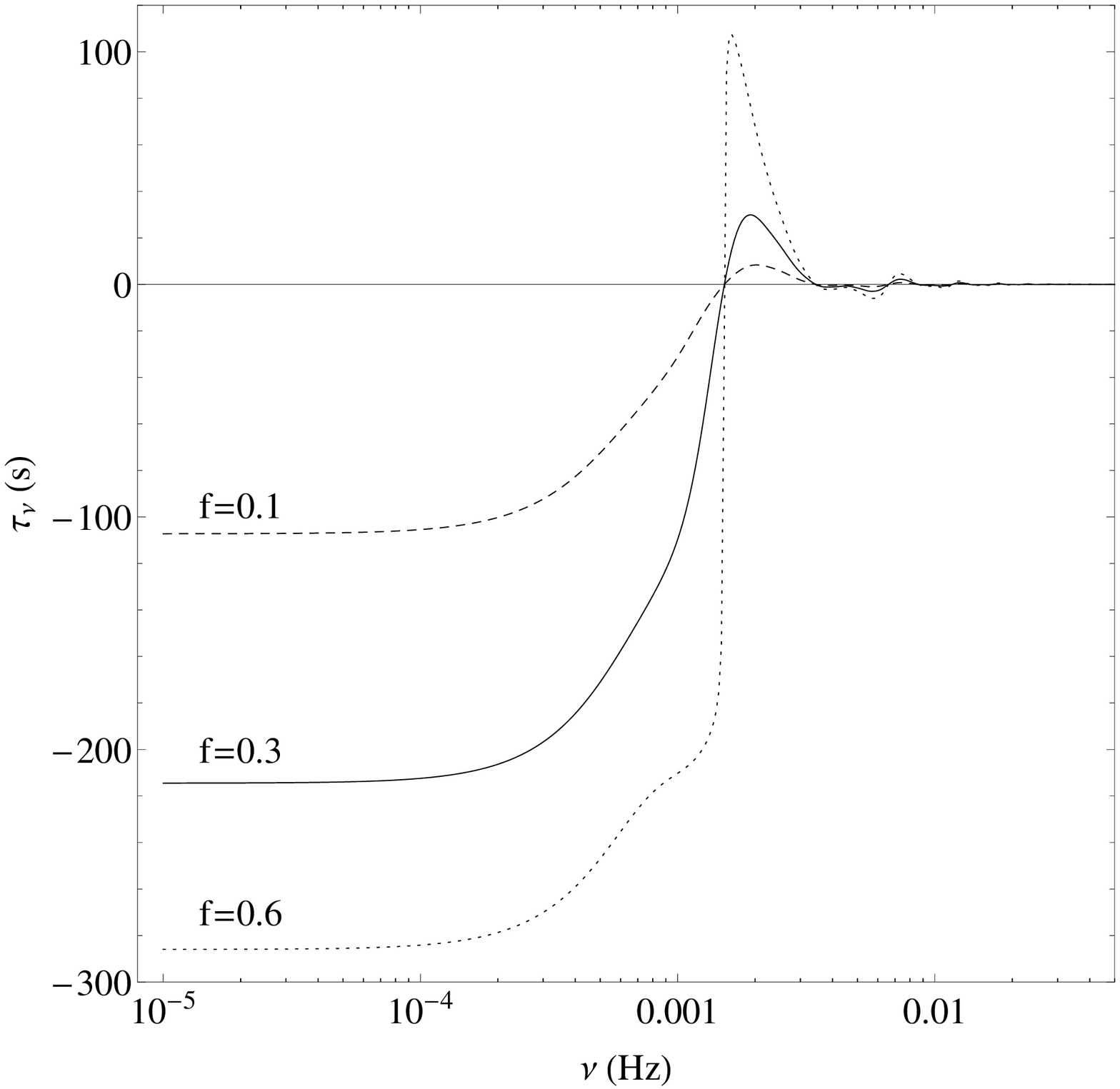}
\caption{Effect of the parameter $f$ during the time-lag spectra estimation. The time-lag spectrum of the lamp-post model with $\alpha=0.676$, $\theta=40\degr$ and $h=3.6$ \rg, for $M=5\times10^6$ \ms, having $f=0.1$ (dashed line), $f=0.3$ (solid line, same as the right panel of Fig.~\ref{fig:tlSpec}) and $f=0.6$ (dotted line).}
\label{fig:TLspecF}
\end{figure}

In this section we show that the value of $f$ does affect the expected time-lag spectra significantly. In Fig.~\ref{fig:TLspecF} we show the time-lag spectra of Fig.~\ref{fig:tlSpec} (right panel, estimated for $f=0.3$) with solid black line, together with two extreme case with $f=0.1$ (dashed line) and $f=0.6$ (dotted line). As we can see, these values can yield differences in the time-lag spectra of the order of 50 to 70 per cent. In practice, this outcome implies that our results regarding the BH mass and/or the source height should be treated with caution.\par
However, as we commented in the text, the fact that our best-fitting BH masses are fully consistent with the BH mass estimates from reverberation studies, and the fact that the best-fitting source heights turned out to be quite similar in all sources, make us believe that our assumption regarding the value of $f$ is rather reasonable. If the value of $f$ were different between different sources, we would expect the best-fitting heights to be rather different among the various objects. 

\section[]{EXPLORATION OF THE MODEL PARAMETER SPACE}
\label{app:param_space}
In this section we explore the model parameter space in order to visualize and understand the effect of the various parameters to the resulting GRIRFs and the corresponding time-lag spectra. Note that colour versions of all the figures in this appendix are available in the online version of the journal.

\subsection{The lamp-post model parameters}
\label{ssect:modPar_effects}
For each case we vary a single model parameter leaving the rest fixed to some typical values: $M=2\times10^6$ \ms, $\alpha=0$, $\theta=40\degr$ and $h=7$ \rg.

\subsubsection{The BH spin parameter, $\alpha$}
\label{ssect:spin_param}
\begin{figure*}
\includegraphics[width=2.58in]{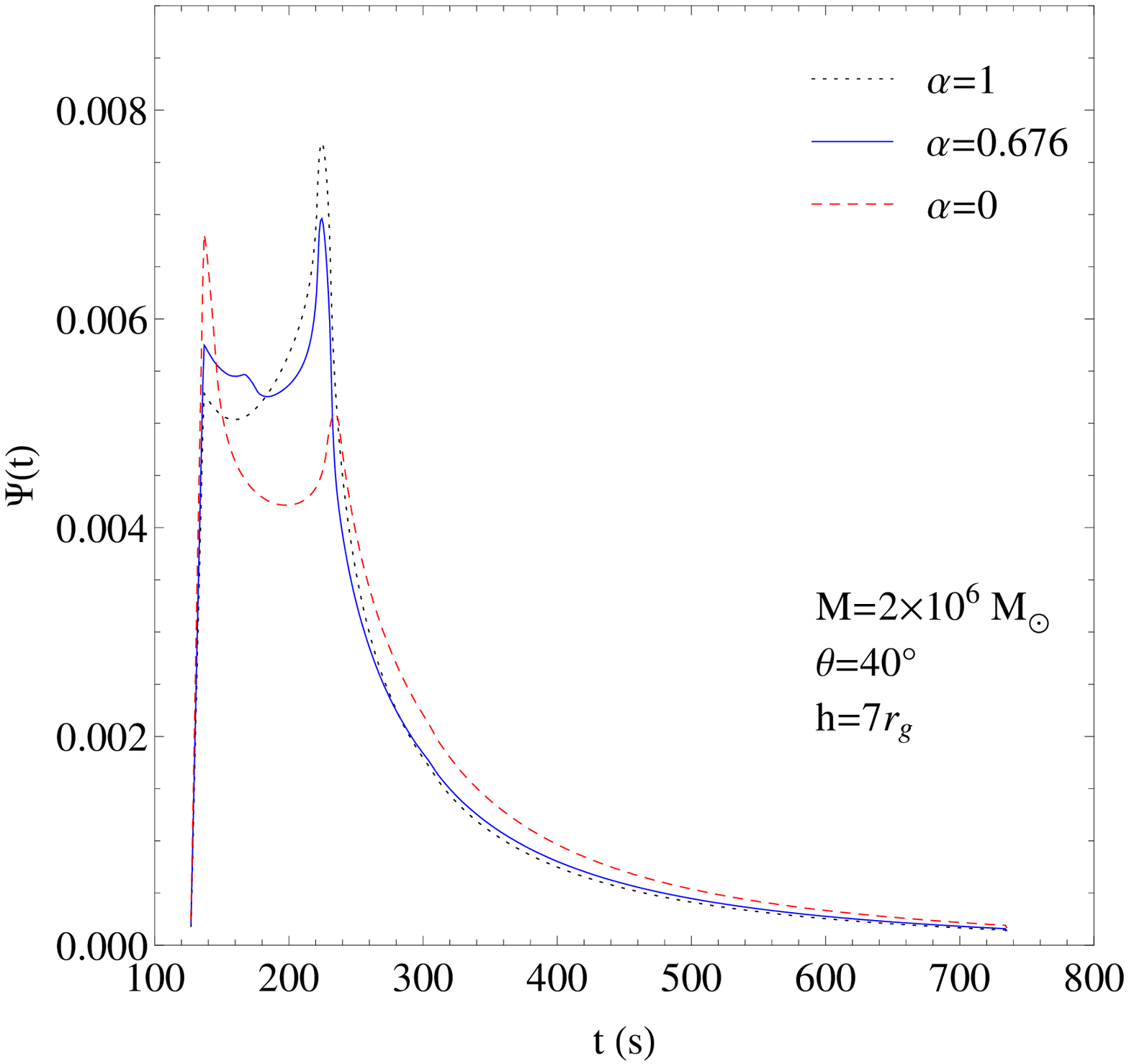}\hspace{4em}
\includegraphics[width=2.5in]{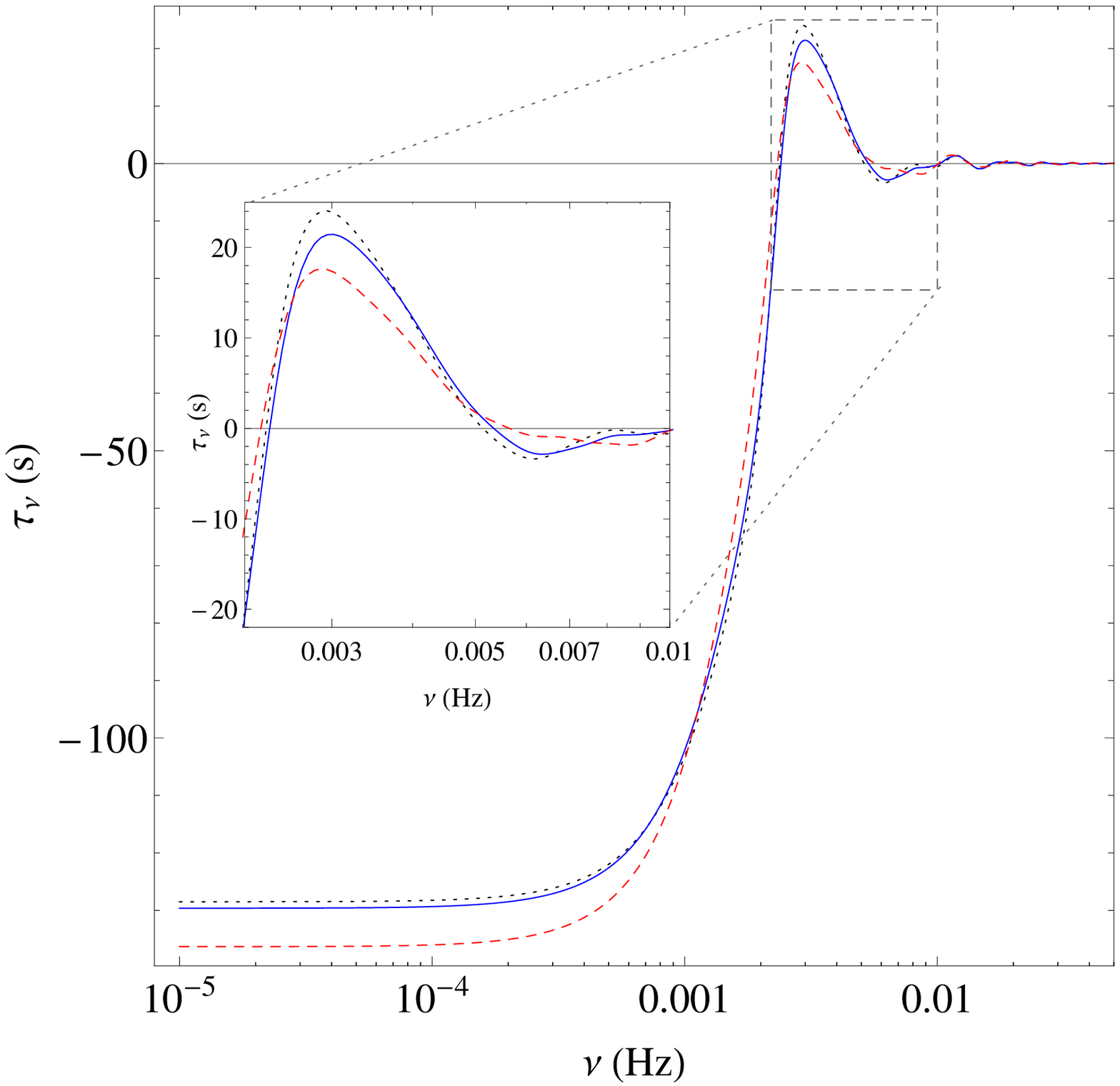}\\[-0.6em]
\caption{BH spin parameter variation for the lamp-post model with $\theta=40\degr$ and $h=7$ \rg, for $M=2\times10^6$ \ms. Left-hand panel: The GRIRFs for three types of spin parameter, $\alpha=$0, 0.676 and 1. Right-hand panel: The corresponding time-lag spectra.}
\label{fig:spin_param}
\end{figure*}

In this section we vary the spin parameter of the BH between the three values (i.e.\ Schwarzschild, intermediate and Kerr). Geometrically this changes the inner radius of the accretion disc $r_{\rm in}$ (see Section~\ref{ssect:model}). As we can see from the left-hand panel of Fig.\ref{fig:spin_param} the GRIRFs differ mainly around the two peaks. The two extreme cases, i.e.\ Schwarzschild ($\alpha=0$) and Kerr ($\alpha=1$), differ significantly from each other but not so much with the intermediate case ($\alpha=0.676$). The corresponding time-lag spectra (Fig.\ref{fig:spin_param}, right-hand panel) differ mainly around the level of the negative plateau between the frequencies $\left(10^{-5}-10^{-4}\right)$ Hz, and the first most positive peak (inset). Again the differences between the three cases are more prominent for the two extreme cases.

\begin{figure*}
\includegraphics[width=2.58in]{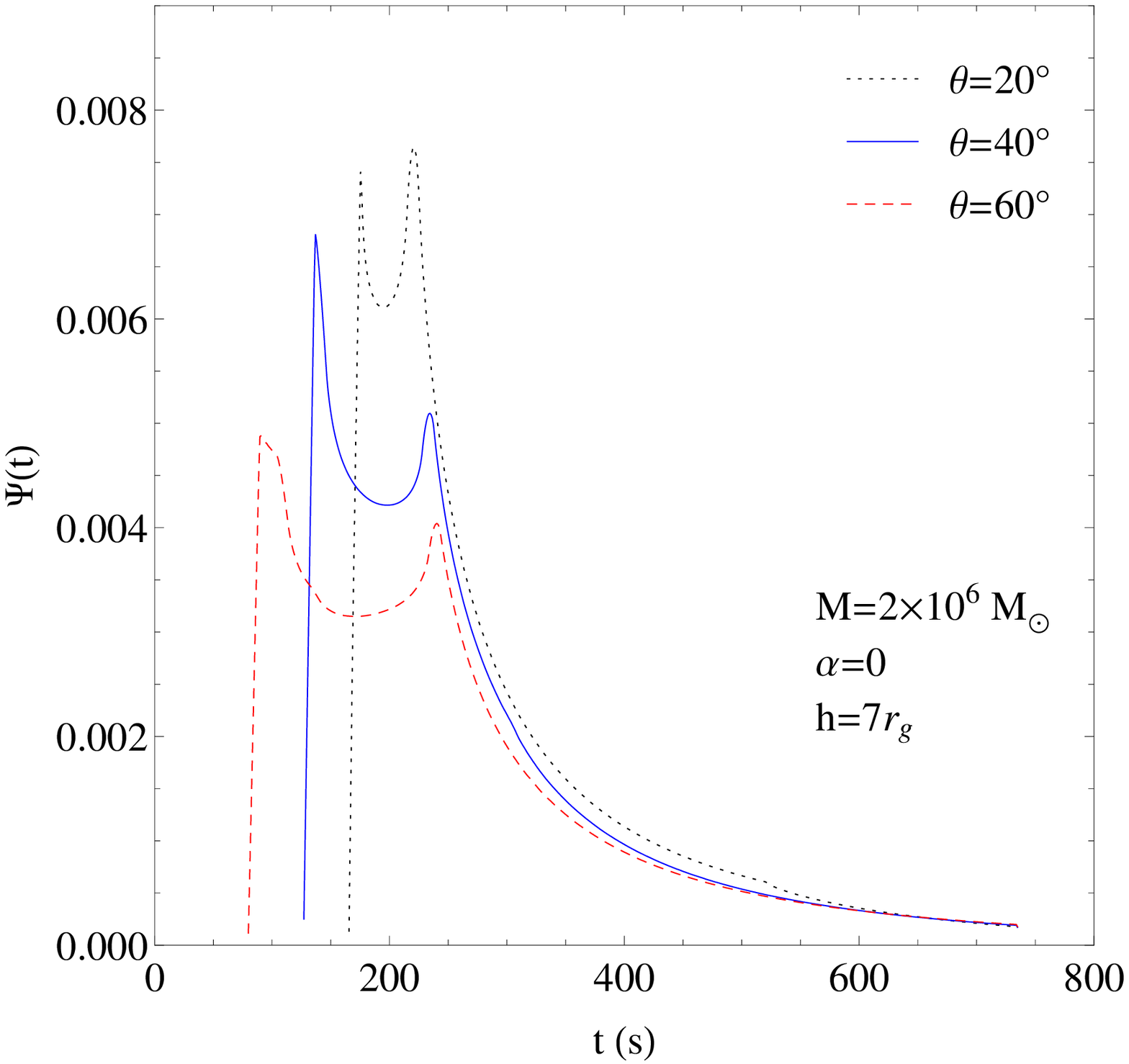}\hspace{4em}
\includegraphics[width=2.5in]{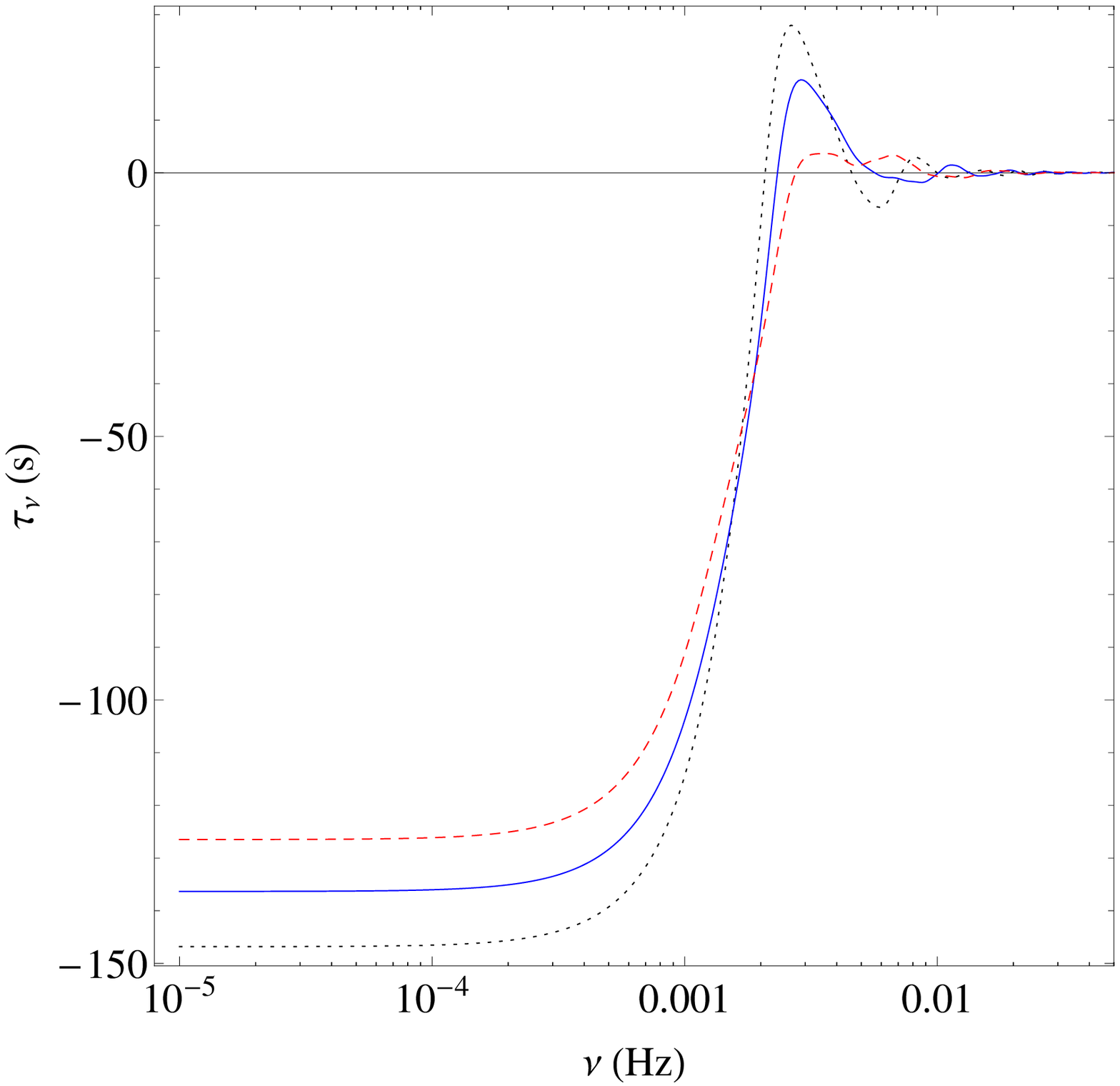}\\[-0.6em]
\caption{Viewing angle variation for the lamp-post model with $\alpha=0$, and $h=7$ \rg, for $M=2\times10^6$ \ms. Left-hand panel: The GRIRFs for three viewing angles, $\theta=$20, 40 and $60\degr$. Right-hand panel: The corresponding time-lag spectra.}
\label{fig:angle_param}
\end{figure*}

\begin{figure*}
\includegraphics[width=2.5in]{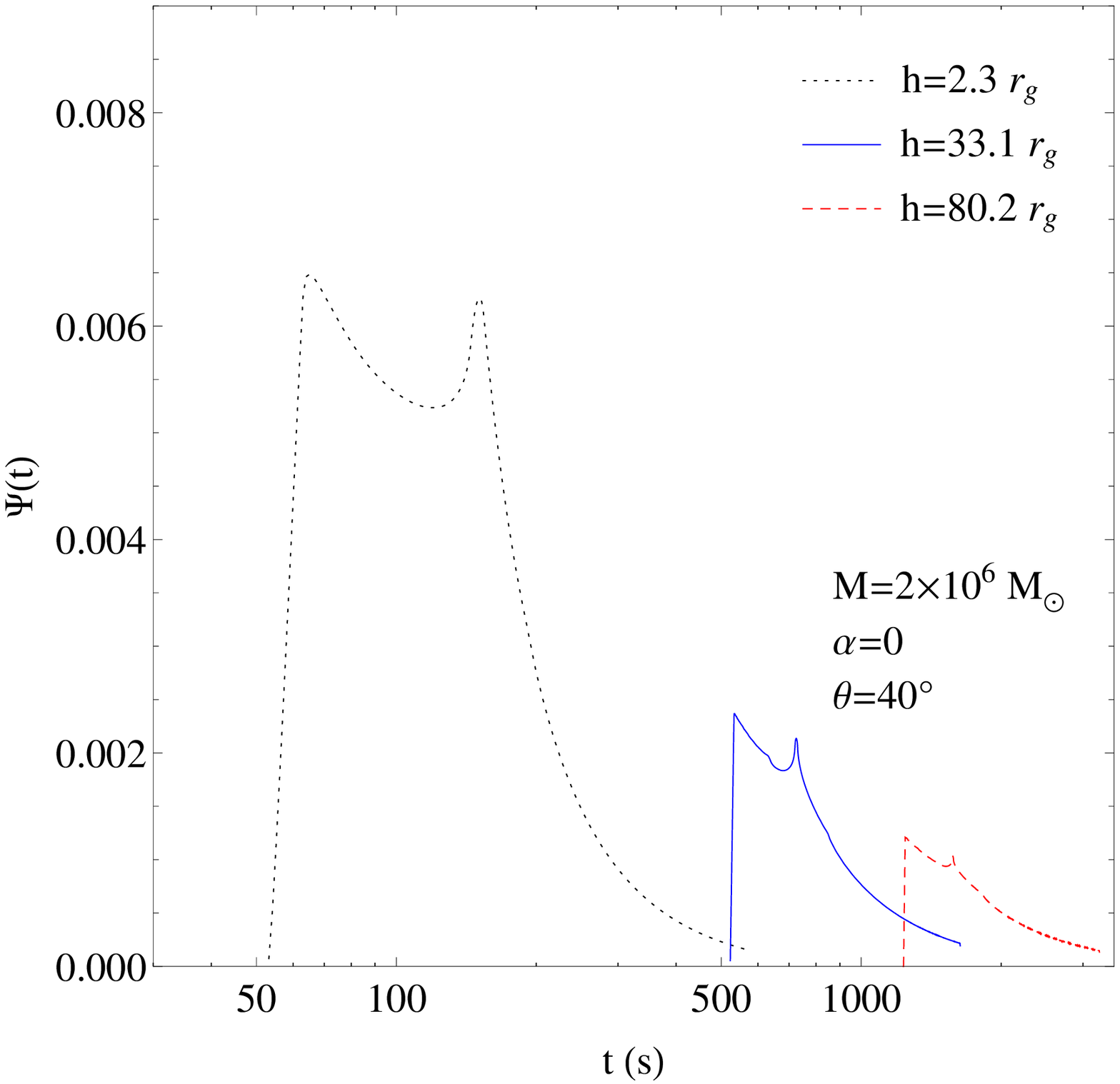}\hspace{4em}
\includegraphics[width=2.5in]{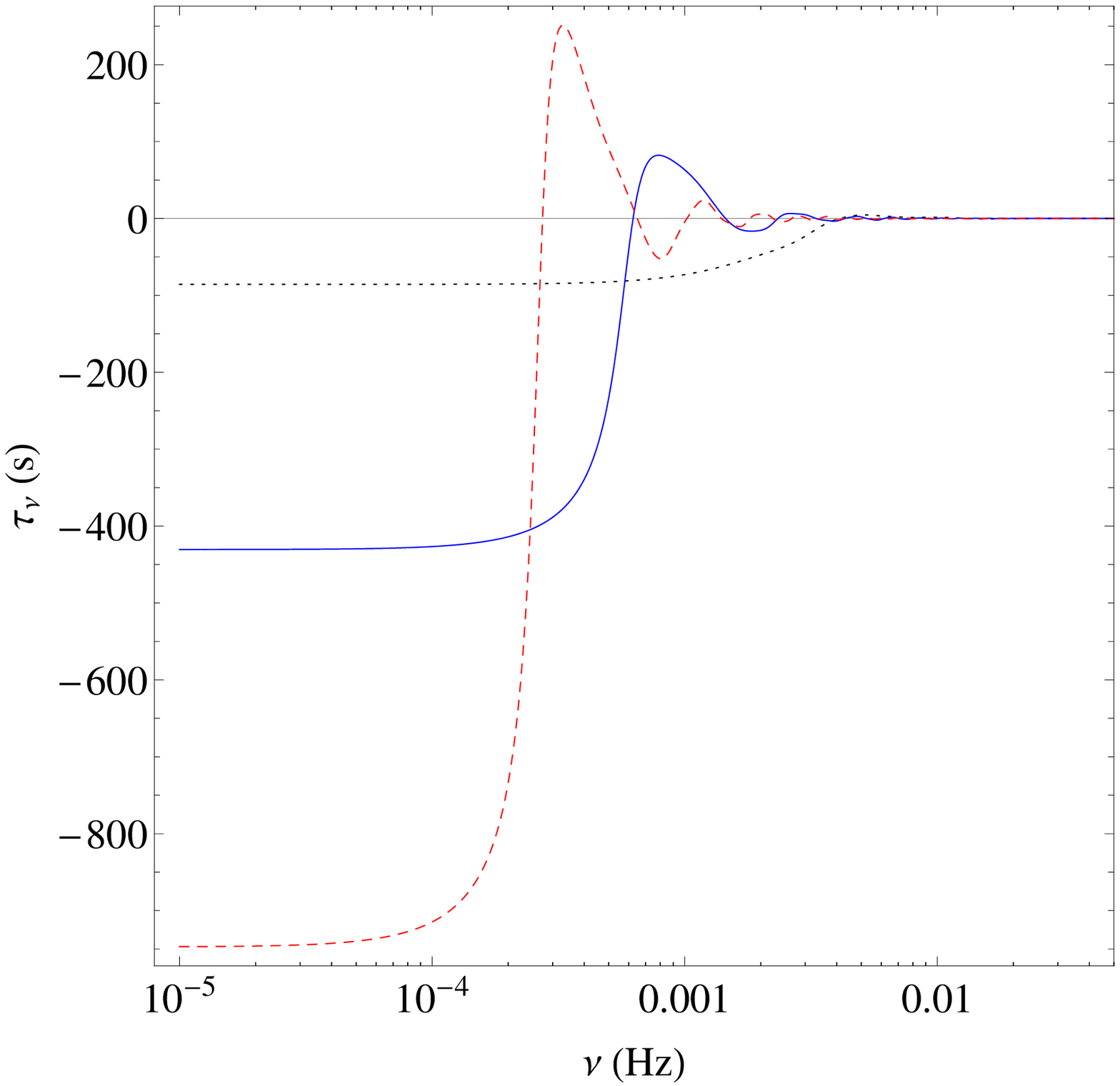}\\[-0.6em]
\caption{X-ray source's height variation for the lamp-post model with $\alpha=0$, $\theta=40\degr$, for $M=2\times10^6$ \ms. Left-hand panel: The GRIRFs for three heights, $h=$2.3, 33.1 and 80.2 \rg. Right-hand panel: The corresponding time-lag spectra.}
\label{fig:height_param}
\end{figure*}

\subsubsection{The viewing angle, $\theta$}
\label{ssect:angle_param}
Here we vary the viewing angle between the three values (i.e.\ $\theta=$20, 40 and $60\degr$). This changes the position of the earliest iso-delay surface that intersects with the accretion disc and thus it affects the onset of the X-ray reverberation phenomenon described by the GRIRF. As we can see in the left-hand panel of Fig.~\ref{fig:angle_param}, the greater the viewing angle the earliest the starting time of the reflection on the disc. Furthermore this changes the ratio height of the two peaks increasing the second peak with decreasing angle. The corresponding time-lag spectra differ significantly from each other, showing distinct features across the the whole frequency range i.e.\ negative plateau and region around the first positive peak.

\subsubsection{The X-ray source height, $h$}
\label{ssect:height_param}
In this section we vary the height of the X-ray source between three values, $h=$2.3, 33.1 and 80.2 \rg. These geometric alterations affect the path-length followed by the hard X-ray photons on their way to the accretion disc. That means that the higher the source the longer the distance (even in the simple Newtonian case) and thus the later the onset of the X-ray reverberation phenomenon. Exactly this behaviour is depicted by the corresponding GRIRFs in the left-hand panel of Fig.~\ref{fig:height_param}. For these cases, the time-lag spectra cover, as expected, different frequency range and they are not `shifted versions' of each other.

\subsection{Interpolated time-lag spectra}
\label{ssect:interpolTLspectra}
\begin{figure}
\scalebox{0.4}{\parbox{1\linewidth}{\includegraphics[trim = 2.3mm 229mm 200.4mm 190mm clip width=1cm]{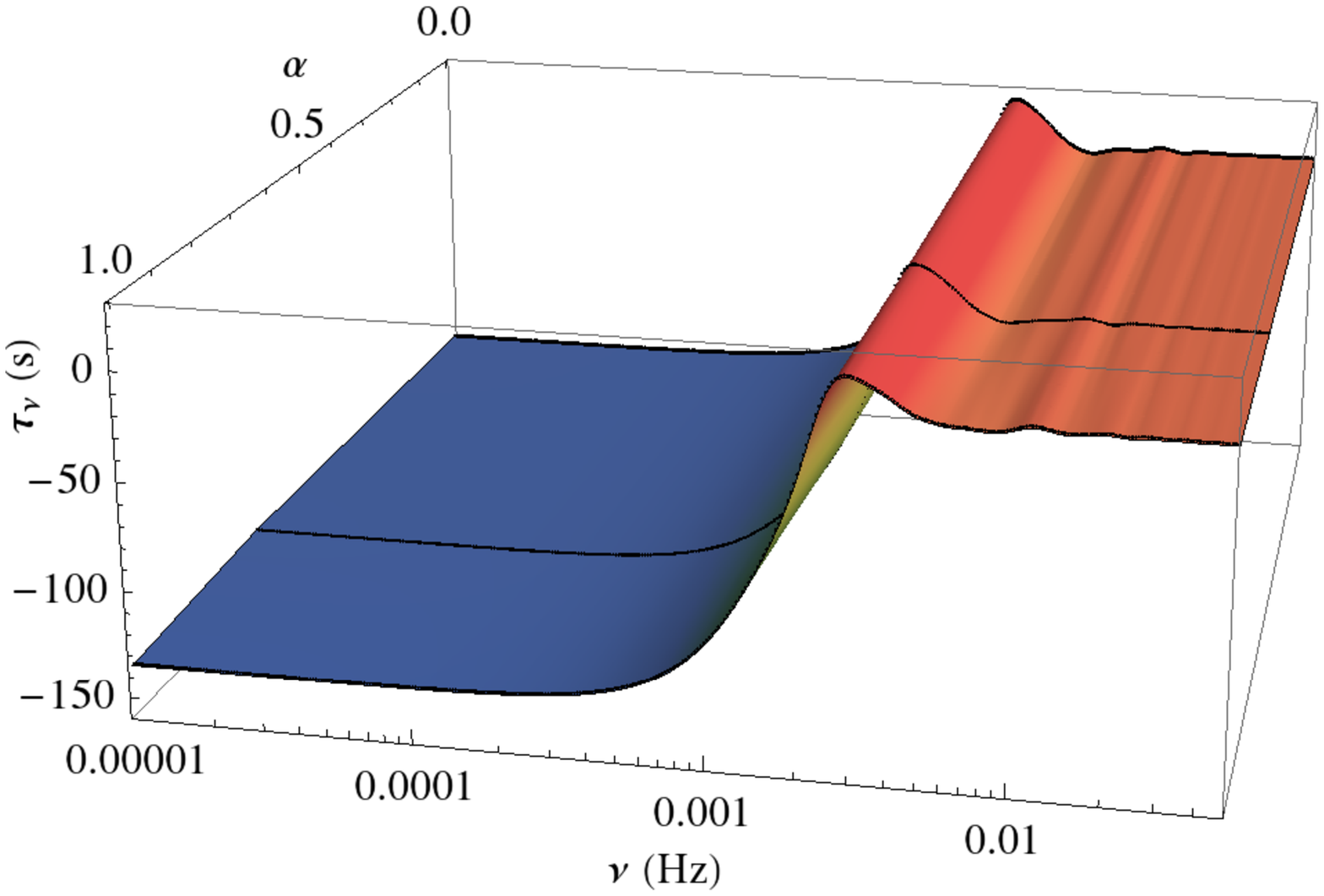}}}\\[18em]
\caption{Spin parameter interpolation for the lamp-post model with $\theta=40\degr$, and $h=7$ \rg, for $M=2\times10^6$ \ms. The cubic interpolation of the various time-lag estimates between the spin parameters 0, 0.676 and 1 indicated by the thick lines on the surface.}
\label{fig:interpolSpin}
\end{figure}

\begin{figure}
\scalebox{0.4}{\parbox{1\linewidth}{\includegraphics[trim = 2.3mm 229mm 200.4mm 190mm clip width=1cm]{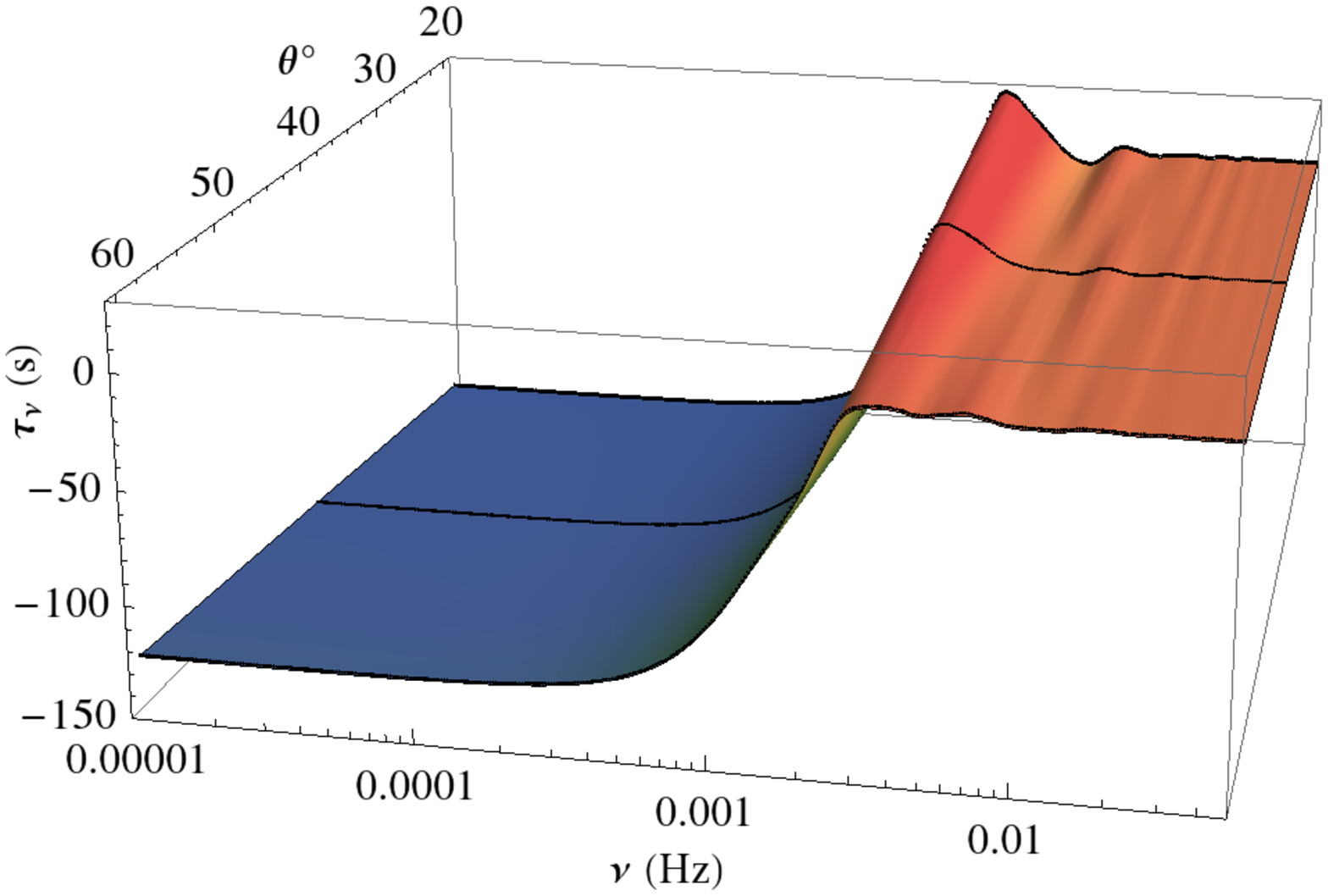}}}\\[18em]
\caption{Viewing angle interpolation for the lamp-post model with $\alpha=0$, and $h=7$ \rg, for $M=2\times10^6$ \ms. The cubic interpolation of the various time-lag estimates between the angles 20, 40 and $60\degr$ indicated by the thick lines on the surface.}
\label{fig:interpolAngle}
\end{figure}

\begin{figure}
\scalebox{0.4}{\parbox{1\linewidth}{\includegraphics[trim = 2.3mm 229mm 200.4mm 190mm clip width=1cm]{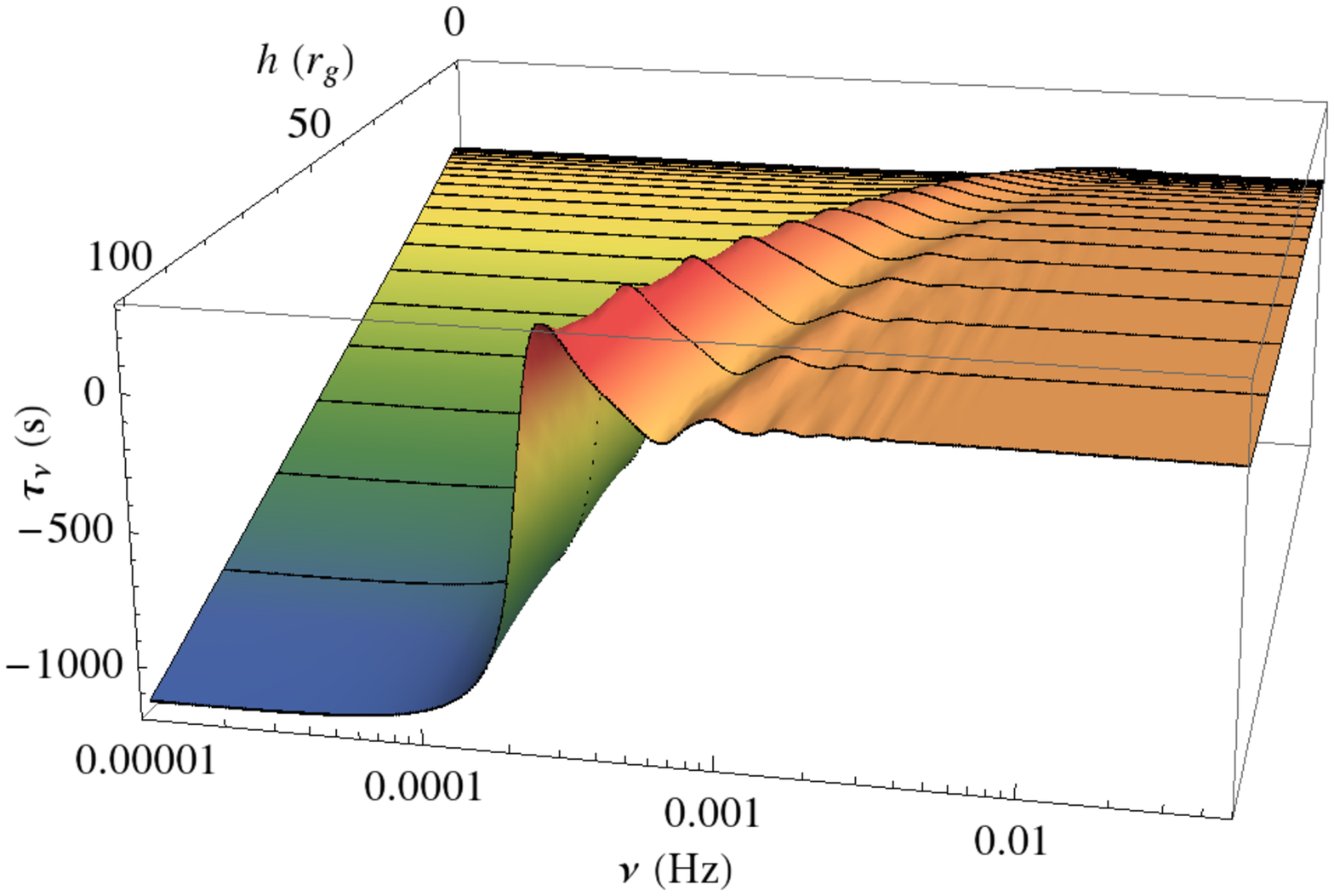}}}\\[18em]
\caption{Height parameter interpolation for the lamp-post model with $\alpha=0$, $\theta=40\degr$, for $M=2\times10^6$ \ms. The cubic interpolation of the various time-lag estimates between the 18 heights from 2.3 to 100 \rg\ indicated by the thick lines on the surface.}
\label{fig:interpolHeight}
\end{figure}

In this section we plot the interpolated versions for the time-lag spectra of the reflected components $\tau_\nu(M,\alpha,\theta,h)$ that are used to create the interpolated version of $\chi^2_k(\mathbf{v})$, $\chi^2(\mathbf{v})$ (Section~\ref{sect:fits}). In Fig.~\ref{fig:interpolAngle}, \ref{fig:interpolSpin} and \ref{fig:interpolHeight} we plot the interpolated versions of the time-lag spectra for the spin, the angle, and the height, respectively.

\bsp
\label{lastpage}
\end{document}